\documentclass[twoside]{ahp}

\usepackage{duplantierBE_h}
\usepackage{amssymb}
\usepackage{amsthm}
\usepackage{amsmath}
\usepackage{amsopn}
\usepackage{amstext}

\input{epsf}
\usepackage{graphicx}

\def\R{\ensuremath{\mathbb R} }

\def\S{\ensuremath{\mathbb S} }

\newcommand{\ds}{\displaystyle}

\newcommand{\be}{\begin{equation}}
\newcommand{\bea}{\begin{eqnarray} \nonumber}
\newcommand{\ee}{\end{equation}}
\newcommand{\eea}{\end{eqnarray}}

\def\(({\left(}
\def\)){\right)}
\def\la{\langle}
\def\ra{\rangle}
\def\[[{\left[}
\def\]]{\right]}

\def \form#1 {eq. (\ref{#1}) }
\def \parziale#1#2  {{\partial {#1} \over \partial {#2}}}

\def \ba#1 {\overline{#1}}

\theoremstyle{plain}

\theoremstyle{definition}

\theoremstyle{remark}

\begin{document}

\title{Brownian Motion, ``Diverse and Undulating''}

\author{Bertrand {\sc Duplantier}\\
}
\maketitle
\vskip-.5cm
\centerline{{\it Translation  by} Emily {\sc Parks} {\it and the} {\sc author} {\it from the original
French text}\footnote{Expanded and updated version (13 May 2007).}}

\begin{abstract}
{\it  Truly man is a marvelously vain, diverse, and undulating object. It is hard
to found any constant and uniform judgment on him.
} {\sc Michel de Montaigne}, Les Essais,
Book I, Chapter 1: ``By diverse means we arrive at the same end''; in {\it The Complete Essays of Montaigne},
Donald M. Frame {\it transl.},
Stanford University Press (1958).\smallskip

{\it Pour distinguer les choses les plus simples de celles qui sont compliqu\'ees et pour les chercher avec ordre,
il faut, dans chaque s\'erie de choses o\`u nous avons d\'eduit directement quelques v\'erit\'es d'autres v\'erit\'es, voir
quelle est la chose la plus simple, et comment toutes les autres en sont plus, ou moins, ou \'egalement \'eloign\'ees.}
{\sc Ren\'e Descartes}, { R\`egles pour la direction de l'esprit,} R\`egle VI.

{\it In order to distinguish what is most simple from what is complex,
and to deal with things in an orderly way, what we must do,
whenever we have a series in which we have directly
deduced a number of truths one from another, is to observe which one
is most simple, and how far all the others are removed from this-whether more,
or less, or equally.} {\sc Ren\'e
Descartes}, {Rules for the Direction of the Mind,} Rule VI.\medskip

{\it Car, supposons, par exemple que quelqu'un fasse quantit\'e de points sur le papier \`a tout hasard,
comme font ceux qui exercent l'art ridicule de la g\'eomance. Je dis qu'il est possible de trouver une ligne
g\'eom\'etrique dont la notion soit constante et uniforme suivant une certaine r\`egle, en sorte que cette ligne passe par tous ces
points, et dans le m\^eme ordre que la main les avaient marqu\'es.

... Mais quand une r\`egle est fort compos\'ee, ce qui luy est conforme, passe pour irr\'egulier.}

{\sc G. W. Leibniz}, Discours de m\'etaphysique, H. Lestienne {\it ed.}, {\sc F\'elix Alcan}, Paris (1907).

{\it Thus, let us assume, for example, that someone jots down a number
of points at random on a piece of paper, as do those  who practice the ridiculous
art of geomancy.\footnote{Note: From {\it g\'eomance}, a way to foretell the future; a form of divination.}
I maintain  that it is possible to find a geometric
line whose notion is constant and uniform, following a  certain
rule, such that this line passes through all the points in the
same order in which the hand jotted them down.

... But, when the rule is extremely complex,
what is in conformity with it passes for irregular.}

{\sc G. W. Leibniz}, Discourse on
Metaphysics.

\medskip

{\it Mens agitat molem}. {\sc Virgil}, AEneid. lib. VI.

\medskip

{\it Un coup de d\'es jamais n'abolira le hasard}. {\sc St\'ephane
Mallarm\'e}, { Cosmopolis}, 1897.

{\it A throw of the dice never will abolish chance}.\medskip

{\it L'antimodernisme, c'est la libert\'e des modernes.} {\sc
Antoine Compagnon}, about his book ``Les antimodernes~: de Joseph de
Maistre \`a Roland Barthes,'' Bi\-blioth\`eque des Id\'ees, Gallimard,
March 2005.

{\it Antimodernism is the liberty of modern men.}\\

Here we briefly describe the history of Brownian motion, as well
as the contributions of Einstein, Sutherland, Smoluchowski, Bachelier, Perrin and Langevin to its
theory.  The always topical importance in physics of the theory of
Brownian motion is illustrated by recent biophysical experiments,
where it serves, for instance, for the measurement of the pulling
force on a single DNA molecule.
     
In the second part, we stress the mathematical importance of the
theory of Brownian motion, illustrated by two chosen examples. The
by-now classic representation of the Newtonian potential
 by Brownian motion is explained in an elementary way. We conclude
with the description of recent progress seen in the geometry of the
planar Brownian curve. At its heart lie the concepts of conformal invariance and
multifractality, associated with the potential
theory of the Brownian curve itself.
\end{abstract}

\section{A brief history of Brownian motion}
Several classic works give a historical view of Brownian motion.
Amongst them, we cite those of Brush,\footnote{Stephen G. Brush, {\it
The Kind of Motion We Call Heat}, Book 2, p. 688, North Holland
(1976).}  Nelson,\footnote{E. Nelson, {\it Dynamical Theories of
Brownian motion}, Princeton University Press (1967), second ed.,
August 2001, http://www.math.princeton.edu/$\sim$nelson/books.html .}
Nye,\footnote{Mary Jo Nye, {\it Molecular Reality: A Perspective on
the Scientific Work of Jean Perrin}, New-York: American Elsevier
(1972).}  Pais\footnote{Abraham Pais, {\it ``Subtle is the Lord...,''
The Science and Life of Albert Einstein}, Oxford University Press
(1982).}, Stachel\footnote{John Stachel, {\it Einstein's Miraculous
Year} (Princeton University Press, Princeton, New Jersey, 1998); {\it
Einstein from `B' to `Z'}, Birkh\"auser, Boston, Basel, Berlin
(2002).} and Wax.\footnote{N. Wax, {\it Selected Papers on Noise and
Stochastic Processes}, New-York, Dover (1954). It contains articles by 
Chandrasekhar, Uhlenbeck and Ornstein, Wang and Uhlenbeck, Rice, Kac,
Doob.}  We also cite a number of essays in
mathematics,\footnote{J.-P. Kahane, Le mouvement brownien : un essai
sur les origines de la th\'eorie math\'ematique, in {\it Mat\'eriaux
pour l'histoire des math\'ematiques au XX\`eme si\`ecle, Actes du
colloque
\`a la m\'emoire de Jean Dieudonn\'e (Nice, 1996)}, volume 3 of {\it
S\'eminaires et congr\`es}, pages 123-155, French Mathematical Society
(1998).}  physics,\footnote{M. D. Haw, {\it J. Phys. C }{\bf 14}, pp. 7769-7779
(2002).} \footnote{R. M. Mazo, {\it Brownian Motion, Fluctuations, Dynamics and Applications},
{International Series of Monographs on Physics} {\bf 112}, Oxford University Press (2002).} especially those which have appeared very recently for the
centenary of Einstein's 1905 articles,\footnote{B. Derrida and \'E. Brunet in {\it Einstein aujourd'hui},
edited by M. Leduc and M. Le Bellac, Savoirs actuels, EDP Sciences/CNRS
Editions (2005); P. H\"anggi {\it et al.}, {\it New J. Phys.} {\bf 7} (2005); J. Renn, {\it Einstein's
invention of Brownian motion}, {\it Ann. d. Phys.} (Leipzig) {\bf 14}, Supplement, pp. 23-37 (2005);
D. Giulini \& N. Straumann,
{\it Einstein's Impact on the Physics of the Twentieth Century},
arXiv:physics/0507107; N. Straumann,
{\it On Einstein's Doctoral Thesis}, arXiv:physics/0504201; S. N. Majumdar, {\it Brownian functionals
in Physics and Computer Science}, {\it Current Science} {\bf 89}, pp. 2075-2092 (2005); J. Bernstein, {\it Einstein
and the existence of atoms}, {\it Am. J. Phys.} {\bf 74}, pp. 863-872 (2006).} and
in biology.\footnote{E. Frey and K. Kroy, {\it Brownian Motion: a Paradigm of Soft Matter and Biological Physics},
{\it Ann. d. Phys.} (Leipzig) {\bf 14}, pp. 20-50 (2005), arXiv:cond-mat/0502602.}

\subsection{Robert Brown}
\subsubsection{Brown's observations and precursors}

Robert Brown (1773-1858), of Scottish descent, was one of the greatest bota\-nists of his time
in Great Britain.  He is renowned for his discovery of the nucleus of plant cells, for being the first to recognize the
phenomenon of cytoplasmic streaming,
and for the classification of several thousands dried plant specimens he brought back to
England from a trip to Australia in 1801-1805.

In 1801 indeed, at the age of twenty-eight, he was
chosen by Sir Joseph Banks as the botanist to accompany Matthew Flinders
in the {\it Investigator} on the first
 circumnavigation of the Australian continent. The voyage
was to extend over 5 years, and Brown used his time well, assembling substantial collections of plants,
animals and minerals, and kept a diary.\footnote{A rivulet south of Hobart in Southern Tasmania, {\it Browns River}, is named after him
(as mentioned by Bruce H. J. McKellar, in
{\it Einstein, Sutherland, Atoms, and Brownian Motion}, Einstein International Year of Physics 2005,
Melbourne AAPPS Conference, July 2005, http://www.ph.unimelb.edu.au/). See also {\it Some aspects of the
work of the botanist Robert Brown (1773-1858) in Tasmania in 1804}, {\it Tasforests}, Vol. 12, pp. 123-146 (2000).}

Brown returned to England with his scientific reputation established. As said by Brown's biographer,
D. J. Mabberley,\footnote{D. J. Mabberley, {\it Jupiter Botanicus: Robert Brown of the British Museum},
Lubrecht \& Cramer Ltd (1985).}
Brown's Australian experiences and connections with the Continental schools of scientific thought moulded his research, with
the result that he was recognized as one of the great European intellectuals of his day.
Brown was called by Humboldt {\it ``Princeps Botanicorum.''}

In a pamphlet, dated July 30th, 1828, first printed privately,\footnote{R. Brown, {\it A Brief Account of Microscopical Observations Made in
the Months of June, July, and August, 1827, on the Particles Contained
in the Pollen of Plants; and on the General Existence of Active
Molecules in Organic and Inorganic Bodies}, July 30th, 1828 [Not Published]. It can be found in: R. Brown, {\it The miscellaneous
botanical works of Robert Brown}, Vol. 1, pp. 464-479, John J. Bennett, ed., R. Hardwicke, London (1866); available online at:
http://sciweb.nybg.org/science2/pdfs/dws/Brownian.pdf .} then published in the {\it Edinburgh New Philosophical Journal}
 later that year,\footnote{R. Brown,
{\it Edinburgh New Phil. J.} {\bf 5}, pp. 358-371 (1828).}
and republished several times elsewhere,\footnote{R. Brown, {\it Ann.
Sci. Naturelles}, (Paris) {\bf 14}, pp. 341-362 (1828); {\it Phil. Mag.} {\bf
4}, pp. 161-173 (1828); {\it Ann. d. Phys. u. Chem.} {\bf 14}, pp. 294-313 (1828).}
entitled {\it ``A Brief Account of Microscopical Observations Made in
the Months of June, July, and August, 1827, on the Particles Contained
in the Pollen of Plants; and on the General Existence of Active
Molecules in Organic and Inorganic Bodies,''}
Brown reported on the random movement of different particles that
are small enough to be in suspension in water.  It is an extremely
erratic motion, apparently without end (see figure 1).\footnote{One can find examples of real Brownian motion at:
www.lpthe.jussieu.fr/poincare/.} A second article, dated July 28th, 1829, was published later and bears the brief title
{\it ``Additional Remarks on Active Molecules.''}\footnote{R. Brown, {\it Additional Remarks on Active Molecules},
{\it Edinburgh Journal of Science}, {\bf 1}, new series, pp. 314-319 (1829); {\it Phil. Mag.} {\bf 6}, pp. 161-166 (1829);
in: R. Brown, {\it The miscellaneous
botanical works of Robert Brown}, Vol. 1, pp. 479-486, John J. Bennett, ed., R. Hardwicke, London (1866); available online at:
http://sciweb.nybg.org/science2/pdfs/dws/Brownian.pdf .}

Brown used the wording {\it active molecule} in these titles in a sense different from its current one. It referred to earlier
 teaching of Georges-Louis Leclerc de Buffon (1707-1788) who introduced this word for the hypothetical ultimate constituents of the bodies of living beings.
 Only later with the acceptance and development of Dalton's 1803 atomic theory the word {\it molecule} was going to take on
 its modern meaning.

The first plant Brown studied was {\it Clarkia pulchella}, whose
pollen grains contain granules varying ``from nearly $\frac{1}{4000}$ to about $\frac{1}{3000}$ of an inch in length, and of a
figure between cylindrical and oblong, perhaps slightly flattened...''
[from about six to eight microns]. It is these granules, not
the whole pollen grains, upon which Brown made his observations. Concerning them, he wrote:\smallskip

{\small ``While examining the form of these particles immersed in water, I observed many of them very evidently in
motion; their motion consisting not only of a change of place in the fluid, manifested by alterations in their
relative positions, but also not unfrequently of a change of form in the particle itself; a contraction or curvature
taking place repeatedly about the middle of one side, accompanied by a correspondong swelling or convexity on the
opposite side of the particle. In a few instances, the particle was seen to turn on its longer axis.
These motions were such as to satisfy me, after frequently repeated observation, that they arose neither
from currents in the fluid, nor from its gradual evaporation, but belonged to the particle itself.''}\medskip

\begin{figure}[tb]
\begin{center}
\includegraphics[angle=0,width=.4\linewidth]{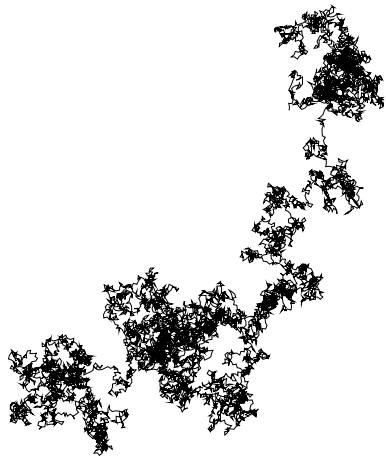}
\end{center}
\caption{{\it Brownian motion described by a pollen
granule in suspension.}}
\label{fig.brownian}
\end{figure}

Brown made his observations just after the introduction of the first compound achromatic objectives for microscopes. Still,
he used a simple microscope with a double convex lens, while he also possessed a pocket microscope with two lenses
having a delicate adjustment:\smallskip

{\small ``The observations, of which it is my intention to give a summary in the following pages, have all been made
with a simple microscope, and indeed with one and the same lens, the focal length of which is about $\frac{1}{32}$nd of
an inch.''}\footnote{It has been sometimes believed that Brown's attention was directed to the movement of pollen grains
themselves, and it has been even claimed that his microscope was not sufficiently developed for
the observation of such a diminutive phenomenon
[D. H. Deutsch, {\it Did Brown observe Brownian Motion: probably not}, {\it Bulletin of the APS} {\bf 36}, 1374 (1991).
Reported in {\it Scientific American}, {\bf 265}, 20, August 1991]. The first observations by Brown were then recreated in 1992 by Brian J. Ford with
Brown's original microscope, pollen grains of {\it Clarkia pulchella}, and also carried out in the month of June!
The phenomenon of
Brownian motion was indeed well resolved by the original microscope lens. See {\it Brownian Movement in Clarkia Pollen:
A Reprise of the First Observation}, {\it The Microscope}, {\bf 40}, pp. 235-241 (1992)
[available online at: http://www.brianjford.com/wbbrowna.htm].}\smallskip

\noindent In fact, it is nowadays sufficient to look in a microscope to see small
objects dancing.
\medskip

Brown may have not been the first, however, to observe Brownian motion. In fact, he discussed in his
second article (1829)\footnote{R. Brown, {\it Additional Remarks on Active Molecules},
{\it op. cit.}} previous observations by others which could have been interpreted as prior to his. 
The universal
and irregular motion of small grains
suspended in a fluid may have been observed soon after the discovery of the
microscope.

The story begins with Anthony van Leeuwenhoek (1632-1723),
a famous constructor of microscopes in Delft, who in 1676 was also
designated executor of the estate of the no-less-famous
painter Johannes Vermeer, who was apparently a personal friend.\footnote{Although
no document exists testifying a relationship between Vermeer and van
Leeuwenhoek, it seems impossible that they did not know one another.
The two men were born in Delft the same year, their respective
families were involved in the textile business and they were both
fascinated by science and optics.  A commonly accepted and probable
hypothesis is that Anthony van Leeuwenhoek was in fact a model for
Vermeer, and perhaps also the source of his scientific information, for
the two famous scientific portraits, {\it The Astronomer}, 1668,
(Louvre Museum, Paris), and {\it The Geographer}, 1668-69,
(St\"adelsches Kunstinstitut am Main, Frankfurt). (See {\it Johannes
Vermeer}, B. Broos et al., National Gallery of Art, Washington,
Mauritshuis, The Hague, Waanders Publishers, Zwolle (1995).)} Leeuwenhoek
built several hundred simple ``microscopes,'' with which he
went as far as to observe living bacteria and discover the existence of spermatozoids.

In his second article, Brown then writes:\smallskip

{\small ``I shall conclude these supplementary remarks to my former Observations, by noticing the degree in which I consider
those observations to have been anticipated.}

{\small That molecular was sometimes confounded with animalcular motion by several of the earlier
 microscopical observers, appears extremely probable from various passages in the writings of Leeuwenhoek, as well as
 from  a very interesting Paper by Stephen Gray, published in the 19th volume of the Philosophical
 Transactions.''}\footnote{S. Gray, {\it Microscopical observations and experiments}, {\it Phil. Trans.} {\bf 19}, 280 (1696).}
\medskip

Next, one meets Needham, Buffon and Spallanzani, the 18th-century
protagonists of the debate on  spontaneous generation.\footnote{Jean Perrin,
in his book {\it Les Atomes} ({\it Atoms}, translated by D. Ll. Hammick, Ox Bow Press, Woodbridge (1990)),
writes: ``Buffon and Spallanzani knew of the
phenomenon but, possibly owing to the lack of good microscopes, they did not
grasp its nature and regarded the ``dancing particles'' as rudimentary animalculae
(Ramsey: Bristol Naturalists' Society, 1881).''}
 Brown continues:\smallskip

{\small ``Needham also, and Buffon, with whom the hypothesis of organic particles originated, seem to have not unfrequently
fallen into the same mistake. And I am inclined to believe that Spallanzani, notwithstanding one
of his statements respecting them, has under the head of {\it Animalculetti d'ultimo ordine} included the active Molecules
as well as true Animalcules.''}\medskip

Brown further cites Gleichen, ``the discoverer of the motions of the Particles of the Pollen,'' Wrisberg and M\"uller,
who, having  ``adopted in part Buffon's hypothesis, state the globules, of which they suppose all organic bodies formed,
to be capable of motion;''
and M\"uller, who ``distinguishes these moving organic globules from real Animalcules, with which, he adds, they have been
confounded by some very respectable observers.''
Lastly, he cites a ``very valuable Paper'' published in 1814 by Dr. James Drummond, of Belfast, which ``gives an account
of the very remarkable motions of the spicula which form the
silvery part of the choroid coat of the eyes of fishes,'' and where ``The appearances are minutely described, and very
ingenious reasoning employed to show that, to account for the motions, the least improbable conjecture is to suppose the
spicula animated.''

However, all these works had confined themselves to the examination of the particles
of some organic bodies. Only
Bywater, of Liverpool, is cited by Brown, in the same second article {\it Additional Remarks on Active Molecules}, for having stated in
1819  that ``not only organic tissues, but also
inorganic substances, consist of what he terms animated or irritable particles,''
and therefore are subject to ``Brownian motion.'' However, Brown adds:\smallskip

{\small ``I believe that in thus stating the manner in which Mr. Bywater's experiments
were conducted, I have enabled microscopical observers to judge of the extent and kind of optical illusion
to which he was liable, and of which he does not seem to have been aware.''}\smallskip

As pointed out by R. M. Mazo,\footnote{R. M. Mazo, {\it op. cit.}} when citing the work
by Van der Pas,\footnote{P. W. Van der Pas, {\it The discovery of Brownian motion}, {\it Scien. Historiae} {\bf 13}, 17 (1971).}
there was, however, one predecessor that Brown overlooked. In July of 1784, Jan Ingen-Housz
published a short article entitled {\it Remarks on the use of the microscope},\footnote{J. Ingen-Housz, in
{\it Vermischte Schriften physisch-medizinschen Inhalts}, C. F. Wappler, Vienna (1789).} that
contains the following lines:\footnote{Van der Pas' translation from the original French.}\smallskip

{\small{`` ... one must agree that, as long as the droplet lasts, the entire liquid and consequently everything which
is contained in it, is kept in continuous motion by the evaporation, and that this motion can give the impression that some of
the corpuscles are living, even if they have not the slightest life in them. To see clearly how one can deceive one's mind
on this point if one is not careful, one has only to place a drop of alcohol at the focal point of a microscope and introduce
a little finely ground charcoal therein, and one will see these corpuscles in a confused continuous and violent motion
as if they were animalcules which move violently around.''}}\medskip

\noindent However, although Ingen-Housz doubtless observed the motion, he ascribed it to evaporation and did not follow up
his observation with any investigation of it.\medskip

Lastly, in 1827, one year before the publication by Brown, similar
observations  were also alluded to in France by the young Adolphe
Brongniart (1801-1876), in  a long Memoir\footnote{Adolphe Brongniart, {\sc M\'emoire} {\it sur la G\'en\'eration et le D\'eveloppement de l'Embryon dans les
v\'eg\'etaux phan\'erogames}, {\it Ann. Sci. Naturelles} (Paris)
{\bf 12}, pp. 41-53, pp. 145-172, pp. 225-296 (1827).} for which he won a Prize in experimental physiology from the
French Academy of Sciences.\footnote{{\it Ann. Sci. Naturelles} (Paris) {\bf 12}, pp. 296-298 (1827).}
  Brongniart's
findings about the motion of particles appear in a particular paragraph, followed by a note later annexed to his
memoir.\footnote{{\it Ann. Sci. Naturelles} (Paris), {\it loc. cit.}, see pp. 42-46 and the added footnote (B) therein.}
 The note below reproduces the original passages.\footnote{In the original text, on p. 44, Brongniart writes:

 ``N'ayant pu d\'ecouvrir ce mouvement dans l'int\'erieur des globules de pollen ou dans leur
 appendice, j'ai cherch\'e \`a l'observer dans les granules r\'epandus dans l'eau apr\`es la rupture des grains de pollen.
 J'avoue que dans plusieurs cas j'ai cru voir de l\'egers mouvemen[t]s dans les granules du pollen du Potiron, des mauves,
 etc. ; mais ces mouvemen[t]s \'etaient si lents, si peu suivis, que [...] je n'ai jamais pu avoir la certitude qu'ils fussent
 spontan\'es. Le mouvement de ces petits corps n'\'etait pas une sorte de tournoiement et de translation comme celui des
 Monades et autres animalcules infusoires ; mais un simple rapprochement ou un l\'eger changement de position relative,
 fort lent, qui cessait bient\^ot pour reprendre quelques temps apr\`es.''

 But in the added footnote (B) one reads:

 ``J'ai fait cette ann\'ee de nouvelles observations sur ce sujet, au moyen du microscope d'Amici, et ces observations me paraissent
 lever presque tous les doutes \`a l'\'egard du mouvement des granules spermatiques. [...] ce m\^eme grossissement permet
 de reconna\^{\i}tre dans les granules spermatiques de plusieurs plantes des mouvemen[t]s tr\`es-appr\'eciables, et
 qu'il para\^{\i}t impossible d'attribuer \`a aucune cause exr\'erieure. [...]

 Dans le Potiron, le mouvement des granules
 consiste dans une oscillation lente, qui les fait changer
 de position respective ou qui les rapproche et les \'eloigne comme par l'effet d'une sorte d'attraction et de r\'epulsion.
 L'agitation du liquide dans lequel ces granules nagent, ne para\^{\i}t pas pouvoir influer en rien sur ce mouvement [...].

 Les mouvemen[t]s de ces granules deviennent bien plus distincts, et ne peuvent plus laisser de doute, lorsqu'on les observe sur les
 Malvac\'ees [...] ; dans ces plantes, les granules spermatiques, beaucoup plus gros, sont oblongs, et ce qui prouve que
 les mouvemen[t]s tr\`es-distincts ne sont pas dus au mouvement du liquide environnant, c'est qu'on les voit souvent
 changer de forme, se courber soit en arc, soit m\^eme en S, comme les Vibrio. Ces mouvemen[t]s \'etaient
 quelquefois si marqu\'es, qu'il m'\'etait impossible de suivre avec la pointe du crayon les contours de ces granules,
 que je voulais dessiner \`a la {\it Camera lucida}, et que je fus oblig\'e pour y parvenir d'attendre que l'eau
 f\^ut presque compl\`etement \'evapor\'ee, ou de saisir des momen[t]s o\`u le mouvement cessait; ce qui a souvent lieu
 pendant des intervalles assez longs.

 Dans une esp\`ece de Rose ({\it Rosa bracteata)}, ces mouvemen[t]s \'etaient d'autant plus distincts, que les granules,
 de forme elliptique et lenticulaire, se pr\'esentaient successivement sous leurs diverses faces.'' }

   It is interesting to notice that Brown actually discussed Brongniart's work in detail in the last
 two pages of his famous 1828 article. There Brown first acknowledges that he was acquainted, before he engaged in his own
  inquiry in 1827, with the abstract of Brongniart's work that was given to him by the author
  himself.\footnote{In {\it  Microscopical
  Observations of Active Molecules}, {\it op. cit.}, Brown writes: ``Before I engaged in the inquiry in 1827, I was acquainted only with
  the abstract  given by M. Adolphe Brongniart himself, of a very elaborate and valuable memoir, entitled
  {\it ``Recherches sur la G\'en\'eration et le D\'eveloppement de l'Embryon dans les V\'eg\'etaux Phan\'erogames,''}
  which he had then read before the Academy of Sciences of Paris, and had since published in
  the {\it Annales des Sciences Naturelles.''}} He nevertheless stresses the lack in this work of observations of importance on the
  motion or form of the particles:\smallskip

  {\small ``Neither in the abstract referred to, nor in the body of the memoir which M. Brongniart has with great candour given
  in its original state, are there any observations, appearing of importance even to the author himself, on the motion or form
  of the particles [...]''}\smallskip

 \noindent  But Brown adds about the note annexed by Brongniart in his article:\smallskip

  {\small ``Late in the autumn of 1827,\footnote{Hence
  {\it after} Brown's own observations during the Summer of the same year [Note of the author].} however, M. Brongniart having at his command a
  microscope constructed by Amici, the celebrated professor of Modena, he was enabled to ascertain many important facts on
  both these points, the result of which he has given in the notes annexed to his memoir. On the general accuracy
  of his observations on the motions, form, and size of the granules, as he terms the particles,
  I place great reliance.''}\smallskip

\noindent This is followed by some criticism of more physiological relevance, to which Brongniart himself
replied
in a note added to the French translation of the same article by Brown in the {\it Annales de Sciences Naturelles}!\footnote{R. Brown, {\it Ann.
Sci. Naturelles} (Paris) {\bf 14}, pp. 341-362 (1828); see pp. 361-362.}\bigskip

\subsubsection{``Active Molecules'' or Brownian motion?}
Brown's first publication
on the erratic motion of the granules of pollen grains garnered much attention, but the use
of the ambiguous terms {\it ``active molecules''} by Brown brought him
criticisms based on some misunderstanding.  Indeed, under the influence
of Buffon, the similar expression {\it ``organic molecules''}
represented hypothetical entities, elementary bricks all living beings
would be made of. Such theories were still around at the
beginning of the 19th century. In his famous first paper, Brown writes:\smallskip

{\small `` Reflecting on all the facts with which I had now become acquainted, I was disposed to believe that
the minute spherical particles or Molecules of apparently uniform size, first seen in the advanced state of the
pollen [...] and lastly in bruised portions of other parts of the same plants, were in reality the supposed constituent or
elementary Molecules of organic bodies, first so considered by Buffon and Needham, then by Wrisberg
with greater precision, soon after and still more particularly by M\"uller, and, very recently, by Dr. Milne Edwards, who has
revived the doctrine and supported it with much interesting detail.''}\medskip

 However, one of the substances he examined,
{\it silicified} wood, once bruised
still produced spherical particles, or molecules, in all respect like those mentioned before, and in such quantity, that,
according to Brown,\medskip

{\small ``the whole substance of the petrifaction seemed
to be formed of them. But hence I inferred that these molecules were not limited to organic bodies, not even
to their products.}

{\small To establish the correctness of the inference, and to ascertain to what extent the molecules existed in mineral bodies,
became the next body of inquiry. The first substance examined was a minute fragment of window-glass, from which, when merely
bruised on the stage of the microscope, I readily and copiously obtained molecules agreeing in size, form, and motion with
those which I had already seen. [...]}

{\small Rocks of all ages, including those in which organic remains have never been found, yielded the molecules
in abundance. Their existence was ascertained in each of the constituent minerals of granite,
a fragment of the Sphinx being one of the specimens examined.''}\smallskip

{\small In a word, in every mineral which I could reduce to a powder, sufficiently
fine to be temporarily suspended in water, I found these molecules more or less copiously ...''}\medskip

\noindent His emphasis leads one to think that Brown's
opinion was that the observed particles themselves were animated. Faraday
himself had to defend him during a Friday night lesson he gave at the
Royal Society on February 21, 1829, about Brownian
motion.\footnote{S. G. Brush, {\it The Kind of Motion We Call Heat},
Book 2, p. 688, North Holland (1976).}

This led Brown in his Supplement {\it Additional Remarks on Active Molecules} to an apology:\smallskip

{\small ``In the first place, I have to notice an erroneous assertion of more than one writer, namely, that I have stated the active
Molecules to be animated. This mistake has probably arisen from my having communicated the facts in the same order
in which they occurred, accompanied by the views which presented themselves in the different stages of the
investigation; and in one case, from my having adopted the language, in referring to the opinion, of another inquirer into the
first branch of the subject.}

{\small Although I endeavoured strictly to confine myself to the statement of the facts observed, yet in speaking of the
active Molecules, I have not been able, in all cases, to avoid the introduction of hypothesis; for such is the supposition
that the equally active particles of greater size, and frequently of very different form, are primary compounds of
these Molecules, --a supposition which, though professedly conjectural, I regret having so much insisted on, especially as it may
seem connected with the opinion of the absolute identity of the Molecules, from whatever source derived.''}\medskip

 Brown's merit was in gradually emancipating himself from this
misconception and in making a systematic study of the ubiquity of ``active molecules,'' hence of the movement named
after him, with grains of pollen, dust and soot, pulverized rock, and
even a fragment from the Great Sphinx! This served to eliminate
 the ``vital force'' hypothesis, where the movement was
reserved to organic particles.

As for the nature of Brownian motion,
even if Brown could not explain it, he eliminated easy explanations,
like those linked to convection currents or to evaporation,
by showing that the Brownian motion of a {\it simple} particle stayed
``tireless'' even in a isolated drop of water in oil!  On the same occasion
he eliminated as well the hypothesis of movements created by
interactions between Brownian particles, a hypothesis that would
nevertheless reappear later. He wrote in his {\it Additional Remarks on Active Molecules}:\smallskip

{\small ``I have formerly stated my belief that these motions of the particles neither arose from currents in the
fluid containing them, nor depended on that intestine motion which may be supposed to accompany its evaporation.}

{\small These causes of motion, however, either singly or combined with others, --as, the attractions and repulsions
among the particles themselves, their unstable equilibrium in the fluid in which they are suspended, their hygrometrical
or capillary action, and in some cases the disengagement of volatile matter, or of minute air bubbles,-- have been
considered by several writers as sufficiently accounting for the appearances. [...] the insufficiency of
the most important of those enumerated may, I think, be satisfactorily shown by means of a very simple experiment.}

{\small The experiment consists in reducing the drop of water containing the particles to microscopic minuteness, and
prolonging its existence by immersing it in a transparent fluid of inferior specific gravity, with which it is not
miscible, and  in which evaporation is extremely slow. If to almond-oil, which is a fluid having these properties, a
considerably smaller proportion of water, duly impregnated with particles, be added, and the two fluids shaken or
triturated together, drops of water of various sizes [...] will be immediately produced. Of these, the most minute necessarily contain
but few particles, and some may be occasionally observed with one particle only. [...] But in all the drops thus formed and protected, the motion of the particles takes place with
undiminished activity, while the principal causes assigned for that motion, namely, evaporation, and their
mutual attraction and repulsion, are either materially reduced or absolutely null.''}\medskip

This ingenious experimental set-up gave him some hope of getting closer to the real cause of Brownian motion:\smallskip

{\small ``By means of the contrivance now described for reducing the size and prolonging the existence of the drops
containing the particles, which, simple as it is, did not till very lately occur to me, a greater
command of the subject is obtained, sufficient perhaps to enable us to ascertain the real cause
of the motions in question.''}\medskip

Still, this real cause always eluded him. The theoretical picture formed
perhaps by Brown, which however he always carefully avoided presenting as
the conclusion of his studies, was that the particles of matter
were animated into a rapid and irregular movement whose source was in the
particles themselves and not in the surrounding fluid.

It is nevertheless fascinating to observe that in some instances he
came close to the truth. One reads indeed in {\it Microscopical Observations of Active Molecules}
the following striking remark:\smallskip

{\small ``In {\it Asclepiade{\ae}}, strictly so called, the mass of pollen filling each cell of the anthera is in
no stage separable into distinct grains; but within, its tesselated or cellular membrane is filled with spherical particles,
commonly of two sizes. Both these kinds of particles when immersed in water are generally seen in vivid motion;
but the apparent motions of the larger particle might in these cases perhaps be caused by the rapid
oscillation of the more numerous molecules.''}\smallskip

\noindent This is precisely the correct explanation of the cause of the movement, if one mentally replaces the latter ``numerous molecules,'' i.e.,
the smaller granules as observed by Brown in this pollen, by the {\it invisible} numerous {\it real} molecules of the surrounding fluid!

 In the same Princeps article, Brown also wondered whether the mobility of the particles existing in bodies was in
 any degree affected by the application of intense heat to the containing substance:\smallskip

{\small ``... and in all these bodies so heated, quenched in water, and immediately submitted to examination, the
molecules were found, and in as evident motion as those obtained from the same substances before burning.''}\smallskip

\noindent  After heating of the substance, instead of a {\it ``quenching''} of the latter in the fluid,
had an {\it ``annealing''} of the whole
system been performed, which would have transferred heat to the surrounding fluid at equilibrium,
an additional increase of Brownian activity with temperature would indeed have occurred!
\medskip

The outstanding scientific stature of Brown brought him elogious comments.  Before leaving
Robert Brown, I cannot refrain from quoting first Mrs Charles Darwin, who said about a dinner party in
1839:\footnote{E. J. Browne, {\it Charles Darwin: Voyaging, Volume 1 of a biography}, Knopf, New York (1950); quoted by
R. M. Mazo, in {\it Brownian Motion, Fluctuations, Dynamics and Applications}, {\it op. cit.}}\smallskip

{\footnotesize ``Mr. Brown, whom Humboldt calls `the glory of Great Britain' looks so shy, as if forced
to shrink into himself, and disappear entirely.''}\smallskip

\noindent Finally, Charles Darwin gave, in his famous autobiographical notes written for his children, his own
recollection from Brown in the late 1830's:\footnote{{\it Charles Darwin: His
Life told in an autobiographical Chapter, and in a selected series of
his published letters}, ed. by his son, Francis Darwin, London
(1892); D. Appleton \& Co., New York (1905), vol. I, chapter 2, pp. 56-57 \& pp. 60-61, available online at:
http://pages.britishlibrary.net/charles.darwin/texts/letters/letters1$_{-}$02.html ; see also {\it The autobiography of Charles Darwin, 1809-1882: with
original omissions restored},
Nora Barlow, ed., W. W. Norton, New York (1969),  available online at:
http://pages.britishlibrary.net/charles.darwin3/barlow.html ; see also Schuman (1950), p. 46;
quoted by S. G. Brush in {\it The Kind of Motion We Call Heat, op. cit.}}
\smallskip

{\footnotesize{``During this time [March 1837-January 1839] I saw also a good deal of Robert Brown; I used often to call and sit
with him during his breakfast on Sunday mornings, and he poured forth a rich treasure of curious
observations and acute remarks, but they almost always related to minute points, and he never with me discussed large or
 general questions in science. [...]''}}\smallskip

\noindent and\footnote{The slight repetition here observable is accounted for by these notes having been added in April,
1881, a few years after the rest of the 'Recollections' were written.}

\smallskip
{\footnotesize{``I saw a good deal of Robert Brown, {\it ``facile Princeps
Botanicorum,''} as he was called by Humboldt. He seemed to me to be
chiefly remarkable by the minuteness of his observations, and their
perfect accuracy. His knowledge was extraordinarily great, and much died
with him, owing to his excessive fear of ever making a mistake. He poured out
his knowledge to me in the most unreserved manner, yet was strangely
jealous on some points. I called on him two or three times before the
voyage of the {\it Beagle} [1831-1836], and on one occasion he asked me to
look through a microscope and describe what I saw. This I did, and
believe now that it was the marvellous currents of protoplasm in some
vegetable cell. I then asked him what I had seen; but he answered me,
{\it ``That is my little secret.''}}}

\noindent {\footnotesize{He was capable of the most generous actions. When old, much out of health, and
quite unfit for any exertion, he daily visited (as Hooker told me) an old man-servant, who lived at a distance
(and whom he supported), and read aloud to him. This is enough to make up for any degree of
scientific penuriousness or jealousy.''}}


\subsection{The period before Einstein}
Between 1831 and 1857 it seems that one can no longer find references to
Brown's observations, but from the 1860's forward his work began to
draw large interest.  It was noticed soon thereafter in literary circles, if
we are to judge by a passage of {\it ``Middlemarch''} published by George
Eliot in 1872, where Rev. M. Farebrother offers to make an exchange to the surgeon Lydgate:
``I have some sea-mice -- fine specimens -- in spirits. And I will throw in Robert Brown's new thing
-- {\it Microscopic Observations on the
Pollen of Plants} -- if you don't happen to have it already."

Jean Perrin wrote in his famous 1909 memoir {\it Brownian Motion and Molecular Reality}:\footnote{J. Perrin,
{\it Mouvement brownien et r\'ealit\'e mol\'eculaire}, {\it Ann. de Chim. et de Phys.}  {\bf
18}, pp. 1-114 (1909). Translated by Frederick Soddy in {\it Brownian Motion and Molecular Reality},
Taylor and Francis, London (1910); facsimile reprint in David M. Knight, ed., {\it Classical scientific papers: chemistry},
American Elsevier, New York (1968).}\smallskip

{\small ``The singular phenomenon discovered by Brown did not attract much attention. It remained,
moreover, for a long time ignored by the majority of physicists, and it may be supposed that those who 
had heard of it thought it analogous to the movement of the dust particles, which can be seen dancing
in a ray of sunlight, under the influence of feeble currents of air which set up small differences
of pressure or temperature. When we reflect that this apparent explanation was able to satisfy even
thoughtful minds, we ought the more to admire the acuteness of those physicists, who have recognised in this,
supposed insignificant, phenomenon a fundamental property of matter.''}
\subsubsection{Brownian motion and the kinetic theory of gases}

It became clear from experiments made in various laboratories that
Brownian motion increases when the size of the suspended particles
decreases (one essentially ceases to observe it when the radius is
above several microns), when the viscosity of the fluid decreases, or when
the temperature increases.  In the 1860's, the idea emerged that
the cause of the Brownian motion has to be found in the internal
motion of the fluid, namely that the zigzag motion of suspended
particles is due to collisions with the molecules of the fluid.

The first name worth citing in this regard is probably that of
Christian Wiener, holder of the Chair of Descriptive Geometry at
Karlsruhe, who in 1863 reaffirmed in the conclusions to his observations
that the motion could be due neither to the interactions between
particles, nor to differences in temperature, nor to evaporation or
convection currents,  but that the cause must be found in the liquid
itself.\footnote{Chr. Wiener, {\it Erkl\"arung des atomischen Wessens des fl\"ussigen
K\"orperzustandes und Best\"atigung desselben durch die sogennanten Molekularbewegungen}, {\it Ann. d. Physik} {\bf 118}, 79
(1863).} That being so, his theory on atomic motion anticipated those of
Clausius and Maxwell, implicating not only the motion of molecules but
also the motion of ``ether atoms''.  The Brownian motion was thus
bound to the vibrations of the ether, to the wavelength corresponding
to that of red light and to the size of the smallest group of
molecules moving together in the liquid. Such an explanation was
criticized by R. Mead Bache, who showed that the motion was
insensitive to the color of light, whether it was violet or
red.\footnote{R. Mead Bache, {\it Proc. Am. Phil. Soc.} {\bf 33}, 163
(1894).} Christian Wiener is nevertheless credited by some authors
as the first to discover that molecular motion could explain the
phenomenon.\footnote{J. Perrin, {\it Mouvement brownien et r\'ealit\'e mol\'eculaire, op. cit.}}

At least three other people proposed the same idea: Giovanni Cantoni
of Pavia, and two Belgian Jesuits, Joseph Delsaulx and Ignace
Carbonelle. The Italian physicist attributed Brownian movement to thermal motions in the liquid:\smallskip

{\small `` In fact, I think that the dancing movement of the extremely minute solid particles in a liquid can be
attributed to the different intrinsic velocities at a given temperature of both such solid particles and
of the molecules of the liquid that hit them from every side.

I do not know whether others have already attempted this way of explaining Brownian motions...''}\smallskip

He concluded that :

{\small ``In this way Brownian motion, as described above, provides us with one of the most beautiful and direct
experimental demonstrations of the fundamental
principles of the mechanical theory of heat, making manifest the assiduous vibrational state that must exist
both in liquids and solids even when one does not alter their temperature.''\footnote{G. Cantoni, {\it Il Nuovo Cim.} {\bf 27}, pp. 156-167 (1867);
quoted by G. Gallavotti in {\it Statistical Mechanics, a Short Treatise}, p. 233, Springer-Verlag, Heidelberg (1999);
 English translation available from G. Gallavotti. See also the reprint with notes by J. Thirion in {\it Revue des
Questions Scientifiques} {\bf 15}, 251 (1909).}}
\medskip

The Belgian physicists published in the {\it Royal Microscopical Society} and in
the {\it Revue des Questions scientifiques}, from  1877 to 1880, various
 Notes on the {\it Thermodynamical Origin of the Brownian
Movement}. In a Note by Father Delsaulx, for example,
one may read:\footnote{``See for this bibliography an article which appeared in the {\it Revue des
Questions Scientifiques}, January 1909, [{\it op. cit.}],  where M. Thirion very properly calls attention to the ideas of these
{\it savants}, with whom he collaborated.'' [original citation and note by J. Perrin in
{\it Brownian Motion and  Molecular Reality, op. cit.}]}\smallskip

{\small ``The agitation of small corpuscles in suspension in liquids truly
constitutes a general phenomenon,''} that it is {\small ``henceforth natural to
ascribe a phenomenon having this universality to some property of matter,''} and that
{\small ``in this train of ideas the internal movements of translation which constitute the calorific state
of gases, vapours and liquids, can very well account for the facts established by experiment.''}\medskip

Such a point of view, parallel to that of the kinetic theory of gases,
faced strong opposition. One opponent, cytologist Karl von N\"ageli of
Switzerland, familiar with the kinetic theory of gases and the orders
of magnitude involved, likewise the British chemist William Ramsey (the
future Nobel laureate in Chemistry), commented that the particles in suspension
have a mass several hundreds of millions of times larger than that of
the molecules in the fluid.  Each random collision with a molecule of
the surrounding fluid produces therefore an effect far too
small to displace the suspended particle. N\"ageli wrote for
example about a supposedly similar motion of micro-organisms in the air:\smallskip

{\small ``The motion which a sun-mote, and on the whole any particle
found in the air, can acquire by the collisions of an individual gas
molecule or a multitude of such molecules is therefore so
extraordinarily small, and the number of simultaneous collisions
against the particle from all sides so extraordinarily large, that
the particle behaves as if it were completely at rest.''}\medskip

He believed instead that the cause of the motion was not the thermal
molecular motion but some attractive or repulsive forces.

Nevertheless, the second part of his proposition about the frequency of
such collisions held the principle of the solution.  Because it is a
collective statistical effect, as described in  perspicacious
manner by Father Carbonelle:\smallskip

{\small ``In the case of a surface
 having a certain area, the molecular
collisions of the liquid, which cause the pressure, would not  produce any
perturbation  of the suspended particles, because these, as a whole, urge
the particles  equally in all directions.  But if the surface is of area
less than necessary to insure the compensation of irregularities, there is no
longer any ground  for considering the mean pressure; the inequal pressure,
 continually varying from place to place, must be recognised, as
the law of large numbers no longer leads to uniformity; and
the resultant will not now be zero but will change continually
in intensity and direction.  Further, the inequalities will become
more and more apparent the smaller the body is supposed to be, and in consequence the
oscillations will at the same time become more and more brisk...''}\smallskip

%
  
 Perrin mentions these authors to conclude:\smallskip

 {\small ``These remarkable reflections unfortunately remained as little known as those of Wiener. Besides
 it does not appear that they were accompanied by an experimental trial sufficient to dispel the superficial
 explanation indicated a moment ago; in consequence, the proposed theory did not impress itself on those who
 had become acquainted with it.''}\smallskip

 He continues:

 {\small ``On the contrary, it was established by the work of M. {\it Gouy} (1888), not only that the hypothesis of
 molecular agitation
 gave an admissible explanation of the Brownian movement, but that no other cause of the movement could be imagined, which
 especially increased the significance of the hypothesis.\footnote{L.-G. Gouy, {\it J. de Physique} {\bf 7}, 561
(1888); {\it C. R. Acad. Sc. Paris}, {\bf 109}, 102 (1889); {\it Revue
g\'en\'erale des Sciences}, {\bf 1} (1895).} This work immediately evoked a considerable response, and it
is only from this time
that the Brownian movement took a place among the important problems of general physics.''}\medskip

Indeed in  1888 the French physicist Louis-Georges Gouy made the best
observations on Brownian motion, from which he drew the following
conclusions:

- The motion is extremely irregular, and the trajectory seems not to 
have a tangent.

- Two Brownian particles, even close, have independent motion from one
another.

- The smaller the particles, the livelier their motion.

- The nature and the density of the particles have no influence.

- The motion is most active in less viscous liquids.

- The motion is most active at higher temperatures.
 
- The motion never stops.
\medskip
 
Gouy seemed, however, to claim again that one cannot explain Brownian
motion by disordered molecular motion, but only by the partially
organized movements over the order of a micron within the liquid.

But somehow he became known as the ``discoverer'' of the cause of 
Brownian motion, as Jean Perrin wrote about his experimental
conclusions:

{\small `` Thus comes into evidence, in what is termed a {\it fluid in equilibrium}, a property
eternal and profound.
This equilibrium only exists  as an
average  and  for large masses; it is  a statistical
equilibrium.  In reality the whole fluid is agitated indefinitely and {\it
spontaneously} by motions the more violent
and rapid the smaller the portion taken into account; the statical
notion of equilibrium is completely illusory.''}\footnote{J. Perrin, {\it Mouvement brownien et
r\'ealit\'e mol\'eculaire,}
{\it
op. cit.}}

\subsubsection{Brownian motion and Carnot's principle}

Brownian agitation continues indefinitely.  It does not 
contradict  the principle of energy conservation, because any
increase in the velocity of a grain, for instance, is accompanied by a local
cooling of the surrounding fluid, and the thermal equilibrium is
statistical.
 
Gouy was the first to note the apparent contradiction between
Brownian motion and {\it Carnot's principle}.  The latter states that
one cannot extract work from a simple source of heat.  However, it
really seems that some work is made, in a {\it fluctuating} manner, by
the thermal motion of the molecules of the fluid.  Gouy mentioned the
theoretical possibility to extract work by a mechanism attached to a
Brownian particle, and he concluded that Carnot's principle perhaps
was no longer valid for dimensions of the size of a micron, in that 
echoing Helmholtz's reservations about the validity of such
principle for living tissues.\medskip
 
These questions sparked the interest of Poincar\'e, who announced at
the following lecture of the Congress of Arts and Sciences in St.
Louis in 1904, about the ``Present Crisis of Mathematical
Physics''\footnote{Henri Poincar\'e, {\it La valeur de la science},
Biblioth\`eque de philosophie scientifique, Flammarion, Paris (1905);
in {\it Congress of Arts and Sciences, Universal Exposition,
St. Louis, 1904}, Houghton, Mifflin and Co., Boston and New York
(1905).}:\smallskip

{\small ``But here the stage changes.  Long ago the biologist, armed
with his microscope, noticed in his specimens disorganized movements
of small particles in suspension; that is the Brownian motion.  He
believed at first that it is a vital phenomenon, but soon he saw that
inanimate bodies did not dance with less fervor than the others, so he
handed it over to physicists.  Unfortunately, physicists have been
uninterested for a long while in this question; one concentrates light
to enlighten the microscopic specimens, they thought; light does not
go without heat, from which inhomogeneities of temperature, and then
internal currents in the liquid that produce the motion we are
speaking about..\smallskip M. Gouy had the idea to look closer and he saw, or
believed he saw that this explanation is unsustainable, that the
motion becomes more lively the smaller the particles, but that they
are not influenced by light.  So, if the motion never stops, or more
exactly is continually reborn without end, without an external source
of energy, what are we to believe?  We must not, without any doubt,
renounce the conservation of energy because of this, but we see before
our own eyes both motion transform into heat by friction, and
inversely heat transform into motion; and all that while nothing is
lost, as the motion lasts forever.  This is the opposite of Carnot's
principle.  If this is the case, to see the world develop in reverse,
we no longer have need of the infinitely subtle eye of Maxwell's
demon, a microscope will suffice.  The largest of bodies, those that
have for example, a tenth of a millimeter, collide with atoms in
motion from all sides, but they do not move at all as the shocks are
so numerous that the law of chance says they compensate one another;
however the smallest particles do not receive enough shocks for the
compensation to be exact and they are unendingly tossed around.  And
{\it voil\`a}, one of our principles already in danger.''}\medskip

It is
rather subtle to prove that the Brownian phenomenon does not infringe
on the impossibility of creating perpetual motion (called of the second
kind), where work is extracted in a coherent manner by the observer
(recalling Maxwell's famous demon). One had to wait for Leo Szilard,
who hinted in 1929 that, because of the amount of information required by
such an attempt, the total produced entropy would compensate the
apparent entropy reduction due to the coherent use of fluctuations. We
shall briefly return to this question later,
after having described Smoluchowski's contributions.

\subsubsection{The kinetic molecular ``hypothesis''}

Nowadays it seems evident to us that the world is made up of
particles, of atoms and of molecules.  However, it was not always the
case, and the hypothesis of a continuous structure of matter was
relentlessly defended until the end of the nineteenth century by
famous names such as Duhem, Ostwald, and Mach.

The intuition or the idea that gases are composed of individual
molecules was already present in the eighteenth century, and in 1738
David Bernoulli was probably the first to affirm that the pressure of
a gas on its container is due to collisions of molecules with the
walls.  Avogadro made the radical affirmation in 1811 that equal volumes of two gases
at the same pressure and same temperature contain the same number of
molecules.  When such conditions are of one atmosphere and of $25^{\rm
o}$ Celsius, the number contained in a volume of 22.412 liters is noted
as ${\mathcal N}$, and called {\it Avogadro's number}.

To understand the stakes surrounding  the determination of Avogadro's number,
 one must recall that the constant $R$ in the perfect gas law has been
experimentally accessible since the eighteenth century, thanks to the
work of Boyle, Mariotte, Charles, and later Gay-Lussac.  It is in fact
associated to the number of moles, $N/{\mathcal N}$, which is an
experimental macroscopic parameter, contrary to the total number of
particles $N$, and  Avogadro's number ${\mathcal N}$, that are
microscopic quantities.

The study of Brownian motion played an essential role in establishing
the ``molecular hypothesis'' definitively. As Jean Perrin observed,
the ``hypothesis'' that bodies, despite their homogeneous appearance,
are made up of distinct mole\-cules, in unending agitation which
increases with temperature, is logically suggested by the phenomenon
of Brownian motion alone, even before providing an explanation.

In fact, according to Perrin, what is really strange and new in Brownian
motion, is, precisely, that it never stops, contrary to our every-day experience with
friction phenomena.  If, for example, we pour a bucket of water into a tub, the initial
coherent motion possessed by the liquid mass disappears,
de-coordinated by the multiple rebounds on the boundaries of the tub, until an
apparent equilibrium settles within the fluid at rest.
Does such a de-coordination of the motion of the
particles proceed indefinitely, as it would in an ideal
continuous medium?   The answer by Perrin is exceptionally convincing:\footnote{Translation by Frederick Soddy, {\it op. cit.}}\smallskip

{\small ``To have information on this point and to follow this de-coordination as far
as possible after having ceased to observe it with the naked eye, a microscope
 will be of assistance, and microscopic powders will be taken as indicators of the movement.
 Now these are precisely the conditions under which the Brownian motion is perceived:
 we are therefore {\it assured} that the de-coordination of motion, so evident on the
 ordinary scale of our observations, does not proceed indefinitely, and on the scale of
 microscopic observation, we {\it establish} an equilibrium between coordination and de-coordination.
 If, that is to say, at each instant, certain of the indicating granules stop, there are some
 in other regions at the same instant, the movement of which is re-coordinated automatically by
 their being given the speed of granules which have come to rest. So that it does not seem possible
 to escape the following conclusion:}

{\small  Since the distribution of motion in a fluid does not progress indefinitely, and is limited by a spontaneous
 re-coordination, it follows that the fluids are themselves composed of granules or {\it molecules},
 which can assume all possible motions relative to one another, but in the interior of which dissemination of motion
 is impossible. If such molecules had no existence it is not apparent how there would be any limit to the
 de-coordination of motion [...] In brief, the examination of Brownian movement alone suffices to suggest
 that every fluid is formed of elastic molecules, animated by perpetual motion.''}\medskip

In 1905 Albert Einstein was the first, actually along with (but
independently from) William Sutherland from Melbourne, to propose a quantitative
theory of Brownian motion. This theory will allow Perrin to determine
the precise value of Avogadro's number ${\mathcal N}$, in his famous
experiments of 1908-1909. Sutherland and Einstein succeeded where many others failed,
because they used an ingenious and global reasoning of statistical
mechanics, that we will explain here.  Marian von Smoluchowski made at
the same time an analysis according to a different ``Gedankenweg,''
more probabilistic, which led him to similar conclusions.
We will came back to this point later in the paper.\medskip


\subsection{William Sutherland, 1904-05}

In his famous biography of Einstein, {\it Subtle is the Lord...} (1982), Abraham Pais noted, while describing Einstein's
route to his well-known diffusion relation, that the same relation had been discovered ``at practically the same time''
by the Melbourne theoretical physicist William Sutherland, following similar reasoning to Einstein's, and had been
submitted for publication to the {\it Philosophical Magazine} in March 1905, shortly before Einstein completed the
doctoral thesis in which he first announced
the relation. Pais, therefore, proposed that the relation be called the ``Sutherland-Einstein relation''.

We follow here the introduction of the essay, {\it Speculating about Atoms in Early 20th-century Melbourne: William Sutherland and the
`Sutherland-Einstein' Diffusion Relation}, written recently by the Australian historian of science
Rod W. Home.\footnote{Most of the material presented in this section originates from the 2005 essay by R. W. Home, {\it
Speculating about Atoms in Early 20th-century Melbourne: William Sutherland and the
`Sutherland-Einstein' Diffusion Relation}, Sutherland Lecture, 16th National Congress, Australian
Institute of Physics, Canberra, January 2005. See also the interesting note by Bruce H. J. McKellar,
{\it The Sutherland-Einstein Equation}, {\it AAPPS Bulletin}, February 2005, 35.} In this section we shall
briefly discuss
Sutherland's work, and  the factors that may have led to his work having been over-shadowed by Einstein's, and soon
forgotten.  When the Einstein International Year of Physics commemorates the hundredth anniversary of the
 {\it Annus Mirabilis} papers' release, focusing also on W. Sutherland's achievements seems to be just fair!

\subsubsection{Sutherland's papers}

Sutherland's paper to which Pais refers was
actually published in June 1905,\footnote{W. Sutherland, {\it A Dynamical Theory for Non-Electrolytes and the
Molecular Mass of Albumin}, {\it Phil. Mag.} S.6, {\bf 9}, pp. 781-785 
(1905).} after Einstein completed his thesis, but shortly before he submitted it for examination.
We seem to be looking here at a perfect example of effectively simultaneous discovery. However, as Rod Home notes, the story is still a
little more complicated, for Sutherland had already reported his derivation over a year earlier, at the congress
of the Australasian Association for the Advancement of Science held in Dunedin, New Zealand,
in January 1904, and his paper had been published in the congress proceedings
in early 1905!\footnote{W. Sutherland, {\it The Measurement of Large Molecular Masses}, Report of the 10th Meeting of the
Australasian Association for the Advancement of Science, Dunedin, pp. 117-121 (1904).} Unfortunately, there was a
 {\it misprint} in the crucial equation giving the diffusion coefficient of a large molecular mass
 in terms of physical parameters: Avogadro's constant was missing!\footnote{As R. W. Home remarks, it is clear that one is
 looking at a genuine misprint in the proceedings, since the preceding line was given correctly.}

 The correct and extended equation, finally published in the {\it Philosophical Magazine}, is
\begin{equation}
\label{Sf}
D=\frac{RT}{\mathcal N}\frac{1}{6\pi \eta\, a}\frac{1+3\eta/\beta a}{1+2\eta/\beta a},
\end{equation}
where $R$ is the perfect gas constant, $T$ the absolute temperature, $\mathcal N$ Avogadro's number, $\eta$ the fluid viscosity, $a$ the radius
of the (spherical) diffusing molecule, and $\beta$ the coefficient of sliding friction if there is slip between the
diffusing
molecule and the solution.\footnote{Sutherland uses the version of Stokes' law,
$F=6\pi \eta\, a\frac{1+2\eta/\beta a}{1+3\eta/\beta a}V$, relating the viscous friction force $F$ to the velocity
of the particle. This relation is generalized here to the case where slip occurs at the boundary between the fluid
and the moving sphere.
For a derivation, see H. Lamb, {\it Hydrodynamics}, pp. 601-602, Cambridge University Press (1932).} To deal with
the available empirical data, Sutherland had indeed to allow for a varying coefficient of sliding friction
between the diffusing molecule and the solution.
By taking $\beta$ to infinity, so there is no slip at the boundary, one recovers the usual form of the equation:
\begin{equation}
\label{Sf2}
D=\frac{RT}{\mathcal N}\frac{1}{6\pi \eta\, a}.
\end{equation}
Since in a fluid the molecules are close packed the molecular radius $a$ should be proportional
to the cube root of the molar volume ${\mathcal B}$, the volume occupied by Avogadro's number of particles.
Hence, from the constancy of the product $aD$ in relation (\ref{Sf2}), should follow that 
 of ${\mathcal B}^{1/3}D$. After having
estimated this constant from experimental data on the diffusion of various dissolved substances, Sutherland could
obtain the molar volume of albumin, and got an estimate of its atomic mass\footnote{The dalton (Da) is
the atomic mass unit; it honors the English chemist John Dalton (1766-1844), who revived the atomic
theory of matter in 1803.} as 32814 Da.\footnote{The present-day value is
 43 kDa  for ovalbumin.}

\subsubsection{Sutherland, Einstein and Besso}

In 1903, Einstein and his friend Michele Besso discussed a theory of dissociation that required the assumption
of molecular aggregates in combination with water, the ``hypothesis of ionic aggregates,'' as Besso called it.
This assumption opens the way to a simple calculation of the sizes of ions in solution, based on hydrodynamical
considerations. In 1902, Sutherland had considered in
{\it Ionization, Ionic Velocities, and Atomic Sizes}\footnote{W. Sutherland,
{\it Ionization, Ionic Velocities, and Atomic Sizes}, {\it Phil. Mag.} S.6, {\bf 4}, pp. 625-645 (1902).}
 a calculation of the sizes of ions on the basis of Stokes' law,
but criticized it as in disagreement with experimental data.\footnote{He wrote:
{\it ``Now this simple theory must have been written down by many a physicist  and found to be wanting, for it makes
the ionic velocities of the different atoms at infinite dilution stand to one another inversely as their radii,
a result which a brief study of data as to ionic velocities and relative atomic sizes shows to be not verified''.}
Sutherland did not use the assumption of ionic hydrates, which can avoid such disagreement by permitting ionic sizes
to vary with temperature and concentration.}
 The very same idea of determining sizes of ions by means of classical hydrodynamics occurred to Einstein in his
 letter of 17 March 1903 to Besso,\footnote{Albert Einstein, Michele Besso, {\it Correspondance 1903-1955},
 translation, notes and introduction by Pierre Speziali, Herrmann, Paris (1979).} where he proposed  what appears to be just the calculation
 that Sutherland had performed:\smallskip

{\small ``Have you already calculated the absolute magnitude of ions on the assumption that they are spheres
and so large that the hydrodynamical equations for viscous fluids are applicable? With our knowledge
of the absolute magnitude of the electron [charge] this would be a simple matter indeed. I would have done it myself
but lack the reference material and time; you could also bring in diffusion in order to obtain information about neutral salt
molecules in solution.''}\medskip

As the editors of Einstein's Collected Papers remark, ``This passage is remarkable, because both key elements of
Einstein's method for the determination of molecular dimensions, the theories of hydrodynamics and diffusion,
are already mentioned, although the reference to hydrodynamics probably covers only
Stokes' law''.\footnote{{\it The Collected Papers of Albert Einstein}, volume 2, {
\sc The Swiss Years: Writings, 1900-1909}, John Stachel ed., pp. 170-182, Princeton University Press (1989).}

It is also striking that an earlier letter of 11-17 February 1903, this time from Besso to Einstein, clearly
indicates  that they had been discussing Sutherland's work together. This letter contains two parts.
The first deals with
 experimental data in connection to the dissociation of bi-ionic molecules. The second discusses what Besso calls
  ``Sutherland's hypothesis,'' in connection to dissociation or dissolution. He states that the theory of
 ``ionic hydrates,'' as he calls it, rescues temporarily this hypothesis
 with regard to Ostwald's dilution law. Since Besso also discusses the role of imperfect semi-permeable membranes
 as a possible experimental test
 of Sutherland's hypothesis, P. Speziali, in the French edition of the Einstein-Besso correspondence, has indicated that
 Besso would have been discussing in this letter another of Sutherland's papers,
 entitled {\it ``Causes of osmotic pressure and
 of the simplicity of the laws of dilute solutions.''}\footnote{{\it Causes of Osmotic Pressure and
 of the Simplicity of the Laws of Dilute Solutions}, {\it Phil. Mag.}, S.5, {\bf 44}, pp. 52-55 (1897).}

 However, upon
 reading these letters of 1903, one cannot refrain from wondering whether
 Besso and Einstein were not also acquainted with and discussing Sutherland's 1902 paper on ionic sizes. In that case, Sutherland
 suggestion to use hydrodynamic Stokes' law to determine the size of molecules would have been a direct inspiration to
 Einstein's dissertation and subsequent work on Brownian motion!

\subsubsection{Sutherland's legacy}
That Sutherland, in spite of his isolation in Melbourne, was well-known in physics circles is
also evidenced by the fact that he was invited to contribute to the Boltzmann Festschrift in 1904
--the only other non-European contributor being J. Willard Gibbs!-- If so, why did Einstein and not Sutherland
 become famous?

  Sutherland had assumed the existence of atoms,
 and attacked a practical question, the measurement of large molecular masses. He was interested in these masses
  because of their role in the chemical analysis of organic substances. While that is what everyone now uses the
 Sutherland-Einstein equation for, it was perhaps not of so widespread interest at the time. However,
 we have just seen from the Einstein-Besso correspondence how extremely important
 Sutherland's idea was of determining the sizes of ions or molecules by means of classical hydrodynamics.\medskip

 On the other hand, as stressed by the editors of {\it The Collected Papers}:\smallskip

{\small  ``In developing in his
 dissertation a new method for the determination of molecular dimensions, Einstein was concerned with
 several problems on different levels of generality. An outstanding current problem of the theory of solutions
 was whether molecules of the solvent are attached to the molecules or ions of the solute. Einstein's dissertation
 contributed to the solution of this problem. He recalled in 1909:}

{\small {\it ``At the time I used the viscosity of the solution to determine the volume of sugar dissolved
in water because in this way I hoped to take into account the volume of any attached water molecule.''}}

{\small  The results obtained in his dissertation indicate that such an attachment does occur.
Einstein's concerns extended beyond this particular question to more general problems of the foundations of the
theory of radiation and the existence of atoms. He later emphasized:}

{\small {\it ``A precise determination of the size of the molecules seems to me of the highest importance
because Planck's radiation formula can be tested more precisely through such a determination than through
 measurements on radiation.''}}

{\small The dissertation also marked the first major success in Einstein's effort to find further evidence
for the atomic hypothesis, an effort that culminated in his explanation of Brownian motion.''}\smallskip

To conclude, it is probably most appropriate to cite R. W. Home:\smallskip

{\small ``Of course, the diffusion-viscosity relation is generally known as the Einstein relation, not
the Sutherland-Einstein relation. Why? In part, I think, this happened because in the early 20th century,
theoretical physics was a largely German affair. In so far as the relation was taken up, and initially it was
not taken up much at all, it was taken up by Continental researchers who had read Einstein's work but failed to
notice that
the relation was also buried in a paper in the Philosophical Magazine entitled {\it ``A dynamical theory
for non-electrolytes and the
molecular mass of albumin.''} In the English-speaking world, where the Philosophical Magazine was one of the leading
journals in the field, there were very few people pursuing theoretical physics in the German style. There is plenty
of testimony that experimentally orientated British physicists were at something of a loss as how to assess
Sutherland's work. His obituary in Nature makes the point very clearly:\footnote{``{\it Nature}, 23 November 1911, p. 116.
The obituary is signed ``J. L.'' [Joseph Larmor?].''[original note]}}

{\small {\it ``His papers are well known to the scientific world. They are distinguished by great width of reading in
the latest phases of the subjects he treated, combined with very bold speculation always brought into
ample comparison with experimental knowledge.}}
{\small {\it His generalisations were, indeed, so numerous that it was often a difficult task
to try to estimate their value.''}}

{\small So in Britain, Surtherland didn't have a readership likely to be alert to the significance of his announcement
of a relationship between diffusion and viscosity, in the way some Continental readers of Einstein's work were.
And, finally, Sutherland's own presentation surely would not have helped, with the relation itself being
almost  submerged by his lengthy computations relating to the molecular mass of albumin. He would have done much better
to highlight the relation, alone, in a paper to itself. But that was not his style! His mind was
firmly fixed on the problem of determining molecular masses of large molecules, and he clearly saw the
diffusion-viscosity relation as an incidental result arrived at on the way to achieving that larger goal, not as
something of particular value in its own right.''}\medskip

In this year 2005, it is definitely time, I think, for the physics community to finally
recognize Sutherland's achievements, and following Pais' suggestion, to re-baptize the
famous relation (\ref{Sf2}) with a double name!

\subsection{Albert Einstein, 1905}


\centerline{{\it Mens agitat molem}}

\medskip
\subsubsection{Einstein's Dissertation}

One finds nowadays in the literature excellent descriptions of Einstein's dissertation.
An outstanding presentation is given in the Editorial Notes of the {\it Collected Papers of Albert Einstein}.\footnote{Editorial notes of the chapter
{\it ``Einstein's dissertation on the determination of molecular dimensions,''} in
{\it  The Collected Papers of Albert Einstein}, volume 2, {\it op. cit.}, pp. 170-182; see also John Stachel,
{\it Einstein's Miraculous
Year}, {\it op. cit.}, pp. 31-43.}
Their presentation is closely followed in this section, which incorporates some material of the
editorial notes of the chapter entitled {\it ``Einstein's dissertation on the determination
of molecular dimensions.''}\footnote{With kind permission of John Stachel, {\sc Editor}.}
The interested reader can also find a detailed  scientific study of Einstein's doctoral thesis
in a recent article by Norbert Straumann.\footnote{Norbert Straumann,
{\it On Einstein's Doctoral Thesis}, arXiv:physics/0504201.}


Einstein completed his dissertation on {\it ``A New Determination of Molecular Dimensions''} on 30 April 1905,
and submitted it to the University of Z\"urich on 20 July.\footnote{Einstein had already
submitted a dissertation in 1901, on ``a topic in the kinematic theory of gases''. By February 1902, he had withdrawn
the dissertation, possibly at his advisor's suggestion to avoid a controversy with Boltzmann. (For a detailed analysis,
see the Editorial Notes of {\it  The Collected Papers of Albert Einstein}, volume 2, {\it op. cit.}, pp. 174-175). Nevertheless, there is no doubt
 that Einstein was a great
admirer of Boltzmann. (For a biography of the latter, see C. Cercignani, {\it Ludwig Boltzmann,
The Man Who Trusted Atoms}, Oxford University Press (1998).)} Shortly
after being accepted there, the manuscript was sent for publication to the
{\it Annalen der Physik}, where it would be published in 1906.\footnote{{\it Eine neue Bestimmung der Molek\"uldimensionen},
{\it Ann. d. Phys.} {\bf 19}, pp. 289-306 (1906).} On 11 May 1905, eleven days after finishing his thesis, Einstein
had also sent the manuscript of his first paper on Brownian motion to the {\it Annalen}, which would publish it on
18 July 1905. 

Einstein's central assumption is the validity of using classical hydrodynamics to calculate the effect
of solute molecules, treated as rigid spheres, on the viscosity of the solvent in a dilute solution.
His method is well suited to determine the size of solute molecules that are large compared to those
of the solvent, and he applied it to solute
sugar molecules. 
As we have seen above, Sutherland published in 1905 a method for determining the masses of large molecules, with which
Einstein's method shares many important elements. Both methods make use of the molecular theory of diffusion
that Nernst\footnote{W. Nernst, {\it Z. Phys. Chem. St\"ochiometrie Verwandschaftslehre}, {\bf 2}, pp. 613-639 (1888).} developed on the basis of \-van 't Hoff's analogy between solutions and gases, and of Stokes'
law of hydrodynamic friction.

The first of the results in the dissertation is a relation between the coefficients of viscosity of a liquid
with and without suspended molecules ($\eta^{*}$ and $\eta$, respectively),
\begin{equation}
\label{visc}
\eta^{*}=\eta \left (1+{{\left[{5}/{2}\right]}}\,\varphi\right ),
\end{equation}
where $\varphi$ is the fraction of the volume occupied by the solute molecules. [The correct coefficient $[\frac{5}{2}]$
appeared later (see below).]

The second result is the famous expression (\ref{Sf2}) for the coefficient of diffusion $D$ of the solute molecules.
Like Loschmidt's method based on the kinetic theory of gases, the expressions
obtained by Einstein give two equations for two unknowns, Avogadro's number $\mathcal N$, and
the molecular radius $a$ of the suspended particles, hence providing a possible determination of molecular dimensions!

The derivation of eq. (\ref{visc}) represents the technically difficult part of Einstein's dissertation. It rests
on the assumption that the motion of the fluid can be described by the hydrodynamical equations for stationary
flow of an incompressible homogeneous fluid, even in the presence of solute molecules; that the inertia of these
molecules can be neglected; that they do not interact; and that they can be treated as rigid spheres moving in the
liquid without slipping, under the sole influence of hydrodynamical stress.

Eq. (\ref{Sf2}) follows from the conditions
of dynamical and thermodynamical equilibrium in the fluid. Its derivation, as does Sutherland's, requires the identification of the force
on a single large molecule, which appears in Stokes' law, with the apparent force due to the osmotic pressure. We shall
return to this derivation in detail in the next section, when describing the content of
Einstein's first paper on Brownian motion. In the dissertation, Einstein's derivation of
eq. (\ref{Sf2}) does not involve yet the theoretical tools he developed in his work on the statistical foundations
of thermodynamics in the preceding years. Here he simply states the osmotic pressure law, while
in his first paper on  Brownian motion, he will instead
derive from first principles the validity
of van 't Hoff's law for large suspended particles.

In 1909, Einstein drew Perrin's attention to his method for determining the size of solute molecules, which allows
one to take into account the volume of any water molecule attached to the latter, and he
suggested its application to the suspensions studied by Perrin  in relation to Brownian motion.
In the following year,
an experimental study of formula (\ref{visc}) for the viscosity coefficient was performed by a pupil of Perrin,
Jacques Bancelin.
 Using the same aqueous emulsions of gum-resin (``gamboge''), he confirmed that the increase in viscosity does not depend on the size
 of the solute molecules, but only on their volume fraction. However, the coefficient of $\varphi$
 in eq. (\ref{visc}) was found to be close to 3.9, instead of the predicted value 1. That prompted Einstein,
 after an unsuccessful attempt to find an error, to ask his student and collaborator Ludwig Hopf to
 check his calculations and arguments:\smallskip

 {\small ``I have checked my previous calculations and arguments and found no error in them. You would be doing a
 great service in this matter if you would carefully recheck my investigation. Either there is an error in the work,
 or the volume
 of Perrin's substance in the suspended state is greater than
 Perrin believes.''}\footnote{{\it The Collected Papers of Albert Einstein}, volume 2, {\it op. cit.}, pp. 180-181.}\medskip

Hopf did find an error in the dissertation, namely in the derivatives of some velocity components,
and obtained for $\varphi$
a corrected coefficient 2.5. The remaining discrepancy between this corrected theoretical
factor and the experimental one led Einstein to suspect that there might be also an
experimental error.\footnote{He asked Perrin: {\it ``Wouldn't it be possible that your mastic particles, like colloids,
are in a swollen state? The influence of such a swelling 3.9/2.5 would be of rather slight influence on
Brownian motion, so that it might possibly have escaped you''}, Einstein to Perrin, 12 January 1911, in
{\it The Collected Papers of Albert Einstein}, volume 2, {\it op. cit.}, p. 181.}

In early 1911 Einstein submitted his correction for publication, and recalculated
Avogadro's number. He obtained a value of $6.56 \times 10^{23}$ per mole, a value that is close to those derived from
kinetic theory and Planck's black-body radiation theory.

The paper published in 1911 by Bancelin in the {\it Comptes rendus de l'Acad\'emie des Sciences} gave an experimental value of 2.9 as the coefficient
of $\varphi$ in eq. (\ref{visc}). Extrapolating his results to sugar solutions, Bancelin recalculated Avogadro's number,
and found a value of $7.0 \times 10^{23}$ per mole.

Einstein's dissertation was at first overshadowed by his more spectacular work on Brownian motion, and it required
an initiative by Einstein to bring it to the attention of the scientists of his time.
The paper on Brownian motion, the first of several on this subject that Einstein published over the course
of the next couple of years, actually included his first published statement of the famous relationship linking
diffusion with viscosity, that he had derived in his thesis.

As
 Abraham Pais points out in {\it Subtle is the Lord...}, this equation has found widespread applications,
 as a result of which Einstein's January 1906 paper in the {\it Annalen der Physik}, the published version of his
 dissertation, later became his most frequently cited paper!\footnote{According to R. W. Home, {\it op. cit.}, it
 became the paper most widely cited in the period 1961-75, the period surveyed for the citation analysis
 of any scientific article published by any author before 1912. According to B. H. J. McKellar, {\it op. cit.},
  the 1905
 citation count is as follows (from World of Science, Dec. 2004):
{\it Ann. d. Phys.} {\bf 17}, 132 (1905): {\bf 325} (photoelectric effect);
{\it Ann. d. Phys.} {\bf 17}, 549 (1905): {\bf 1368} (Brownian motion);
{\it Ann. d. Phys.} {\bf 17}, 891 (1905): {\bf 664} (special relativity);
{\it Ann. d. Phys.} {\bf 18}, 639 (1905): {\bf 91} ($E=mc^2$);
{\it Ann. d. Phys.} {\bf 19}, 289 (1906): {\bf 1447} (molecular dimensions, Einstein's thesis).}
 As stressed by R. H. Home in his essay on Sutherland, Pais also goes on to argue that the thesis was also one
 of Einstein's {\it ``most fundamental papers''}, of comparable intrinsic significance to the other papers Einstein
 wrote in that year of 1905. {\it ``In my opinion,''} Pais writes, {\it ``the thesis is on a par with [Einstein's]
 Brownian motion article''}: indeed, {\it ``in some if not all respects, his results are by-products of his thesis work.''}

It is  now time to turn to this famous 1905 Brownian motion article.

\subsubsection{The 1905 article on Brownian motion}
The 1905 article is entitled: {\it ``On the Motion of Small Particles Suspended
in Liquids at Rest, Required by the Molecular-Kinetic Theory of
Heat.''}\footnote{A. Einstein, {\it Ann. d. Physik} {\bf 17}, pp. 549-560
(1905).} There, Einstein tried to establish the existence and the size
of molecules, and to determine a theoretical method for computing
Avogadro's number precisely, by using the molecular kinetic theory of
heat. In fact, he concluded:\smallskip

{\it ``M\"oge es bald einem Forscher gelingen, die hier aufgeworfene,
f\"ur die Theorie der W\"arme wichtige Frage zu
entscheiden~!''}\footnote{``Let us hope that
a researcher will soon succeed in solving the problem presented here, which is so important for the theory of heat!''}\medskip

Astonishingly enough, he was not yet certain that one could apply it to
Brownian motion.  In fact, his introduction opens with: {\it ``In this
paper it will be shown that, according to the molecular-kinetic theory of heat,
bodies of a microscopically visible size suspended  in
liquids must, as a result of thermal molecular motions,  perform motions of such magnitude that they can be
easily observed with a microscope. It is possible that the
motions to be discussed here are identical with so-called Brownian molecular
 motion; however the data available to me on the latter are so imprecise that
I could not form a  judgement on the question.''}\medskip

Einstein relied on the results of his thesis, that he completed
eleven days before submitting his famous article on the
suspensions of particles.  Only later would his predictions be progressively
confirmed by refined experimental data on Brownian motion.\footnote{This led J. Renn, {\it op. cit.}, to speak of
``Einstein's invention of Brownian motion''.}

\subsubsection{The Einstein-Sutherland derivation}

The demonstration is based  on two distinct elements from
apparently contradicting domains.

It seemed initially natural to use a {\it hydrodynamic}
representation for particles in suspensions with size much greater
than that of the liquid's molecules.  A substantial amount of
knowledge on the subject was available, in particular the famous
``Stokes' formula,'' which gives the force of friction opposing to a
sphere moving in the liquid.

But at the same time it was necessary for Einstein to exploit the
kinetic theory of heat, pulling it away from the original context of
the theory of gases and bringing it closer to the context of liquids,
where the state of the theory was much less advanced. It was the crucial 
notion of {\it osmotic pressure}, developed by van 't Hoff, that made the passage possible.
It is based on the concept of
kinetic molecular disorder, where solute molecules, with a size
comparable to that of the liquid's molecules, participate
to the general motion like in a dilute gas.

Einstein was in possession of two theories about particles in
a fluid. The first: Stokes' hydrodynamic theory, based on the hypothesis
that a liquid is a continuous medium which adheres to a large solid surface
moving through it, without any turbulence, and where the 
molecular agitation does not seem to play any role. The other:
van 't Hoff's osmotic theory, based on the hypothesis that a particle in
solution is similar to any other fluid molecule, and therefore 
is subjected to the same laws of molecular agitation.

One needed Einstein's perspicacity and his profound knowledge of
statistical mechanics to understand and to prove that the two points
of view were simultaneously valid for particles as big as Brownian
particles.

Einstein first studied the osmotic pressure created in the
solution by solute molecules.  This notion was developed by
J. H. van 't Hoff\footnote{J. H. van 't Hoff, {\it Kongliga Svenska
Vetenskaps-Academiens Handlingar}, Stockholm, {\bf 21}, 1 (1884).}
who, for dilute solutions, showed the identity between the pressure exerted
on  semi-permeable walls by molecules in solution and the
partial pressure exerted by a gas.  For sufficiently dilute solutions,
this additional pressure $p$ due to the 
molecules in solution satisfies  the law of perfect gases

\begin{equation}
\label{gp}
p=\frac{n}{\mathcal N} R\, T,
\end{equation}
where $R$ is the ideal gas constant, $T$ is the absolute
temperature, and $n$ is the number of solute particles per unit
volume, or particle density.\medskip

In his thesis, Einstein considered the effect of the density of such
molecules on the viscosity, such as in the case of sugar in water.
This time the particles in suspension are much larger so as to be
observable under a microscope.  Einstein right away affirms that the
difference between solute molecules and particles in suspension is
only a matter of size, and that van 't Hoff's law must be applied
as well to particles in suspension.  Next, he proves this fact and
formula (\ref{gp}), by determining the free energy of an ensemble of
such particles in suspension.  In fact, he calculates the associated partition
function by the phase space method.\medskip

Einstein then imagines that the numerous particles of the suspension
are subjected to an external force $F$, which may depend on their
positions but not on time.\footnote{This force can be, for example
gravitational, as in the sedimentation experiments by Jean Perrin, but
the beauty of the argument is that the result does not depend on the
nature of the force, that can even be virtual, as in the notion of
``virtual work'' of the eighteenth century Mechanics.}  This force,
acting along the $x$ axis for instance, moves each particle of the
solute, and generates a gradient of concentration.  Let $n(x,y,z;t)$
be the number of particles in suspension per unit volume around the
point $x,y,z$ at the instant $t$.  From (\ref{gp}), a non-uniform
osmotic pressure corresponds to a gradient of concentration of
particles in suspension.  By considering the resultant of all pressure
forces on an elementary interval ${\rm d}x$, one also obtains the
force of the osmotic pressure {\it per unit volume}:
\begin{equation}
\label{fpo}
\Pi=-\frac{\partial p}{\partial x}=-{\rm grad}\, 
p=-\frac{R}{\mathcal N}\, T {\rm grad}\, n(x,y,z;t),
\end{equation}
where here the gradient is the spatial derivative in the direction $x$
of the force.

In addition, the quantity $\Pi_{F}=n\, F$ represents the total
external force per unit volume acting on the Brownian particles in
suspension.  From both a {\it hydrostatic} and {\it thermodynamic}
point of view, one imagines {\it a priori} that the equilibrium of a
unit of volume of the suspension is established when the force
$\Pi_{F}$ is balanced by the osmotic pressure force $\Pi$.  In fact, 
by using arguments of equilibrium invariance of the free energy with
respect to virtual displacements, Einstein demonstrates that actually
the sum of the external and osmotic forces per unit volume cancels:
\begin{eqnarray}
\label{pression}
\Pi_{F}+\Pi&=&0,\\
\label{forces}
n\, F&=&\frac{R}{\mathcal N}\, T {\rm grad}\, n.
\end{eqnarray}
One notices that he directly obtained the explicit formula
(\ref{forces}) from the free energy of the particles in suspension,
without relying on the result (\ref{gp}), which shows the two results
come from the same approach.

The second part of this argument focuses on the {\it dynamics of the
flux equilibrium}.  Equilibrium in the fluid is actually just an
apparent effect: while the force $F$ moves the particles in
suspension, these are also subjected to Brownian motion which reflects
 the
kinetic nature of heat.

By moving in the liquid under the force $F$, each particle in
suspension experiences an opposing force of viscous friction.  This
brings the particle to a limit velocity $V=~F/\mu$, where $\mu$ is the
coefficient of viscous friction for each particle in suspension.  The
result is a flux of particles
\begin{equation}
\label{vis}
\Phi_{F}=n\, V=n\, F/\mu,
\end{equation}
that is the number of particles crossing a unit surface perpendicular
to the direction $x$ of the force.

The particle density $n(x,y,z;t)$ satisfies the local diffusion equation
\begin{equation}
\label{diffusionbis}
\frac{\partial n}{\partial t}=D \Delta n,
\end{equation}
where $\Delta$ is the Laplacian
$\Delta=\frac{\partial^2}{\partial x^2}+\frac{\partial^2}{\partial
y^2}+\frac{\partial^2}{\partial z^2},$ and where $D$ is a coefficient,
called the coefficient of diffusion, measured in square meters per second units.
To this equation is naturally associated a diffusion
flux  
$\Phi_{D}$, which is the number of particles diffusing across a
unit surface per unit of time.  This flux is directly connected to the
concentration gradient
by\footnote{Einstein, like Sutherland, writes this equation
directly, without passing through the diffusion equation he will prove
farther along. This is indeed the celebrated  Fick's law
(A. Fick, {\it \"Uber Diffusion}, {\it Ann. Phys. Chem.} {\bf 4}, 59-86 (1855)). For mathematically inclined readers, let us
recall that the Laplacian is also $\Delta= {\rm div}({\rm grad})$, where
the divergence is the operator of derivation of a vector $\vec A$: ${\rm
div} \vec A=\vec \nabla .\,
\vec A=\frac{\partial A_x}{\partial x}+\frac{\partial A_y}{\partial y}+
\frac{\partial A_z}{\partial z},$ 
and where the gradient is the vector operator of derivation ${\rm
grad}=\left(\frac{\partial}{\partial x},\frac{\partial}{\partial
y},\frac{\partial}{\partial z}\right).$ From the diffusion equation, 
$\frac{\partial n}{\partial t}=D \Delta n,$ by counting the number of 
particles crossing an arbitrary closed surface and by applying the
Green-Ostrogradski theorem, one immediately finds the existence 
across the  surface of a diffusion flux $\Phi_D=-D\, {\rm grad}\,
n$.}
\begin{equation}
\label{fluxdif}
\Phi_{D}=-D\, {\rm grad}\, n.
\end{equation}

At equilibrium, here both local and dynamic, the force-driven flux
$\Phi_{F}$ (\ref{vis}) and the flux of diffusion $\Phi_{D}$
(\ref{fluxdif}), cancel:
\begin{eqnarray}
\label{eqflux}
\Phi_{F}+\Phi_{D}&=&0,\\
\label{flux}
 n\, F/\mu&=&D\, {\rm grad}\, n.
\end{eqnarray}

By comparing the static equation (\ref{forces}) and the dynamic equation
(\ref{flux}), one sees that  they have identical structures for
the dependence on $n$ and its gradient, from which we obtain the required identity
between the coefficients:
\begin{equation}
\label{ES}
D=\frac{1}{\mu}\frac{RT}{\mathcal N}.
\end{equation}
By supposing that the particles in suspension are all spheres of radius
$a$, Einstein uses at last Stokes' relation which gives the coefficient of
friction $\mu$ of a sphere immersed in a (continuous) fluid with viscosity
$\eta$:
\begin{equation}
\label{stokes}
{\mu}=6\pi \eta \, a,
\end{equation}
from which he finally deduced:
\begin{equation}
\label{ESf}
D=\frac{RT}{\mathcal N}\frac{1}{6\pi \eta\, a}.
\end{equation}
This is Einstein's famous relation, which is already in his
thesis. In fact, as mentioned above, the same relation was
discovered earlier in Australia and, by a remarkable coincidence,  published
 at practically the same moment as Einstein was
working on his thesis.  William Sutherland published his Philosphical Magazine
article in March of 1905.  One should therefore definitely call this relation the Sutherland-Einstein relation.\medskip

In the 1905 article, Einstein completes these results by means of
mathematical and probabilistic considerations.  Let $P(x,y,z; t)$ be
the probability density of finding a Brownian particle at a point
$x,y,z$ at the time $t$.  This density satisfies the diffusion
equation:
\begin{equation}
\label{diffusion}
\frac{\partial P}{\partial t}=D\, \Delta P.
\end{equation}
Let us follow Einstein in his demonstration. 

He starts by introducing a time interval $\tau$, small compared to the
duration of the observation, but large enough for the motions made by
a particle during two consecutive intervals of time $\tau$ to be
considered as independent events.
Let us suppose then that in a liquid suspension there is a total number
of particles $N$.  During the time interval $\tau$, the coordinates of
each particle along the $x$ axis will change by an amount $\Delta$,
where $\Delta$ takes a different value (positive or negative) for each
particle.  A probability distribution governs $\Delta$: the number
${\rm d}N$ of particles with a displacement between $\Delta$ and
$\Delta+{\rm d}\Delta$ is:
$${\rm d}N=N\varphi_{\tau}(\Delta){\rm d}\Delta,$$
where
\begin{equation}
\int_{-\infty}^{+\infty} \varphi_{\tau}(\Delta) {\rm d}\Delta=1,
 \label{consphi}
\end{equation}
and where, for small $\tau$, $\varphi_{\tau}(\Delta)$ differs
from zero only for very small values of $\Delta$.  This function also
satisfies the symmetry condition
\begin{equation}
 \varphi_{\tau}(\Delta)=\varphi_{\tau}(-\Delta).
 \label{symphi}
\end{equation}

Einstein tries then to determine how the coefficient of diffusion
depends on $\varphi$, once again by considering only the
unidimensional case where the particle density $n$ depends only on $x$
and $t$.  We can thus write $n=f(x,t)$ (the number of particles per
unit volume) and we calculate the particle distribution at the time
$t+\tau$ given the distribution at the time $t$. From the definition
of the function $\varphi_{\tau}(\Delta)$, we obtain the number of
particles between two planes in $x$ and $x+{\rm d}x$ at the time
$t+\tau$:
\begin{equation}
f(x,t+\tau){\rm d}x={\rm d}x\int_{-\infty}^{+\infty}
f(x+\Delta,t)\varphi_{\tau}(\Delta) {\rm d}\Delta.
 \label{equf}
\end{equation}
Since $\tau$ is very small, we can assume
 that
\begin{equation}
f(x,t+\tau)=f(x,t)+\tau \frac{\partial f}{\partial t}.
 \label{dift}
\end{equation}
Moreover, expand $f(x+\Delta,t)$ in powers of $\Delta$:
$$f(x+\Delta,t)=f(x,t)+\Delta\frac{\partial f(x,t)}{\partial x}
+\frac{\Delta^2}{2}\frac{\partial^2 f(x,t)}{\partial x^2}+\cdots$$ We
can then substitute such an expansion inside the integral in
(\ref{equf}) as only very small values of $\Delta$ contribute to 
the latter.  We obtain: $$f+\tau \frac{\partial f}{\partial t}= f
\times\int_{-\infty}^{+\infty} \varphi_{\tau}(\Delta) {\rm d}\Delta
+\frac{\partial f}{\partial x}\times \int_{-\infty}^{+\infty}
\Delta\varphi_{\tau}(\Delta) {\rm d}\Delta +\frac{\partial^2
f}{\partial x^2}\times\int_{-\infty}^{+\infty}
\frac{\Delta^2}{2}\varphi_{\tau}(\Delta) {\rm d}\Delta +\cdots$$
On the right side, the second term, fourth term, etc., cancel out
because of the parity property (\ref{symphi}), while each of the other
terms is very small in relation to the preceding one.  From this
equation, taking into account the conservation property
(\ref{consphi}), defining
\begin{equation}
\frac{1}{\tau}\int_{-\infty}^{+\infty} \frac{\Delta^2}{2}
\varphi_{\tau}(\Delta) {\rm d}\Delta=D
\label{Dphy}
\end{equation}
and keeping only the first and the third terms on the right hand side,
we obtain
\begin{equation}
\frac{\partial f}{\partial t} =D\frac{\partial^2 f}{\partial x^2}.
\label{Diff}
\end{equation}
This is the famous diffusion equation, where the diffusion coefficient
$D$ is given by (\ref{Dphy}).

We comment now on the method of Einstein.  The definition (\ref{Dphy})
of the diffusion coefficient $D$ can be rewritten as
\begin{equation} 
\label{D2} 
\langle \Delta^2\rangle_{\tau}\equiv\int_{-\infty}^{+\infty}  
\Delta^2 \varphi_{\tau}(\Delta) {\rm d}\Delta=2D\tau, 
\end{equation}
which is the average quadratic variation produced by the thermal
agitation during the time $\tau$.  Formally identical to formula
(\ref{x2}) (see below), which gives the law of the average quadratic
displacement as a function of time, it somehow contains the latter
tautologically. Moreover, as $\tau$ is assumed to be small, this
definition implies the existence of the limit (\ref{Dphy}) for $\tau
\to 0$, if one requires  $D$ to be independent of $\tau$.\footnote{For a discussion of the
range involved for this auxiliary time parameter $\tau$, its physical meaning and the logical and
mathematical intricacies related to its formal limit $\tau \to 0$,
 see: C. W. Gardiner, {\it Handbook of Stochastic Methods},
2nd ed., Springer, Berlin (1985); N. G. van Kampen, {\it Stochastic
Processes in Physics and Chemistry}, North-Holland, Amsterdam (1992); D. T. Gillespie, {\it Markov Processes},
Academic, Boston (1992); and in particular G. Ryskin, {\it Phys. Rev. E} {\bf 56}, pp. 5123-5127 (1997).}\medskip

Einstein continues by noting that until then all particles have been
considered with respect to a common origin on the $x$ axis, but that
their independence also allows us to consider each particle with respect
to the position it occupied at the time $t=0$.  Therefore $f(x,t)\, {\rm
d}x$ is also the number of particles (per unit area) whose abscissa
 $x$  has changed by an amount
 comprised  between $x$ and
$x+{\rm d}x$, over the time interval from $0$ to $t$.  The
function $f$ then obeys the diffusion equation (\ref{Diff}). Einstein
also says that evidently one must have, for $t=0$,
 
$$f(x,t=0)=0, \forall x \neq 0\, ;\,\,\,\, {\rm and}
\int_{-\infty}^{+\infty} f(x,t){\rm d}x=N.$$

The problem thus coincides with that of diffusion from a given point
(neglecting the interactions between diffusing particles), and is now
 entirely determined mathematically; its solution is:
\begin{equation}
\label{gaussf} 
f(x;t)=\frac{N}{(4\pi D t)^{1/2}}
\exp{\left(-\frac{x^2}{4Dt}\right)}. 
\end{equation}

The probability density $P(x,t)=f(x,t)/N$ for a Brownian particle to
be within ${\rm d}x$ of $x$, assuming it was at $x=0$
at the instant $t=0$, is thus the normalized {\it Gaussian}
distribution
\begin{equation} 
\label{gauss1} 
P(x;t)=\frac{1}{(4\pi D t)^{1/2}}
\exp{\left(-\frac{x^2}{4Dt}\right)}. 
\end{equation} 
 
In three dimensions, if the Brownian particle is at $\vec 0$ at the
instant $t=0$ then the solution of equation (\ref{diffusion}) is still
Gaussian and written as:
\begin{equation} 
\label{gauss} 
P(x,y,z;t)=\frac{1}{(4\pi D t)^{3/2}}
\exp{\left(-\frac{x^2+y^2+z^2}{4Dt}\right)}. 
\end{equation} 
One clearly finds the previous density $P(x,t)$ by integrating over
the variables $y$ and $z$.\medskip

From these results one can evaluate the integral of the average
quadratic displacement along, say, the $x$ axis. One finds
\begin{eqnarray} 
\nonumber 
\langle x^2\rangle_t&=&\int_{-\infty}^{+\infty}x^2P(x;t) \, {\rm d}x 
=\frac{1}{(4\pi D t)^{1/2}} \int_{-\infty}^{+\infty}x^2
\exp{\left(-\frac{x^2}{4Dt}\right)}\, {\rm d}x\\ 
\label{fluc} 
&=&2Dt.
\end{eqnarray} 
As already pointed out above, this result for $\langle x^2\rangle_t$
is absolutely identical to the result (\ref{D2}) for $\langle
\Delta^2\rangle_{\tau}$, which is just a reflection of the {\it scale
invariance} of Brownian motion, a notion perhaps not yet appreciated in 1905!\medskip

From the Sutherland-Einstein relation (\ref{ESf}), one finally obtains
the average Brownian displacement as a function of time
\begin{equation} 
\label{x2} 
\langle x^2\rangle_t=2Dt=\frac{RT}{\mathcal N}\frac{1}{3\pi \eta\, a} t. 
\end{equation} 
This is the first appearance of a {\it
fluctuation-dissipation} relation, linking position
fluctuations and a property of dissipation (the viscosity). As stressed by Ryogo Kubo in his essay
{\it Brownian Motion and Nonequilibrium Statistical Mechanics},\footnote{R. Kubo, {\it Brownian Motion and Nonequilibrium Statistical Mechanics},
{\it Science} {\bf 233}, pp. 330-334 (1986).} fluctuation-dissipation relations are at the heart
of the so-called {\it linear response theory}, which is, in a sense, the most natural extension of the Sutherland-Einstein
theory of Brownian motion. In particular, the so-called Green\footnote{M. S. Green,
{\it J. Chem. Phys.} {\bf 20}, 1281 (1952); {\it ibid.} {\bf 22}, 398 (1954).}--Kubo\footnote{R. Kubo,
{\it J. Phys. Soc. Jpn.} {\bf 12}, 570 (1957).} formulae there provide the generalizations of
relation
(\ref{x2}).

In this
fundamental equation for Brownian motion, $\langle x^2\rangle$,
$t$, $a$ and $\eta$ are measurable quantities and thus Avogadro's
number can be determined.  This is an astonishing result: first
prepare a suspension of small spheres, but large however with respect
to molecular dimensions, then take a chronometer and a microscope, and
finally measure $\mathcal N$~!  Einstein gave this example: for water
at $17^{\rm o}$C,\footnote{According to John Stachel in {\it Einstein's
Miraculous Year} (Princeton University Press, New Jersey, 1998), the
data Einstein uses on the viscosity of water is taken from his 
thesis, and in fact corresponds to the temperature $9.5^{\rm o}$C.}
$a \approx 0.001\, {\rm mm}=1\, \mu{\rm m}$, $\mathcal N \approx
6\times 10^{23}$, one finds a displacement of $\sqrt{\langle x^2\rangle
}
\approx 6 \, \mu{\rm m}$ for $t=1\, {\rm mn}$.

One can ask to what extent does the Sutherland-Einstein formula
(\ref{ES}) or (\ref{ESf}) prove the existence of molecules.
In other words, what would be the limit of the diffusion coefficient
$D=\frac{RT}{\mu\mathcal N}$ if Nature were continuous, i.e., if
Avogadro's number was infinite?  Then $D$ would cancel out, and the
displacement of Brownian diffusion (\ref{x2}) would simply {\it
disappear} in this limit, but one should verify, for the sake of
rigour, the simultaneous existence of a finite continuous limit of the
friction coefficient $\mu$ or of the viscosity $\eta$ when $\mathcal N
\to \infty$. We will come back to this point in section (\ref{modele})
where the study of a microscopic model allows for an explicit
calculation of $\mu$, and for concluding that Brownian motion is
surely a manifestation of the existence of molecules!

\subsubsection{Einstein, 1906, general theory of Brownian motion}

In another article written in December 1905 and received on the 19th
of the same month by {\it Annalen der Physik},\footnote{A. Einstein,
{\it Ann. d. Physik} 19, pp. 371-381 (1906); translated in A. Einstein, {\it Investigations on the
Theory of the Brownian Movement}, R. F\"urth {\it Ed.},  A. D. Cowper {\it Transl.},
Dover Publications, pp. 19-35 (1956).} this time entitled: {\it ``On
the Theory of Brownian Motion,''} Einstein mentions that {\it
``Soon after the appearance of my paper on the movements of particles suspended in liquids required by the
 molecular theory of heat,
Siedentopf (from Jena) informed me that he and other physicists --firstly, Prof. Gouy (of Lyons)--
 had been convinced by direct observation that the so-called Brownian motion is caused by the
irregular thermal
 movements of the molecules of the liquid.

 Not only the qualitative properties of Brownian motion,
but also the order of magnitude of the paths
described by the particles correspond completely with the results of the theory.''}\medskip

This time Einstein is convinced that Brownian motion is the phenomenon he just
described.  He then gives another, more
general, theoretical approach.  It can be applied not only to the  translational,
but also  rotational diffusion motion of particles in suspension, or to charge
fluctuations in an electric resistance.  We briefly describe such a
general and, from our standpoint, very enlightening approach.  It
shows the central role of the Boltzmann's distribution at
thermodynamic equilibrium, and shows that its stationarity in time
requires the existence of Brownian motion and its link to the
molecular nature of heat.
	
Einstein considers a quantity $\alpha$, which has a Boltzmann
distribution

\begin{equation}
 {\rm d}n= A e^{-\frac{\mathcal N}{R T}\varPhi(\alpha)} {\rm
 d}\alpha=F(\alpha){\rm d}\alpha, \label{Boltzphi}
\end{equation}
where $A$ is a normalization coefficient and $\varPhi(\alpha)$ is the
potential energy associated to the parameter $\alpha$. Here ${\rm d}
n$ is proportional to the probability density of $\alpha$ and gives
the number of systems ({\it \`a la} Gibbs) identical to the present
system taken in the same state.

Einstein uses that relation for determining the irregular changes of the
parameter $\alpha$ produced by thermal phenomena.  He states that the
function $F(\alpha)$ does not change during a time interval $t$ under
the combined effect of the force corresponding to the potential
$\varPhi$ and the irregular thermal phenomena; $t$ is so small that
all changes of the variable $\alpha$ can be considered as infinitesimally
small in the arguments of the function $F(\alpha)$.

We consider the real line representing all $\alpha$ values and take
an arbitrary point $\alpha_0$ on it.  During the time interval $t$, the same
number of systems must pass through the point $\alpha_0$ in one
direction as in the other.  The force $-\frac{\partial
\varPhi}{\partial \alpha}$ corresponding to the potential $\varPhi$
induces a change of the parameter $\alpha$ per unit of time:
\begin{equation}
\frac{{\rm d}\alpha}{{\rm d}t}=-B\frac{\partial \varPhi}{\partial \alpha},
 \label{vitesse}
\end{equation} 
where $B$ is, according to Einstein's words, the ``mobility of the
system with respect to $\alpha$''.  This is an equation of
viscous-friction type, like equation (\ref{vis}) with $B=1/\mu$.
According to (\ref{Boltzphi}), the variation of the number of systems
passing through the point $\alpha_0$ during the time interval $t$ is:
\begin{equation}
n_1=-B\left(\frac{\partial \varPhi}{\partial
\alpha}\right)_{\alpha=\alpha_0}
\times \, t F(\alpha_0),
 \label{n1}
\end{equation}
where the number of systems is counted algebraically (positive or
negative) according to the side of $\alpha_0$ they are moving from,
i.e., according to the sign of the velocity (\ref{vitesse}).

Let us suppose that the probability that the parameter $\alpha$
changes of an amount between $\Delta$ and $\Delta + {\rm d}\Delta$,
during the time $t$ and under the effect of the irregular thermal
processes, is equal to $\psi_t(\Delta){\rm d}\Delta$, where
$\psi_t(\Delta)=\psi_t(-\Delta)$ is independent of $\alpha$.  This
last assumption reflects the intrinsic nature of thermal agitation.
The number of systems passing through the point $\alpha_0$ during the
time $t$ in the positive direction is given by
\begin{equation}
n_2=\int_{0}^{+\infty} F(\alpha_0-\Delta) \chi_t(\Delta) {\rm
d}\Delta,
 \label{n2}
\end{equation}
where $\chi_t(\Delta)$ is the cumulative probability that the system
makes a jump to the right of size at least $\Delta$ during the time
$t$:
\begin{equation}
\chi_t(\Delta)=\int_{\Delta}^{+\infty} \psi_t(\Delta') {\rm d}\Delta'.
\label{chi}
\end{equation}
Analogously, the number of systems that, under the effect of thermal
fluctuations, pass through the value $\alpha_0$ in the negative
direction during the same time is (taking into account the algebraic
sign),
\begin{equation}
n_3=-\int_{0}^{+\infty} F(\alpha_0+\Delta) \chi_t(\Delta) {\rm
d}\Delta,
\label{n3}
\end{equation}
where we have used the symmetry property
\begin{equation}
\chi_t(\Delta)=\int_{\Delta}^{+\infty} \psi_t(-\Delta') {\rm d}\Delta'.
\label{chi-}
\end{equation}

The equation which mathematically states the invariance of the
equilibrium distribution  $F(\alpha)$ is thus the law of
algebraic conservation of the number of ensembles
\begin{equation}
n_1+n_2+n_3=0.
\label{dyn}
\end{equation}
By substituting the expressions for $n_1$, $n_2$, and $n_3$, by
remembering that $t$ is infinitesimally small, as well as the values of
$\Delta$ for which $\psi_t(\Delta)$ is different from $0$, and by performing a
first order expansion, one finds the essential equation\footnote{In
fact, we find that for the part concerning the thermal fluctuations
$$n_2+n_3=\int_{0}^{+\infty} {\rm d}\Delta \left[F(\alpha_0-\Delta)-
F(\alpha_0+\Delta)\right]\chi_t(\Delta)
=-2F'(\alpha_0)\int_{0}^{+\infty}{\rm d}\Delta\,
\Delta\,\chi_t(\Delta),
$$ where the integral is explicitly written
$$2\int_{0}^{+\infty}\Delta\,{\rm d}\Delta
\int_{\Delta}^{+\infty}  \psi_t(\Delta') {\rm d}\Delta'=
\int_{0}^{+\infty}  (\Delta')^2 \psi_t(\Delta') 
{\rm d}\Delta'=\frac{1}{2}\langle \Delta^2\rangle_t,$$ after having
exchanged the order of integrations or again integrated by
parts.}:
\begin{equation}
B\left(\frac{\partial \varPhi}{\partial
\alpha}\right)_{\alpha=\alpha_0}
\times\, t F(\alpha_0)+\frac{1}{2}F'(\alpha_0) \langle \Delta^2\rangle_t=0.
\label{equil}
\end{equation}
Here $$\langle \Delta^2\rangle_t=\int_{-\infty}^{+\infty} \Delta^2
\psi_t(\Delta) {\rm d}\Delta$$ 
represents the average quadratic variation of the quantity $\alpha$
due to thermal agitation during time~$t$.

Then, by using Boltzmann's distribution $F(\alpha)\propto
\exp\left[-\frac{\mathcal N}{RT} \varPhi(\alpha)\right]$ which
automatically satisfies equation (\ref{equil}) for any potential,
Einstein obtains the average quadratic fluctuation
\begin{equation}
\langle \Delta^2\rangle_t=2 B\frac{RT}{\mathcal N} t.
 \label{delta2}
\end{equation}
Here, as before, $R$ is the perfect gas constant, $\mathcal N$ is
Avogadro's number, $B$ is the system mobility with respect to the
parameter $\alpha$, $T$ is the absolute temperature, and $t$ is the
time interval during which $\alpha$ changes due to thermal agitation.

Einstein's study shows that Boltzmann's equilibrium distribution, {\it
dynamically} interpreted as in the conservation equation (\ref{dyn}),
implies the existence of Brownian diffusion
for any physical quantity $\alpha$ for which the system possesses a mobility.

This idea is so rich that one can reverse the point of view and
consider the equilibrium equation (\ref{equil}) as an equation for
$F(\alpha)$, where $\langle \Delta^2\rangle_t$ is independent of
$\alpha$ and where $t$ is arbitrary.  It is then remarkable that the
solution of (\ref{equil}) necessarily has the exponential form of
Boltzmann's distribution (\ref{Boltzphi}), where $\frac{RT}{\mathcal
N}$ appears as a parameter connected with Brownian diffusion,
according to the identity (\ref{delta2}).  In other words, Einstein's
study of the general dynamics of Brownian motion implies equally well
the particular form of the Boltzmann-Gibbs equilibrium
distribution\footnote{This strongly suggests introducing, in
courses on Statistical Physics, Einstein's demonstration of Brownian
motion, in order to clarify the statistical and dynamical nature of
thermodynamic equilibrium.  In fact, in the usual approach,
Brownian motion is not taught at first, and even when it is, it
appears more as a curiosity.  The approach that one usually takes
consists in introducing Boltzmann's distribution, either via the
microcanonical ensemble and the associated Boltzmann entropy, and by
evaluating the latter for a small system in contact
with a thermostat, or via Shannon statistical entropy and the canonical
ensemble.  In these formal approaches, the emphasis is put on the
probabilities and one does not see the necessity of the thermal
agitation process for keeping the equilibrium distribution
dynamically.  After all, molecules or particles in suspension, even
when initially distributed according to Boltzmann's statistics,
will always fall to the bottom of the container under the effect of
gravity in the absence of thermal agitation!}.\medskip

Einstein applies the result (\ref{delta2}) to translational and
rotational Brownian motions.  For translational motions, the
parameter $\alpha$ is any spatial coordinate $x$, and one needs to
insert the corresponding value of the mobility $B$.  For a sphere of
radius $a$ in suspension in a liquid of viscosity $\eta$, Stokes'
formula, for which he cites Kirchhoff's course\footnote{G. Kirchhoff,
{\it Vorlesungen \"uber Mechanik}, 26. Vorl., \S 4, Teubner, Leipzig (1897); available on http://gallica.bnf.fr/.}, gives
$$B=\frac{1}{\mu}=\frac{1}{6\pi \eta a},$$ and we find the famous
formula (\ref{x2}) again:
\begin{equation}
\label{delta2'}
\langle x^2\rangle_t=\frac{RT}{\mathcal N}\frac{1}{3\pi \eta\, a} t.
\end{equation}

Next, Einstein considers for the first time the Brownian motion of the
{\it rotation} of a sphere suspended in a liquid, and he considers the
squared fluctuations $\langle \vartheta^2\rangle$ of any rotation
angle $\vartheta$ resulting from the thermal agitation.

 If one then defines $\varGamma=-\frac{\partial \varPhi}{\partial
\vartheta}$ the moment of the forces acting on a sphere suspended in a
liquid with viscosity $\eta$, then the associated angular limit
velocity is (again from Kirchhoff):
\begin{equation}
\frac{{\rm d}\vartheta}{{\rm d}t}=\frac{\varGamma}{8\pi\eta a^3},
 \label{vitesseang}
\end{equation}
and in this case, one has: $$B=\frac{1}{8\pi\eta a^3}.$$ One deduces
\begin{equation}
\label{delta2''}
\langle \vartheta^2\rangle_t=\frac{RT}{\mathcal N}\frac{1}{4\pi \eta\, a^3} t.
\end{equation}
The angular motion produced by the molecular thermal agitation
decreases with the radius of the sphere much faster than the translational motion does.

For $a=0.5\, {\rm mm}$, and with water at $17^{\rm o}$ C, the formula
gives, for $t=1\, {\rm s}$, an angular shift of roughly 11 seconds of an
arc, while for $a=0.5 \, \mu{\rm m}$ it gives for the same time duration
roughly $100^{\rm o}$ of arc.\medskip

Finally Einstein mentions that the same formula (\ref{delta2}) for
$\langle \Delta^2\rangle_t$ can be applied to other situations.  For
example, if $B$ is chosen as the inverse of the electric resistance
$\rho$ of a closed circuit, the formula indicates the average squared total charge $$\langle
e^2\rangle_t=2\frac{RT}{\mathcal N}\frac{1}{\rho} t$$ which moves
through any section of the circuit during time $t$.\medskip

Einstein concludes his article by assessing the limits of applicability
of his formula at very short time scales, for which memory effects can
occur.  He arrives  thereby at the estimate that  the formula is valid for $t$ large compared to a
characteristic time $\tau'=m'B$, where $m'$ is the mass of the fluid
displaced by the sphere.

\subsubsection{The problem of measuring the velocity}

In subsequent articles, published in 1907 and 1908 in the {\it
Zeitschrift f\"ur Elektrochemie}, Einstein tries to draw
experimenters' attention to his results and to explain them in a
simpler manner.  He comes back to the average velocity of a particle
in suspension, which must follow the equipartition law
$$\frac{1}{2}m\langle v^2\rangle=\frac{3}{2}\frac{RT}{\mathcal N}.$$

For Svedberg's colloid solutions of platinum, of mass $m
\approx 2.5 \times 10^{-15}$ g, it gives an average velocity of
8.6 cm/s.  However Einstein says that there is no possibility to observe such
a velocity because of the effectiveness of viscous friction, which
reduces the velocity to 1/16 of its initial value in $3.3 \times
10^{-7}$ s.  He continues:\footnote{A. Einstein, {\it Zeit. f. Elektrochemie}, {\bf 13}, pp. 41-42 (1907); translated
in A. Einstein, {\it Investigations on the
Theory of the Brownian Movement}, {\it op. cit.}}\smallskip

{\small ``But, at the same time, we must assume that the particle
gets new impulses to movement during this time by some process that is the inverse of
viscosity, so that it retains a velocity which on average is equal to $\sqrt{\langle
v^2\rangle}$. But since we must imagine that direction and magnitude  of
these impulses are (approximately) independent of  the
original direction of motion and velocity of  the particle, we must conclude that
the velocity and direction of motion of the particle will be already very greatly altered
 in the extraordinarily  short time $\theta$ [$=3.3 \times 10^{-7}$
s] and, indeed, in a totally irregular manner.

It is therefore
impossible --at least for ultra-microscopic particles-- to ascertain $\sqrt{\langle v^2\rangle}$ by
observation.''}\medskip

According to Einstein's result (\ref{x2}), the apparent velocity in a
time interval $\tau$ is inversely proportional to $\sqrt{\tau}$ and
therefore grows without limit when this time interval becomes shorter.
Any attempt to measure the instantaneous velocity of a particle brings
one to erratic results.  This explains experimenters' repeated
failures to obtain well defined conclusions for the velocity of
particles in suspension.  They simply were not measuring the correct
quantity, and they had to wait for Einstein to show that only the
ratio of the quadratic displacement over time has a theoretical limit
for the experiments to connect to the theory.

As Brush remarked,\footnote{S. G. Brush, {\it op. cit.}, pp. 682-683.} it was not
the first time that the particular nature of a motion governed by a
diffusion equation pointed out something right under one's nose.  In
1854, William Thomson (who would go on to become Lord Kelvin) applied
the diffusion equation (i.e.,  Fourier's equation for heat conduction)
in his study of motion of electricity in telegraph lines.  After
having carried out almost exactly the same mathematical analysis that
Einstein would do fifty years later, Thomson wrote:\smallskip

{\small ``We may infer that the  signal delays are
proportional to the squares of the distances, and not to the distances
simply; and hence different observers, believing they have found a
``velocity of electric propagation,'' may well have obtained widely
discrepant results; and the apparent velocity would, {\it caetaris
paribus}, be the less, the greater the length of wire used in the
observation.''}\medskip

A better estimate of the very short time behavior of particles
in suspension follows from subsequent work made by many
physicists,\footnote{P. Langevin, {\it C. R. Ac. Sci. Paris} {\bf
146}, 530 (1908); L. S. Ornstein, {\it Proc. Amst.} {\bf 21}, 96
(1918); L. de Haas-Lorentz, {\it The Brownian Mouvement and some
Related Phenomena}, {\it Sammlung Wissenschaft}, B. 52, Vieweg
(1913); R. F\"urth, {\it Zeit. f. Physik} {\bf 2}, 244 (1920).} among
which those of Langevin, through his stochastic equation that we will
see later, and that culminated with the
Ornstein-Uhlenbeck analysis.\footnote{G. E. Uhlenbeck and L. S. Ornstein, {\it
On the Theory of Brownian Motion}, {\it Phys. Rev.} {\bf 36}, pp. 823-841
(1930).}

A more complete formula is actually
\begin{equation}
\label{delta2comp}
\langle \Delta^2\rangle_t =2 D \left[t-mB\left(1-e^{-\frac{t}{mB}}\right)\right],
\end{equation}
where $D=B\frac{RT}{\mathcal N}$ is the diffusion coefficient, and $m$
this time is the mass of the particle.  Therefore we clearly get the
formula (\ref{delta2}) for $t$ large compared to the microscopic time
\begin{equation}
\label{tau}
\tau=mB=\frac{m}{\mu},
\end{equation}
of the same order of magnitude as the time $\tau'$ estimated by
Einstein.
	
For $t$ smaller than $\tau$, we find a {\it ballistic} regime
\begin{equation}
\label{ball}
\langle \Delta^2\rangle_t = D \frac{t^2}{mB}=\frac{RT}{\mathcal N}
\frac{1}{m}t^2,\,\,\,\,\, \tau \gg t,
\end{equation}
independent of the viscosity of the medium, and which remarkably can
be interpreted as corresponding to the energy equipartition theorem,
this time in the form: $$\frac{1}{2} m\frac{\langle
\Delta^2\rangle_t}{t^2}=\frac{1}{2}\frac{RT}{\mathcal N}\,\,\,\,\,
\tau \gg t.$$

 \subsubsection{Einstein's third derivation  of Brownian motion}
A third approach to Brownian motion was incidentally offered by Einstein in a lecture given in front of the
Z\"urich Physical Society, on 2 November 1910, which was entitled: {``On Boltzmann's Principle
and Some Immediate Consequences Thereof.''}\footnote{{\it \"Uber
das Boltzmann'sche Prinzip und einige unmittelbar aus demselben
fliessende Folgerungen}, {\it Vorlesungen f\"ur die
 Physikalische Gesellschaft Z\"urich}, 2 November 1910, Zangger Nachla\ss, Zentral Bibliothek Z\"urich. English
 translation by B. Duplantier \& E. Parks: {\it On Boltzmann's Principle and Some Immediate Consequences Thereof}, in: {\it Einstein, 1905-2005},
Poincar\'e Seminar 2005, Eds. T. Damour, O. Darrigol, B. Duplantier and V. Rivasseau, pp. 183-199 (Birkha\"user Verlag, Basel, 2006).
}
 This text seems not to have appeared in print before, so
an English  translation, followed by a commentary, is included in this volume.

In this fascinating lecture, Einstein describes his point of view on Statistical Physics at that time.
He illustrates it
 by stressing  the role of
fluctuations, in relation to Boltzmann's formula for the entropy. This text is of particular importance,
since Einstein asks more generally
whether a complete causal connection can always be found between physical events; this epistemological
interrogation takes place 
 at the dawn of Quantum Mechanics.

 Among
other examples,  Einstein considers the case of  a suspended particle
 in a gravitational field, and performs a calculation of
 the mean square position of the  particle. From the simple
  assumption of the stationarity of that average, he rederives the famous Sutherland-Einstein formula
 (\ref{ESf}). This is perhaps the most direct and illuminating derivation of the Brownian diffusion formula!


\subsection{Marian von Smoluchowski}


\centerline{{\it ``A throw of the dice never will abolish chance.''} (St\'ephane Mallarm\'e, 1897)}

\medskip

\subsubsection{Probabilities and stochasticity}

Smoluchowski's name is closely attached to Brownian motion and the
theory of diffusion, as we will show here.  Moreover, as Marc Kac
wrote about Smoluchowski,\footnote{{\it Marian Smoluchowski, His Life
and Scientific Work}, S. Chandrasekhar, M. Kac, R. Smoluchowski,
Polish Scientific Publishers, PWN, Warszawa (2000).} the latter showed through a true
intellectual {\it tour de force}, that the notion of a game of
chance lies at the heart of our comprehension of physical phenomena.  We are
indebted to him for his original and bold introduction of the calculus
of probability in statistical physics, and he deserves a place
beside the great names of Maxwell, Boltzmann, and Gibbs.\medskip

Marian von Smoluchowski was born in 1872, the same year as Paul
Langevin, and the year Boltzmann published the great memoir
containing the equation that bears his name, as well as the famous
``$H$ theorem''. There, Boltzmann derives the irreversible increase of
entropy linked to the second principle of thermodynamics, in the area
of classic Newtonian mechanics, with the help of a hypothesis of
molecular chaos, which Smoluchowski thought should have been instead a
consequence in this framework.  This brought about serious paradoxes
(Loschmidt,\footnote{J. Loschmidt, {\it Wien. Ber.} {\bf 73}, 139 (1876); {\bf 75}, 67 (1877).}
Zermelo\footnote{E. Zermelo, {\it Ann. d. Physik} {\bf 57}, 485 (1896); {\bf 59}, 793 (1896).}), because the equations of classical mechanics are
reversible and have recurring cycles, called Poincar\'e recurrence cycles. So
this forbade {\it a priori} the monotonic growth of a function of
positions and the momenta, as seen for Boltzmann's $H$ function which is
directly connected to entropy.  Each time on the defensive,\footnote{L. Boltzmann,
{\it Wien. Ber.} {\bf 75}, 62 (1877); {\bf 76}, 373 (1877); see also {\it Nature} {\bf 51}, 413 (1895) and
{\it Vorlesungen \"uber Gas Theorie} {\bf I}, 42, Leipzig (1895) (or the reprinted edition of 1923).}
\footnote{L. Boltzmann, {\it Ann. d. Physik} {\bf 57}, 773 (1896); {\bf 60}, 392 (1897).} Boltzmann had to
introduce probabilistic and statistical arguments to justify his
results, often by completely changing his point of view about the true
nature of the probabilities involved.  The situation became so
confused that Paul and Tatyana Ehrenfest, for example, tried to
clarify Boltzmann's ideas by banishing the term (but not the concept)
``probability'' from their famous 1911 Encyclopedia memoir!\footnote{P. and T. Ehrenfest,
{\it Begriffliche Grundlagen der statistischen Auffassung in der Mechanik}, {\it Encyklop\"adie der 
mathematischen Wissenschaften} {\bf 4}, 4 (1911).}\medskip

As S. G. Brush noted,\footnote{S. G. Brush, {\it loc. cit.}} the
research line of the kinetic theory of gases that Smoluchowski
pursued was a continuation of that of Clausius, Maxwell, O.E. Meyer,
Tait and Jeans, according to which one describes the effects of
collisions on the trajectory of a molecule, and therefore on the
properties of the gas.  Einstein, on the contrary, followed the
path opened by Boltzmann, Maxwell (in his subsequent articles) and
Gibbs, where the objective was to obtain more general laws starting
from statistical distributions postulated for molecular ensembles,
without making any assumption about intramolecular forces and
collision mechanisms.  It is thus extremely interesting to see these
two ``Gedankenwege,'' kinetic theory and statistical mechanics, meet
up in relation to Brownian motion, {\it terra incognita} for both
theories.\medskip

In this context, by working in the same pragmatic spirit as Maxwell,
Smoluchowski courageously showed how to use the theory of probability
in physics as an efficient instrument, during an era when
mathematicians looked down on it, and physicists mostly ignored it.
Without knowing it, Smoluchowski opened a new sub-field of statistical
physics, that nowadays bears the name Stochastic
Processes.\footnote{From the Greek word $\sigma\tau o\chi\alpha\sigma\tau\iota\kappa$\'o$\varsigma$
{\it (stokhastikos)}, ``to aim well,''
``capable of making conjectures,'' already used by Jacob Bernoulli in
1713 in {\it Ars Conjectandi}: {\it ``We define the art of conjecture, or stochastic art, as the art of
evaluating as exactly as possible the probabilities of things, so that in our judgments and actions we can always
base ourselves on what has been found to be the best, the most appropriate, the most certain, the best advised; this is the
only object of the wisdom of the philosopher and the prudence of the statesman.''}}

\subsubsection{Brownian motion and random walks}

This probabilistic point of view is clearly present in Smoluchowski's
first article on Brownian motion, {\it ``Essay on the theory of
Brownian motion and disordered media''}\footnote{M. R. von Smolan
Smoluchowski, {\it Rozprawy Krak\'ow} {\bf 46 A}, pp. 257-281 (1906); French
translation: ``Essai d'une th\'eorie du mouvement brownien et de
milieux troubles,'' {\it Bull. International de l'Acad\'emie des
Sciences de Cracovie}, pp. 577-602 (1906); German translation: {\it
Ann. d. Physik} {\bf 21}, pp. 755-780 (1906).} published in 1906 (very
likely under the pressure of Einstein's publication of his first two
articles), as well as in another article, about the mean free path of
molecules in a gas.\footnote{M. R. von Smolan Smoluchowski, {\it Sur le
chemin moyen parcouru par les mol\'ecules d'un gaz et sur son rapport
avec la th\'eorie de la diffusion},
{\it Bulletin International de l'Acad\'emie des Sciences de Cracovie},
pp. 202-213 (1906).}  In these remarkable articles he was seemingly the
first to establish the relation between {\it random walks} and
Brownian diffusion, even though in 1900 Louis Bachelier had already
introduced the model of a random walker in his thesis {\it The Theory
of Speculation}. We shall return to this later.

Smoluchowski begins by citing Einstein's work from 1905 and writes
that the latter's results {\it ``completely agree with those I obtained a few
years ago by an entirely different path of reasoning, and that since
then I have considered an important argument in favor of the kinetic
nature of these phenomena.''}  However, he adds further along that his
own method {\it ``seems more direct, simpler, and perhaps more
convincing than that of Einstein.''}

\begin{figure}[htbp]
\begin{center}
\includegraphics[angle=0,width=.3\linewidth]{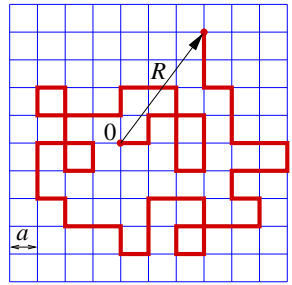}
\end{center}
\caption{{\it Random walk on a square lattice with elementary lattice step $a$.
We choose each step at random.  In two dimensions, two equivalent
methods exist.  In the first one, we draw heads or tails (with a
probability of $1/2$) for a direction, vertical or horizontal, and next
the orientation along the chosen direction.  In the second method, we
draw with the same probability (with probability of $1/4$) one of the
four possible directions.  In the continuous limit where the lattice
step goes to 0, a very long random walk will take the appearance of the
Brownian motion of figure \ref{fig.brownian}.}}
\label{fig.RW}
\end{figure}

While Einstein (like Sutherland) avoids all treatment of collisions in favor of a
general thermodynamic approach, Smoluchowski has a clear
kinetic vision and treats the Brownian motion as a random walk or a
game of heads or tails (see figure \ref{fig.RW}).

The newness and the originality of Smoluchowski's approach is in the
replacement of an incredibly difficult problem (a Brownian particle
which collides within a gas or liquid) by a relatively simple
stochastic process.  Each dynamic event like a collision is considered
as a random event similar to a game of heads or tails, or to the throw
of a dice, where the elementary probabilities are (to a certain
extent) determined by underlying mechanical laws.  This way of
reasoning plays a fundamental role in mechanics and statistical
physics today and, as Marc Kac noticed, it is difficult for us today
to imagine the degree of Smoluchowski's intellectual boldness in starting
this subject during the early years of the last century.

\subsubsection{Smoluchowski's contributions}

Smoluchowski\footnote{In this section we follow Brush's presentation of Smoluchowski's work.} knew about the most recent studies on Brownian motion and
in particular the work of Felix Exner. The latter sent Smoluchowski diagrams
made from memory, called ``Krix-Krax'' because of the several
inter-crossing ``jumps'' apparently made by a Brownian particle
observed under a microscope over a set of discrete instants of time.\medskip

Smoluchowski began by criticizing N\"ageli's arguments which claimed
that a collision of a molecule of water with a sphere 0.001 mm in
diameter would give a velocity of $3 \times 10^{-6}$ cm/s, which would
be impossible to observe under a microscope, and  that the
collision effects would cancel out on average.  He compared this way
of thinking to that of a player who believed himself never to be able to lose
more than a single bet, despite repeated draws!  By continuing the
analogy further, he calculated for the heads or tails game how the
positive (or negative) cumulated gains grow with the number $n$ of
draws (``time'').

Let $p_{n,m}$ be the probability to have met $m$ favorable outcomes in the
total of $n$ draws, with a net gain of $m -(n-m)= 2m-n$.  This
probability can be written as
\begin{equation}
\label{binomial}
p_{n,m}=\frac{1}{2^n}\frac{n!}{m!(n-m)!}=\frac{1}{2^n} \left( n\atop m\right),
\end{equation}
where the number of combinations $\left( n\atop m\right)$ is the number of ways of choosing
$m$ out of $n$ objects.

The positive or negative mean deviation from zero, $\delta_n$,
 i.e., the average of the absolute value of a gain or of a loss after
$n$ turns, can be calculated as $$\delta_n=\langle
|2m-n|\rangle=2\sum_{m={n}/{2}}^{n}(2m-n)p_{n,m}=2\sum_{m={n}/{2}}^{n}(2m-n)\frac{1}{2^n}
\left( n\atop m\right)=\frac{n}{2^n} \left( n\atop {\frac{n}{2}}\right),$$ where $n$ is supposed an
even number, to simplify the notation.  For large $n$, we then use
Stirling's formula $n!\simeq n^n e^{-n}
\sqrt{2\pi}$, to evaluate $\delta_n$:
$$\delta_n \simeq \sqrt{\frac{2 n}{\pi}},\,\,\, n\gg 1.$$ The
(arithmetical) average of successive gains (or losses) with respect to $0$
increases as $\sqrt n$, even when the total (algebraic) average is zero.
The analogous number $n$ of molecular collisions per second on a
sphere was estimated by Smoluchowski as $10^{16}$ in a gas and
$10^{20}$ for a liquid.  If the gain in velocity is of the order
$10^{-6}$ cm/s at each collision, one obtains a mean cumulated
velocity of from $10^2$ to $10^4$ cm/s per second.  However Smoluchowski
immediately reduces this conclusion, remarking that each individual
gain of velocity will fluctuate, and that a high velocity value
decreases the probability of one more positive gain.

He shows next that a ``true'' velocity could be
obtained from the equipartition of kinetic energy, which gives a
velocity of 0.4 cm/s, again much too large in relation to experimental
observations!  In fact, Exner's diagrams in ``Krix-Krax'' gave a
velocity of about $3\times 10^{-4}$ cm/s, an apparently irreconcilable
disagreement.  As Smoluchowski says, {\it ``this contradiction,
already seen by F. Exner, seems at first to be a decisive objection to
kinetic theory.  Nevertheless the explanation is very simple.''}

He presents the following simple and clear explanation: such a
velocity is too large to be observed with a microscope magnifying 500
times.  What one observes is the average position of a particle having
this velocity, but hit $10^{20}$ times per second, each time in a
different direction, such that one cannot observe the instantaneous
velocity.  Each zig-zag displacement is incomparably smaller than the
particle's size, and it is only when the geometric sum of these
elements reaches a certain value that one can observe a displacement.
This is clearly the substance of
Einstein's argument, here supported by the concrete image of kinetic
theory: the average displacement is the observable physical quantity,
while velocity is not.\medskip

After such qualitative, but illuminating, considerations, Smoluchowski
develops his model of random collisions. Let $m$ and $v$ ($m'$ and $
v'$ respectively) be the mass and the velocity of a particle in
suspension (of molecules in the liquid, respectively). From the
equipartition of energy, one has on average:
\begin{equation}
\frac{v}{ v'}=\sqrt\frac{m'}{m}.
\label{equipa}
\end{equation}
He affirms that from {\it ``the laws of collision of elastic
spheres,''} the change of velocity of the sphere in suspension is, at
a collision, on average given by a small transverse component $\alpha \,
m' v'/m$, where $\alpha=3/4$.  The result is a random change of the
velocity direction of a small angle $\varepsilon=\alpha \, m' v'/mv$.
(According to (\ref{equipa}), one also has $\varepsilon=\alpha v/ v'$
on average.)  He assumes also that the molecular impacts occur after
equal intervals of time, which makes the particle trajectory a
chain made of constant-length segments.

In other words, Smoluchowski adapts the idea of the mean free path of a
molecule in a gas, even though here the persistence of motion is
shortened by the presence of numerous molecules of the surrounding fluid.
 
The problem of Brownian motion is thus mathematically mapped onto the
one of finding the end-to-end average distance $\Delta_n^2$, of a
chain of $n$ segments, all of length $\ell$, each randomly turned by a
small angle $\varepsilon$ with respect to the preceding one.  He
then obtains the general solution by a complicated recurrence
relation, containing multiple angular integrals of trigonometric
functions, of the form:
\begin{equation}
\Delta_n^2=\ell^2\left\{\frac{2n}{\delta}+1-n-2\frac{(1-\delta)^2-(1-\delta)^{n+2}}{\delta^2}\right\},
\label{recurr}
\end{equation}
where $\delta=1-\cos \varepsilon\simeq \varepsilon^2/2$.

In the limit where $n\delta$ is small, one finds
\begin{equation}
\Delta_n=n\ell\left(1-\frac{n\delta}{6}\right),
\label{recurr1}
\end{equation}
which represents a quasi-ballistic trajectory.

In the opposite case of a large number of collisions per unit of time
$n\delta\gg 1$ , the first term of (\ref{recurr}) dominates and one
finds the expected result:
\begin{equation}
\Delta_n^2=\ell^2 \frac{2n}{\delta}=\ell^2 \frac{4n}{\varepsilon^2}.
\label{recurr2}
\end{equation}

If we call $\bar n$ the number of collisions per unit of time, such
that there are $n=\bar n\, t$ collisions over the time $t$, we have
for a free path $\ell=v/\bar n$, and by using $\varepsilon=\alpha v/
v'$, we find an average quadratic displacement at time $t$,
\begin{equation}
\Delta_n^2\equiv \Delta_t^2=\frac{4}{\alpha^2}\frac{ v'^{\,2}}{{\bar n}} { t}.
\label{recurr3}
\end{equation}
The momentum $mv$ of the particle in suspension changes on average by
a quantity $\alpha'm'v$ per collision, where, from Smoluchowski,
$\alpha'=2/3$, which means the friction force $F=-\bar n \alpha' m'v$, and
thus the friction coefficient $\mu=\bar n \alpha' m'$.  Substituting
$\mu$ in $\bar n$ one obtains: $\Delta_t^2=\frac{4{\alpha'}}{\alpha^2}\frac{{m'}
v'^{\,2}}{{\mu}} { t}.$ From the equipartition of kinetic energy of the
molecules in the surrounding fluid: $\langle {m'}{
v'^{\,2}}\rangle=3RT/\mathcal N$, and the result of Smoluchowski
finally becomes:
\begin{equation}
\Delta_t^2=\frac{2\alpha'}{\alpha^2}\,6\frac{RT}{{\mu}\mathcal N} { t}.
\label{recurr4}
\end{equation}
One finds again the Sutherland-Einstein result (\ref{ESf}), (\ref{x2}),
this time in three dimensions, with a supplementary numerical factor
of kinetic origin $2\alpha'/{\alpha^2}=(4/3)^3=64/27$. Because
of the various physical and geometrical approximations involved, this factor should perhaps not
come as a surprise!  The experiments of The (Theodor) Svedberg in 1907 seemed to
support this result, but Langevin mentioned later in 1908, in his
article in the Comptes Rendus, that once these approximations were
corrected, Smoluchowski's stochastic method gave the same formula
(\ref{x2}) as Einstein's method.  Smoluchowski himself adopted this formula in
his subsequent articles.

Afterwards he gave the complete theory of density fluctuations
within an ensemble of Brownian particles, as well as that of their
sedimentation in a gravitational field and of the coagulation of
colloids.\footnote{M. von Smoluchowski, {\it Drei Vortr\"age
\"uber Diffusion, Brownsche Molekularbewegung und Koagulation von
Kolloidteilchen, Physikalische Zeitschrift}, {\bf Jg. 17}, pp. 557-571,
pp. 585-599 (1916). The English translation from German can be found in: {\it Marian Smoluchowski, His Life
and Scientific Work}, S. Chandrasekhar, M. Kac, R. Smoluchowski,
Polish Scientific Publishers, PWN, Warszawa (2000), pp. 43-127.} The content of this reference is described in detail by S. Chandrasekhar in his famous review article
on {\it Stochastic Problems in Physics and Astronomy}\footnote{S. Chandrasekhar,
{\it Rev. Mod. Phys.} {\bf 15}, pp. 1-89 (1943), see in Chap. III the enlightening discussion of Smoluchowski's theory
of fluctuations and its experimental verification, as well as of the limits of validity of the Second Law of Thermodynamics.} and praised as follows:\smallskip

{\small ``In [this] reference [...] we have an extremely valuable account of the entire subject of Brownian motion and
molecular fluctuations; there exists no better introduction to this subject than these lectures by Smoluchowski.''}

He adds:
{\small { ``The theory of density fluctuations as developed by Smoluchowski represents one of the most
outstanding achievements in molecular physics. Not only does it quantitatively account for and clarify a wide
range of physical and physico-chemical phenomena, it also introduces such fundamental notions as the
`probability after-effect' which are of great significance in other connections.''}}\smallskip

We should also mention that we owe to Smoluchowski (and to Einstein) the theory of critical
opalescence as well.

Smoluchowski's name is  traditionally attached to
the generalization of the diffusion equation (\ref{diffusion}) governing the probability density $P(\vec r,t)$ in presence of a
force field $\vec F(\vec r)$:
\begin{equation}
\label{diffusionforce}
\frac{\partial P}{\partial t}=D \Delta_{\vec r} P -\frac{1}{\mu} {\rm div}_{\vec r} (\vec F P),
\end{equation}
where $\mu$ is the same as in (\ref{stokes}). This equation applies directly
 to the case of a uniform gravitational field.  In one
dimension it is simply written as
\begin{equation}
\label{FP}
\frac{\partial P(x,t)}{\partial t}=D \frac{\partial^2}{\partial x^2} P(x,t) +\frac{1}{\mu}
\frac{\partial}{\partial x} \left
(\frac{\partial V(x)}{\partial x} P(x,t)\right),
\end{equation}
for a force field $F(x)$ derived from a potential $V(x)$.

The passage to such a differential equation in configuration space was first achieved by Smoluchowski in
1915.\footnote{M. von Smoluchowski, {\it Ann. d. Physik} {\bf 48}, pp. 1103-1112 (1915).}
This equation, as the standard  ``free field'' diffusion equation, are valid only if we
ignore effects which happen in time intervals of the order of the viscous damping time, $\tau=m/\mu$, introduced
in eq. (\ref{tau}).

When such effects are of interest, as in eq. (\ref{delta2comp}), one should use a more general differential equation, the
so-called {\it Fokker-Planck equation}. The passage to such a differential equation for the description in velocity space
of the Brownian motion of a free particle
 was indeed achieved by Fokker,\footnote{A. D. Fokker, Thesis, Leiden (1913); {\it
Ann. d. Physik} {\bf 43}, 810 (1914).} while a more general discussion of this problem
is due to Planck.\footnote{M. Planck, {\it
Sitzungsber. Preuss. Akad. Wissens.} p. 324 (1917); in {\it
Physikalische Abhandlungen und Vortr\"age} II, p. 435, Vieweg,
Braunschweig (1958).} Let us also mention the pionnering work by Rayleigh in one dimension as early as
1891!\footnote{J. W. S. Rayleigh, {\it Phil. Mag.} {\bf 32}, pp. 424-445 (1891);
in {\it Scientific Papers by Lord Rayleigh}, Vol. III, {\bf 183}, pp. 473-490, Dover Publications, New-York (1964).}

The Fokker-Planck equation is a differential equation governing the time evolution of the probability density
$\mathcal P(\vec p,t)$
in {\it velocity} ($\vec v$)  or {\it momentum} ($\vec p=m\vec v$) space, valid for all time intervals. In the absence of an
external force field, it has the form
\begin{equation}
\label{diffFP}
\frac{\partial \mathcal P}{\partial t}= {\mu^2} D\Delta_{\vec p} \mathcal P +{\mu}\, {\rm div}_{\vec p}
\left(\frac{\vec p}{m}\mathcal P\right).
\end{equation}
This is fully equivalent to the description of Brownian motion
by the {\it stochastic} Langevin equation, described in \S\ref{langevin} below.

Their solution gives the full Ornstein-Uhlenbeck process.
For times shorter than the damping time $\tau$, a ballistic regime dominates,
while asymptotically one recovers the so-called ``overdamped'' regime, i.e.,  standard diffusion. [See eqs. (\ref{delta2comp}-\ref{ball}) and \S\ref{OrnsteinUhl}.]

When an external force field is present, a more general probabilistic description in phase space, involving the
probability density function $\mathbb P (\vec p,\vec r, t)$ for both
the momentum $\vec p$ and the position $\vec r$ of the Brownian particle, is
 required.
Its time evolution, valid for all time intervals, is then given by
\begin{equation}
\label{diffFPcomp}
\frac{\partial \mathbb P}{\partial t}+\frac{\vec p}{m}\cdot\vec \nabla_{\vec r} {\mathbb P}+\vec F\cdot\vec\nabla_{\vec p}{\mathbb P}=
{\mu^2} D\Delta_{\vec p} \mathbb P +{\mu}\, {\rm div}_{\vec p}
\left(\frac{\vec p}{m}\mathbb P\right).
\end{equation}
The foregoing equation represents the complete generalization of the Fokker-Planck
equation (\ref{diffFP}) to the phase space. At the same time eq. (\ref{diffFPcomp}) represents also the generalization of
 Liouville's equation of
classical mechanics to include Brownian motion; more particularly, on the right-hand side of eq. (\ref{diffFPcomp}) we have
the terms arising from Brownian motion while on the left-hand side we have the usual Stokes differential operator $D/Dt$
acting on $\mathbb P$.

The earliest attempts to generalize Liouville's equation of
classical mechanics to include Brownian motion were made by O. Klein\footnote{O. Klein,
{\it Arkiv for Matematik, Astronomi, och Fysik} {\bf 16}, No. 5 (1921).}
and H. A. Kramers,\footnote{H. A. Kramers, {\it Physica} {\bf 7}, 284 (1940).}
culminating with the work by S. Chandrasekhar.\footnote{S. Chandrasekhar,
{\it Rev. Mod. Phys.} {\bf 15}, pp. 1-89 (1943), Chap. II.}

\subsubsection{Brownian motion and the second principle}

Another aspect of Smoluchowski's work concerns the correct
statistical formulation of the second principle of thermodynamics.\footnote{For a recent discussion
 of the physics and mathematics behind the Second Law, see: E. Lieb and J. Yngvason, {\it Phys. Today} {\bf 53}-4, pp. 32-37 (2000);
{\it The physics and mathematics of the Second Law of Thermodynamics}, {\it Phys. Rep.} {\bf 310}, pp. 1-96 (1999);
Erratum {\bf 314} (1999); arXiv: cond-mat/9708200. See also:
G. Gallavotti, {\it Statistical Mechanics, a Short Treatise}, Springer-Verlag, Heidelberg (1999).}
With The Svedberg's recent data on Brownian motion,\footnote{The Svedberg, {\it Zeits. f. physik. Chemie} {\bf 77}, 147 (1911).} Smoluchowski
had experimental results which permitted him, armed with his own
theory of fluctuations near-to-equilibrium, to estimate the {\it persistence} and {\it recurrence
 times} of a system slightly out of equilibrium, and to
check these results against experiments.\footnote{M. von Smoluchowski, {\it Wien. Ber.} {\bf 123}, pp. 2381-2405 (1914); 
see also {\it Phys. Z.} {\bf 16}, pp. 321-327 (1915) and {\it Kolloid Z.} {\bf 18}, pp. 48-54 (1916).}  He used neither phase space,
nor Liouville's theorem as in classical statistical mechanics {\it \`a
la} Boltzmann.  He introduced simply the calculus of probability.
By incorporating the theory of fluctuations he gave a correct
formulation of the second principle of thermodynamics, where this
principle appeared valid only in a {\it statistical} sense, and
was
therefore
susceptible to multiple twists at the microscopic level.\footnote{M. von Smoluchowski,
{\it Phys. Z.} {\bf 13}, pp. 1069-1080 (1912);
{\it G\"ottinger Vortr\"age \"uber die kinetische Theorie der
Materie u. Elektrizit\"at}, Leipzig, {pp. 89-121} (1914).} These considerations are relevant to and bear on
Loschmidt's reversibility paradox and Zermelo's recurrence paradox. Discussing these paradoxes in the context of
Boltzmann's views, Smoluchowski concludes that {\it ``a process appears irreversible if the initial
state is characterized by a long average time of recurrence compared to the times during which the system is
under observation.''}

Further precision experiments carried out with expressed
intention of verifying Smoluchowski's theory are those of A. Westgren,\footnote{A. Westgren,
{\it Arkiv for Matematik, Astronomi, och Fysik} {\bf 11}, Nos. 8 and 14 (1916) and {\bf 13}, No. 14 (1918).}
as described in the survey article by S. Chandrasekhar.\footnote{S. Chandrasekhar
{\it Rev. Mod. Phys.} {\bf 15}, pp. 1-89 (1943), Chap. III, \S\S  2-3.}\medskip

A  modern discussion of Smoluchowski's ideas was given by
Richard Feynman in his famous elementary physics lectures.\footnote{R. P. Feynman, R. B. Leighton and M. Sands,
{\it The Feynman Lectures on Physics I}, Chap. 46, Addison-Wesley, Reading MA (1963).} 
He compared Maxwell's demon with a ratchet and pawl and an electric rectifier, neither of
which can systematically transform internal energy from a single reservoir to work. He wrote:

{\it ``If we assume that the specific heat of the demon is not infinite, it must heat up. It has but
a finite number of internal gears and wheels, so it cannot get rid of the extra heat that it gets
from observing the molecules. Soon it is shaking from Brownian motion so much that it cannot tell
whether it is coming or going, much less whether the molecules are coming or going, so it does not work.''}

Modern day computer simulations strikingly reveal the fluctuation phenomena envisaged by
Smoluchowski and Feynman.\footnote{See, e.g., P. A. Skordos and W. H. Zurek, {\it Am. J. Phys.} {\bf 60},
876 (1992).}

Smoluchowski's observation suggested that Maxwell's demon ought to be buried and forgotten.\footnote{In his contribution to
{\it Marian Smoluchowski, His Life
and Scientific Work}, by S. Chandrasekhar, M. Kac \& R. Smoluchowski
[Polish Scientific Publishers, PWN, Warszawa (2000)], Smoluchowski's son Roman recalls the interesting following anecdote. In his book
{\it Inferno} [Mercure de France, Paris (1898)], August Strindberg recalls an instance that occurred when he lived at the
Hotel Orfila in Paris. On the day after his arrival the addresses of several letters waiting by the board of
room keys caught his eye. He mused on one from Vienna which was of particular interest because it bore what he
referred to later as the Polish pseudonym {\it ``Smulachowsky''} and he wrote that he considered the name to be a disguise
and that it was the {\it devil himself} that now interfered in his affairs. Strindberg's speculation was certainly
inspired by the simultaneous presence of Smoluchowski 
at the same Hotel Orfila. The later indeed stayed there from 1895 to early 96 when he worked in
Lippmann's laboratory in Paris, while Strindberg's stay extended from February to July 1896.  Nowadays there are two tablets
 on that house (60-62 rue d'Assas, Paris 6) commemorating separetely Smoluchowski's and Strindberg's stays!} But
that did not happen, apparently because Smoluchowski's approach left open the possibility that somehow,
a perpetual motion machine operated by an ``intelligent'' being might be achievable.\footnote{Max Jammer points
out in {\it The Conceptual Development of Quantum Mechanics}, New York (1966), that Smoluchowski's 1913 Wolfskehl
lectures in G\"ottingen influenced decisively Leo Szilard in his well-known 1929 paper about entropy and information.
Jammer writes: ``Smoluchowski's conception of an intellect that is constantly cognizing of the intantaneous state
of a dynamical system and thus able to invalidate the Second Law of Thermodynamics without performing work was
probably the earliest logically unassailable speculation about a physical intervention of mind on matter.'' [Quoted
by R. S. Ingarden, {\it ed.}, in {{\it Marian Smoluchowski, His Life
and Scientific Work}, 
Polish Scientific Publishers, PWN, Warszawa (2000).]}} It was this fascinating idea
of using intelligence that captured Leo Szilard's interest, in his classic 1929 paper, {\it ``On the decrease of entropy
in a thermodynamic system by the intervention of intelligent beings.''}\footnote{L. Szilard,
{\it Z. Phys.} {\bf 53}, pp. 840-856 (1929); transl. reprinted in {\it The Collected Works of Leo Szilard, Scientific
Papers}, B. T. Feld and G. Weiss Szilard, eds., The MIT Press, Cambridge, Mass. (1972).}

The feature associated with intelligence that is needed by a demon is memory: it must remember what
it measures, even if only briefly. Notably, Szilard discovered with his heat engine, with a one-molecule working fluid,
 the idea of a ``bit'' of information with entropy $k_B \ln 2$, now central
in computer science, and
established the connection between entropy and information.

At this stage, rather than fully opening Pandora's box which contains the Protean Maxwell's demons,
we prefer to recommend the survey, {\it Maxwell's Demon 2}, by H. S. Leff and A. F. Rex and in particular
the thoughtful introduction of the second edition.\footnote{H. S. Leff and A. F. Rex,
{\it Maxwell's Demon 2}, Adam Hilger, Bristol (2003).} Let us only mention a
few historical landmarks that are described in their presentation.\medskip

After a hiatus of 20 years, L\'eon Brillouin, assuming the use
of (quantum) light signals in the demon's attempts to defeat the second law, concluded that
information acquisition, like measurement, is dissipative. This led him to break new ground by developing an extensive
mathematical theory connecting measurement and information. The impact of
Brillouin's and Szilard's work was far reaching and the result was a proclaimed, but temporary, ``exorcism'' of the demon.

A new life began for the demon when Rolf Landauer made the important discovery that memory erasure in computers
feeds entropy to the environment.\footnote{R. Landauer, {\it IBM J. Res. Dev.} {\bf 5}, pp. 183-191 (1961).} This is now called ``Landauer's principle''. It states that
the erasure of one bit of information stored in a memory device requires sending an amount of entropy of at least $k_B \ln 2$
to the environment, i. e., a minimal heat generation of $k_BT \ln 2$.

Charles Bennett, after his important demonstration in 1973 that reversible computation, which avoids the erasure of information,
is possible in principle, argued in 1982 that erasure of a demon's memory is the fundamental act that saves the second
law because of Landauer's principle.\footnote{C. H. Bennett, {\it Int. J. Theor. Phys.} {\bf 21}, pp. 905-940 (1982).}
This was a turning point in the history of Maxwell's demon.

In his 1970 book {\it Foundations of Statistical Mechanics},
Oliver Penrose independently recognized the importance of ``resetting'' operations that bring all members of a
statistical ensemble to the same observational state. Applied to Szilard's heat engine, this is nothing else than
 memory erasure, which sends an amount of entropy of at least $k_B \ln 2$ to the environment.

Among recent proofs of Landauer's principle we cite here, somehow arbitrarily, the one by K. Shizume,
who uses a
solvable model of memory based on a Brownian particle in a time-dependent potential
well;\footnote{K. Shizume, {\it Phys. Rev. E} {\bf 52}, pp. 3495-3499 (1995).} the one by M. Magnasco with
a detailed analysis of Szilard's heat engine;\footnote{M. O. Magnasco, {\it Europhys. Lett.} {\bf 33}, pp. 583-588 (1996).} and
the one by B. Piechocinska, who assumes the
decoherence of the states of the thermal
reservoir\footnote{B. Piechocinska, {\it Phys. Rev. A} {\bf 61}, 062314 (2000).}.

Let us finally mention that, despite several attempts to argue against its validity, the
Landauer-Penrose-Bennett framework seems to be generally accepted as providing the solution to the Max\-well's demon--second principle
 puzzle, at least in
classical mechanics, and in a thermodynamical limit of some
sort.\footnote{For possible violations of Thompson's formulation of the second principle
for a mesoscopic work source,
 see A. Allahverdyan, R. Balian and T. M. Nieuwenhuizen, {\it Entropy} {\bf 6}, pp. 30-37 (2004); see also
 {\it Europhys. Lett.} {\bf 67}, pp. 565-571 (2004).}

However, there are now indications that  Landauer's principle, as well as the second principle, might not hold in
the (strong) quantum regime. The source of the violation is  {\it quantum entanglement} between the system
and the constant-temperature reservoir, which then act as a single
entity.\footnote{A. Allahverdyan and T. M. Nieuwenhuizen, {\it Phys. Rev. Lett.} {\bf 85}, pp. 1799-1802 (2000);
{\it Phys. Rev. E} {\bf 64}, 056117 (2001);
T. M. Nieuwenhuizen and A. Allahverdyan, {\it Phys. Rev. E} {\bf 66}, 036102 (2002); T. D. Kien, 
{\it Phys. Rev. Lett.} {\bf 93}, 140403 (2004). See also M. O. Scully, {\it Phys. Rev. Lett.} {\bf 88}, 050602 (2002);
L. S. Schulman and B. Gaveau, {\it Physica E} {\bf 29}, pp. 289-296 (2005).}\medskip

In close relation to Brownian motion and the second principle,  the topic of  Brownian motors
 has recently received considerable attention.\footnote{P. Reimann, {\it Phys. Rep.} {\bf 361}, 57 (2002);
 R. D. Astumian and P. H\"anggi, {\it Physics Today} {\bf 55}, 33 (2002); H. Linke (ed.),
  {\it Ratchets, Experiments and Applications},
 {\it Appl. Phys.} A {\bf 75} (2002).}
C. Van den Broeck {\it et al.}\footnote{C. Van den Broeck,
R. Kawai and P. Meurs, {\it Phys. Rev. Lett.} {\bf 93}, 090601 (2004); C. Van den Broeck, P. Meurs and R. Kawai, {\it From Maxwell Demon to
Brownian Motor}, {\it New Journal of Physics} {\bf 7}, 10 (2005).}  were able to find a solvable model for a {\it thermal Brownian motor}.
They show that immersed in  two different thermal baths, two rigidly coupled Brownian particles with
a geometrical asymmetry
can function as a microscopic engine able to rectify Brownian fluctuations. As expected, when the temperatures of the two baths
are equal, the drift motion ceases, and one is left only with a standard Brownian displacement, which obeys Gauss' distribution law.
The drift speed can be computed exactly for convex bodies, in the limit of dilute gases. Extremely precise
molecular dynamics simulations with hard disks confirm the theory. In effect, this is a
microscopic and soluble Feynman's ratchet.

In a recent work,\footnote{C. Van den Broeck and R. Kawai,
{\it Phys. Rev. Lett.} {\bf 96}, 210601 (2006).} Van den Broeck and Kawai propose a model for a {\it Brownian refrigerator},
with a cooling mechanism based on such a Brownian motor submitted to an external force. A heat flow is generated between
the two components of the motor. Such a marvellously simple microscopic model would have certainly greatly pleased Einstein,
 Smoluchowski and Sutherland!\smallskip

It is necessary to note here that these discussions are
current research topics of intense interest.  In fact today there exist new theoretical
results, known as the Gallavotti-Cohen fluctuation theorem,\footnote{D. J. Evans and D. J. Searles,
{\it Phys. Rev. E} {\bf 50}, pp. 1645-1648 (1994);
G. Gallavotti and E. G. D. Cohen, {\it Phys. Rev. Lett.} {\bf 74},
pp. 2694-2697 (1995); {\it J. Stat. Phys.}  {\bf 80}, pp. 931-970 (1995); see also D. J. Evans,
E. G. D. Cohen and G. P. Morris,
{\it Phys. Rev. Lett.} {\bf 71}, pp. 2401-2404 (1993); G. M. Wang, E. M. Sewick, E. Mittag, D. J. Searles and D. J. Evans,
{\it Phys. Rev. Lett.} {\bf 89}, 050601 (2002).}
 Jarzynski's equality,\footnote{C. Jarzynski, {\it
Phys. Rev. Lett.} {\bf 78}, pp. 2690-2693 (1997).} or Crooks' fluctuation theorem.\footnote{G. E. Crooks,
{\it Phys. Rev. E} {\bf 60}, pp. 2721-2726 (1999).} They quantify the
spontaneous average work provided by a source of heat during
irreversible phenomena.  The manipulations of single biological
molecules like DNA and RNA, which are mesoscopic objects, allow
the experimental testing of these relations.  The interpretations of these
results and experiments are currently the topic of a lively debate,\footnote{See, e.g, E. G. D.
Cohen and D. Mauzerall, {\it J. Stat. Mech. Theor. Exp.}
 P07006
 (2004), and the reply by C. Jarzynski, {\it J. Stat. Mech. Theor. Exp.} P09005
 (2004), arXiv:cond-mat/0407340.} just as at the dawn of Brownian motion!\footnote{ The interested
reader can consult the texts by Ch. Maes and F. Ritort in the Poincar\'e Seminar on {\it
Entropy} (2003), available on the website
www.lpthe.jussieu.fr/poincare/, and published in: J.~Dalibard,
B.~Duplantier \& V.~Rivasseau, {eds.}, {\it Poincar\'e Seminar
2003}, Progress in Mathematical Physics, Vol. 38, Birkh\"auser, Basel
(2004). See also C. Bustamante, J. Liphard and F. Ritort in {\it Physics Today}, July 2005, pp. 43-48.}

\subsubsection{Brownian motion and the mathematical aspects of irreversibility}
Let us open here a brief mathematical parenthesis.\footnote{The material of this section
is  borrowed from the introduction of the recent paper by L. Erd\"os, M. Salmhofer and H.-T. Yau,
{\it Towards the quantum Brownian motion}, arXiv:math-ph/0503001 (2005).} Einstein's and Smoluchowski's theories, based upon a Newtonian dynamics of the particles, in fact postulated
the emergence  of  Brownian motion from a classical non-dissipative reversible dynamics, a point of
view which was far from being physically obvious or, {\it a fortiori},  mathematically rigorous. This
 led to the heated controversy about the second principle.  The key difficulty is similar to the justification
of Boltzmann's molecular chaos assumption ({\it ``Stosszahlansatz''}) standing behind the derivation of
the Boltzmann equation. Mathematically, the dissipative character can only emerge in a scaling limit, as the number of degrees
of freedom goes to infinity.

As we shall see below, the first mathematical definition of Brownian motion was given only
in 1923 by Wiener. But the derivation of Brownian motion from Hamiltonian dynamics was not seriously investigated
until the end of the seventies. Kesten and Papanicolaou\footnote{H.  Kesten, G. Papanicolaou, {\em Commun. Math. Phys.}
\textbf{78}, pp. 19-63 (1980).} proved that the velocity distribution of a particle moving in a random scatterer environment
(the so-called Lorenz gas with random scatterers) converges to Brownian motion in dimension $d\geq 3$.
The same result was obtained in $d=2$ dimensions by D\"urr, Goldstein and
Lebowitz.\footnote{D. D\"urr, S. Goldstein, J. Lebowitz, {\em Commun. Math. Phys.}
\textbf{113}, pp. 209-230 (1987).} A very recent work establishes
the convergence to Brownian motion in position space
as well.\footnote{T. Komorowski, L. Ryzhik, \textit{Diffusion in a weakly random
Hamiltonian flow}, arXiv:math-phys/0505082 (2005);
\textit{The stochastic acceleration problem in two dimensions}, arXiv:math-phys/0505083 (2005).}

Bunimovich and Sinai proved the convergence to Brownian motion
of  the periodic Lorenz gas with a hard-core
interaction.\footnote{L. Bunimovich, Y. Sinai, {\em Commun. Math. Phys.}  \textbf{78}, pp. 479-497
(1980).} The only source of randomness there is the distribution of the initial conditions. Finally,
D\"urr, Goldstein and Lebowitz\footnote{D. D\"urr, S. Goldstein, J. Lebowitz,
 {\em Commun. Math. Phys.} \textbf{78},
pp. 507-530 (1981).} established rigorously that the velocity process of a heavy particle in an ideal gas converges in three
(actually an arbitrary number of)
dimensions to
the Ornstein-Uhlenbeck process, that is the version of Brownian motion described by Langevin's equation (see below).
The same result in one dimension was first established by
R. Holley.\footnote{R. Holley, {\it Z. Warscheinlichkeitstheorie verw. Geb.} {\bf 17}, pp. 181-219 (1971).}

Brownian motion was discovered and theorized in the context of  classical mechanics, and it postulates a
microscopic reversible Newtonian world for atoms and molecules. Nowadays, it is thus natural
to replace Newtonian dynamics with Schr\"odinger dynamics and investigate if  Brownian motion
still correctly describes the motion of a quantum particle in a random environment. For a discussion
of this fundamental and difficult question, we refer the reader to a
recent work by  Erd\"os, Salmhofer and Yau\footnote{L. Erd\"os, M. Salmhofer and H.-T. Yau,
{\it op. cit.}; see also arXiv:math-ph/0502025, math-ph/0512014, math-ph/0512015, and L. Erd\"os, M. Salmhofer, math-ph/0604039.} and to the references therein.
\subsubsection{Smoluchowski's legacy}
With Einstein, Smoluchowski shares the credit for having shown the
importance of microscopic fluctuations in statistical physics, at the
same time promoting the probabilistic approach.  In this sense he appears
as a great master inheritor in physics of the {\it Doctrine of
Chance} of Abraham de Moivre.

In 1917, Marian von Smoluchowski had just been elected rector of the Jagellonian University in Krak\'ow (Cracow) University, but he
was never to fulfill his new task. During the summer he succumbed to an epidemic of dysentery. During his illness
he complained to his wife that he could have done so much more. He died prematurely in September of 1917 at the age of forty
five.

In 1973 Chandrasekhar was awarded the Marian von Smoluchowski Medal of the Polish Physical Society in appreciation of his
contributions to stochastic methods in physics and astrophysics and, especially, the Review of Modern Physics 1943
article 
which covered Smoluchowski's contributions. In his speech at the award ceremony, Chandrasekhar noted that
the Nobel prizes in chemistry awarded to R. Zsigmondy in 1925 and to T. Svedberg in 1926 were for experimental
confirmation of Smoluchowski's theoretical predictions on colloidal and disperse systems and that if Smoluchowski had
been still alive he would
certainly have been a Nobel laureate himself.\footnote{{\it Marian Smoluchowski, His Life
and Scientific Work}, S. Chandrasekhar, M. Kac, R. Smoluchowski,
Polish Scientific Publishers, PWN, Warszawa (2000), see the preface by the editor R. S. Ingarden.}

\subsection{Louis Bachelier}

\subsubsection{Bachelier and mathematical finance}
Louis Bachelier is nowadays considered as having laid the foundations for mathematical finance, and is further credited
with the first mathematical study of the continuous Brownian process, including a random walk
approach to the latter. A detailed and very interesting presentation of Bachelier's life and scientific achievements
 was given in 2000 in an essay, entitled
{\it Louis Bachelier on the Centenary of Th\'eorie de la Sp\'eculation}, for the
centenary of the publication of his thesis.\footnote{J.-M. Courtault, Y. Kabanov, B. Bru, P. Cr\'epel, I. Lebon and A. Le Marchand,
{\it Louis Bachelier on the Centenary of Th\'eorie de la Sp\'eculation}, {\it Mathematical Finance}, Vol. {\bf 10}, No. 3,
pp. 341-353 (2000).} This section is essentially based on this presentation, and  a significant part
of it
incorporates material in the cited article.

The importance of Bachelier's work was not properly recognized during his time. As Beno\^{\i}t Mandelbrot
writes in {\it The Fractal Geometry
of Nature},\footnote{B. B. Mandelbrot, {\it The Fractal Geometry of Nature}, Freeman, New-York (1982).} it was
Kolmogorov in 1931 who re-discovered his name in an article in {\it
Mathematische Annalen}.\footnote{A. N. Kolmogorov, {\it \"Uber die analytischen Methoden in der
Warscheinlichkeitsrechnung}, {\it Math. Annalen} {\bf 104}(3), pp. 415-458 (1931).}

Bachelier was interested in the theory of
speculation at the Paris stock market. He successfully defended his thesis, entitled
{\it Th\'eorie de la sp\'eculation}, on 29 March 1900 at the Sorbonne, in front
of a jury composed of Paul Appell, Joseph Boussinesq and  Henri Poincar\'e, his thesis advisor.
As a work of exceptional merit, stongly supported by
 Poincar\'e,  his thesis was published in the
{\it Annales Scientifiques de l'\'Ecole Normale Sup\'erieure}.\footnote{L. Bachelier,
{\it Ann. Sci. \'Ecole Normale Sup\'erieure}
{\bf 17}, pp. 21-86 (1900).}
\subsubsection{The Thesis}
Bachelier begins with the mathematical modeling of stock price movements,
and  formulates the principle that ``the expectation of the speculator is zero,'' by which he
means that the conditional expectation given the past information is zero. In other words, he assumes that the market
evaluates assets  according to a martingale measure. The further hypothesis is that the price evolves
as a continuous Markov process (with no memory), homogeneous in time and space. Bachelier then shows that
the density of one-dimensional distributions of this process satisfies an integral relation, now known as the
Chapman-Kolmogorov equation. Bachelier, without addressing the question of uniqueness, shows that
the Gaussian density, with a linearly increasing variance, solves this equation.

He also considers a discrete version
of the problem, where the price process is the continuum limit
of random walks, and where the binomial formula (\ref{binomial}) appears.
He then proceeds to show that the distribution functions of the process satisfy Fourier's heat equation, as in
the similar eq. (\ref{Diff}) in Einstein's article. Bachelier then introduces a novel expression: ``the radiation of the probability''.\medskip

One finds indeed many
of the well-known results for Brownian motion: On p. 37 of his memoir, one reads that: {\it ``On voit
 que la probabilit\'e est r\'egie par la loi de Gauss
d\'ej\`a c\'el\`ebre dans le Calcul des probabilit\'es;''} on p. 38, that {\it `` L'esp\'erance math\'ematique}
$$
\int_0^{\infty}pxdx=k\sqrt t
$$
{\it est proportionnelle \`a la racine carr\'ee du temps.''} Bachelier also calculates the probability that the
Brownian motion does not exceed a fixed level and finds the distribution of the supremum of that motion.\medskip

 He therefore  developed in his first thesis a theory of continuous stochastic
processes close to the modern mathematical theory of Brownian motion. As stressed by the authors of the essay
{\it Louis Bachelier on the Centenary of Th\'eorie de la
Sp\'eculation}, ``more than one hundred years
 after the publication of the thesis, it is quite easy to appreciate the importance of Bachelier's ideas. The thesis can
 be viewed as the origin of mathematical finance, and of several branches of stochastic calculus such as the theory of
 Brownian motion, Markov processes, diffusion processes, and even weak convergence in functional spaces.''

It is also quite interesting to read Poincar\'e's original report,
translated in the essay cited above. Poincar\'e's report
shows that Bachelier's thesis was highly appreciated by the outstanding mathematician. In contrast to the legend that the
evaluation note {\it ``honorable''} means somehow that the examiners were dissatisfied with the thesis, it can perhaps
be argued
that it might have been the highest grade possible for a thesis which was addressing a problem
not in the realm of standard mathematics,
and that in addition had  a number of non-rigorous arguments.\medskip

The official report of the Thesis Committee states:

{\it In the presentation of his First Thesis, M. Bachelier showed mathematical intelligence and insight. He has added
some interesting results to those already contained in the printed version of the thesis, in particular an application
of the image method.

As for the Second Thesis, he proved to possess a complete knowledge of Boussinesq's work on the motion of a sphere in an
indefinite fluid.

The Faculty gave him the degree of Doctor with honors.}

\rightline{Paul Appell,  President}

\medskip

It is indeed very intriguing  that the ``Proposal given by the
Faculty,'' subject of his Second Thesis, was entitled: {\it Resistance of an indefinite liquid mass
with internal frictions, described by the formulae of Navier, to small
translational motions of a solid sphere, submerged inside the fluid
and adhering to it.}

But there is of course no mention in his first thesis, published in 1900, about
any link between the speculation problem and the motion of a sphere in
a viscous fluid! However, we saw above Poincar\'e's early interest in Brownian motion in relation to
Carnot's principle.  We also saw that Einstein's (as well as Sutherland's) application of
hydrodynamical laws to the motion of a sphere
 suspended in a fluid, was key to the solution of Brownian motion. We now observe
the amazing coincidence that the thesis
subject proposed by the Faculty, if joined with the subject of the first thesis, could have led
 Poincar\'e and Bachelier
to establish the quantitative theory of
Brownian motion, before any of Einstein, Sutherland or Smoluchowski! All necessary mathematical equations
were indeed present for that, if
 only a little spark of physical intuition would have struck these eminent mathematicians!

\subsubsection{Further Studies}

Louis Bachelier continued to develop the mathematical theory of diffusion processes in a series of memoirs and books.
In his 1906 memoir on the {\it Th\'eorie des probabilit\'es
continues},\footnote{{\it Th\'eorie des probabilit\'es continues}, J. Math. Pures et Appl., pp. 259-327 (1906).}
he defined new classes of stochastic processes,
which are now called processes with independent increments
and Markov processes, and he derived the distribution of the Ornstein-Uhlenbeck process.

He was aware of the
importance of his contributions. He wrote in his 1924 ``Notice de Travaux'' that {\it ``this theory has no relation to
the geometrical theory of probability, the range of application of which is quite limited. We are concerned here with a
science of a different order of generality, compared to classical probability calculus. Among the new concepts,
one can cite the assimilation to an energy of the probability which is an abstraction. That original concept was
noticed  by Henri Poincar\'e, and it made much progress possible.''} One also reads about his 1912 book {\it Calcul des probabilit\'es},\footnote{L. Bachelier,
{\it Calcul des probabilit\'es}, Gauthier-Villars, Paris (1912).} that {\it ``it is the first that
surpassed the great treatise by Laplace.''}

We shall not describe in detail here  the very unfortunate misunderstanding with Paul L\'evy, which in 1926 prevented
Bachelier from becoming a full professor at the University of Dijon. We refer the interested reader to the essay
mentioned above for a thorough and well-documented
analysis of this dramatic event.

Later, L\'evy, under the influence of Kolmogorov's fundamental paper (1931)
on diffusion processes, which referred to Bachelier's work, realized that a number of
properties of Brownian motion had been discovered by Bachelier several decades earlier. He revised
 his opinion, and wrote him a
letter with apologies.

 Bachelier's ideas receive nowadays a widespread recognition. Famous probability treatises,
like the ones by W. Feller, {\it An Introduction to Probability Theory and its Applications} (1957), or by K. It\^o
and H. McKean, {\it Diffusion Processes and their Sample Paths} (1965), refer to Bachelier's seminal work.

In the literature written by economists, one finds reference to him in Keynes (1921), and more recently in the work of
other famous economists, like that of the 1997 Nobel laureates in Economic Sciences, Robert Merton and Myron Scholes. It is perhaps appropriate here
to reproduce Merton's tribute to Bachelier:

{\it `` The origin of much of the mathematics in modern finance can be traced to Louis Bachelier's 1900 dissertation on
the theory of speculation, framed as an option-pricing problem. This work marks the twin births of both the
continuous-time mathematics of stochastic processes and the continuous-time economics of derivative-security pricing.''}

 No doubt that today Bachelier would have been awarded a Nobel Prize in Economic Sciences for his work of 1900!

\subsection{Paul Langevin}

Knowing the great interest in the theory of Brownian motion, signalled
by the works of Gouy, Einstein, and Smoluchowski, Langevin took the
next steps in 1908. He first said that the factor of 64/27 of Smoluchowski's
results, due to the approximations made, was erroneous and that the
result coincided with Einstein's formula (\ref{x2}) after his
correction.  Next, he provided another demonstration of this fact, in
which was contained the first mathematical example of a {\it
stochastic equation}.

\subsubsection{Langevin's equation}
\label{langevin}

Langevin's argument is enlightening and we follow his
demonstration faithfully.\footnote{P. Langevin, {\it C. R. Ac. Sci. Paris} {\bf
146}, 530 (1908).}  The starting point is the Maxwell equipartition
theorem of kinetic energy.  It states that the energy of a particle in
suspension inside a fluid in thermal equilibrium has, for instance in
the $x$ direction, an average kinetic energy
$\frac{1}{2}\frac{RT}{\mathcal N}$, equal to that of any gas molecule,
in a given direction, at the same temperature.  This is directly
related to  van 't Hoff's law seen above, which affirms the identity
between diluted solutions and perfect gases.  If $v=\frac{{\rm d}
x}{{\rm d}t} $ is the particle velocity in a chosen direction at a
given moment, then the average over a large number of identical
particles with mass $m$ is
\begin{eqnarray}
\label{equi}
\frac{1}{2}m\langle v^2\rangle=\frac{1}{2}\frac{RT}{\mathcal N}.
\end{eqnarray}
A particle which is large compared to the molecules of a liquid, and is 
moving at speed $v$ with respect to this liquid, experiences a viscous resistance force equal to
$-6\pi \eta a v$, according to Stokes' formula.  In reality this is
only an average value, and because of the irregular shocks of the
surrounding molecules, the action of the fluid on the particle fluctuates
around the average value. The equation of motion along the direction
$x$, given by Newtonian dynamics, is
\begin{eqnarray}
\label{newton}
m\frac{{\rm d} v}{{\rm d}t}=m\frac{{\rm d^2} x}{{\rm d}t^2}=-6\pi \eta
a v +X.
\end{eqnarray}
The complementary force $X$, introduced by Langevin,
is {\it random}, and also called {\it stochastic}.  In reality we know little about
it, apart from that it is equally likely to be positive or negative, and that its
magnitude is such that it maintains the particle's agitation which,
without it, would end by stopping because of the viscous
resistance.

By multiplying equation (\ref{newton}) by $x$, one has\footnote{Since
$v=\frac{{\rm d} x}{{\rm d}t}$, we use the identities between derivatives $xv
=x\frac{{\rm d} x}{{\rm d}t}=\frac{1}{2}\frac{{\rm d} x^2}{{\rm d}t}$,
and $x\frac{{\rm d} v}{{\rm d}t}=x\frac{{\rm d^2} x}{{\rm
d}t^2}=\frac{1}{2}\frac{{\rm d^2} x^2}{{\rm d}t^2}- v^2.$}
\begin{eqnarray}
\nonumber
m x\frac{{\rm d} v}{{\rm d}t}&=&\frac{1}{2}m\frac{{\rm d^2} x^2}{{\rm
d}t^2}-m v^2\\
\label{viriel}
&=&-\mu \, xv +xX=-\mu \frac{1}{2} \frac{{\rm d} x^2}{{\rm d}t} +xX,
\end{eqnarray}
where the friction coefficient $\mu$ represents $\mu=6\pi \eta a$ as
before.  If we consider a large number of identical particles and take
the average of equations (\ref{viriel}) written for each of them, then
the average value of the term $xX$ is ``evidently'' zero because of
the irregularity of the random forces $X$, and one finds\footnote{One
should note that the force $X$ disappears from the calculation thanks
to that observation. The only under-lying role of $X$ is therefore to
ensure the physical possibility of a kinetic average $\la v^2\ra \neq
0$. On the other hand, the equality $\la xX\ra =0$ does not appear as
evident, because there could have existed a subtle correlation 
between
the position $x$ and the stochastic force $X$, as it exists between
velocity and stochastic force.  The existence of two types of
stochastic calculations, {\it \`a la} It\^o and {\it \`a la}
Stratonovitch, illustrates this difficulty. (See for example N. G. van
Kampen, {\it Stochastic Processes in Physics and Chemistry}, Elsevier,
Amsterdam (1992).) Einstein made the same hypothesis in his third
demonstration of Brownian motion; see in this volume the
translation of his
lecture on November 2, 1910
for the Z\"urich Physical Society.}
\begin{eqnarray}
\label{virielmoy}
\frac{1}{2}m\frac{{\rm d^2} \langle x^2\rangle }{{\rm d}t^2}-m
\langle v^2\rangle=-\mu\frac{1}{2}\frac{{\rm d} \langle x^2\rangle}{{\rm d}t}.
\end{eqnarray}

One puts $u=\frac{1}{2}\frac{{\rm d} \langle x^2\rangle}{{\rm d}t}$,
and uses the equipartition theorem of kinetic energy (\ref{equi}) to
get a simple differential equation of first order:
\begin{eqnarray}
\label{equadif}
m\frac{{\rm d} u}{{\rm d}t}-\frac{RT}{\mathcal N}=-\mu u.
\end{eqnarray}
The general solution is
\begin{eqnarray}
\label{solution}
u=\frac{RT}{ \mu {\mathcal N}}+C\exp\left(-\frac{\mu}{m}t\right),
\end{eqnarray}
where $C$ is an arbitrary constant.\footnote{Here, there seems to be a
contradiction between the existence of an exponential term and the
hypothesis of the equipartition of energy, $m\langle
v^2\rangle=\frac{RT}{\mathcal N}$, made {\it for every} $t$ by
Langevin, because it is only at large $t$ that memory effects are
exponentially suppressed.  This hypothesis, as well as a solution of
the form (\ref{solution}), can however be correct for all $t$,
provided that one imposes the same condition for the initial velocity,
which in fact fixes the value of the constant $C$ to be equal to
$C=-\frac{RT}{\mathcal \mu N}$. We will come back to this point
further along in a more detailed study of the solution of Langevin's
equation.}  The exponentially decreasing term rapidly fades away, and the
result goes to the constant value of the first term, in a limiting
regime after a time $\tau$ of order $\frac{m}{\mu}$ or $10^{-8}$
seconds, for all Brownian particles.

Thus, we have
\begin{eqnarray}
\label{perm}
u=\frac{1}{2}\frac{{\rm d} \langle x^2\rangle}{{\rm
d}t}=\frac{RT}{\mu\mathcal N},
\end{eqnarray}
from which, for the time interval $t$,
\begin{eqnarray}
\label{as}
 \langle x^2\rangle=\frac{2RT}{\mu \mathcal N} t=\frac{RT}{\mathcal N}
 \frac{1}{3\pi \eta a} t,
\end{eqnarray}
if one supposes that the particle was observed at the origin $x=0$ at time
$t=0$. Langevin's method indeed reproduces Einstein's result
(\ref{x2}). In this paper (published in 1908 in the Comptes Rendus of
the Academie de Sciences) Langevin introduced, without knowing it, the
first element (the random force $X$) of what was to become 
stochastic calculus.\footnote{J. L. Doob, The Brownian Motion and
Stochastic Equations, {\it Ann. of Math.}, {\bf 43}, pp. 351-369
(1942), reprinted in [Wax 1954, pp. 319-337], {\it op. cit.}}

\subsubsection{Boltzmann's constant}

Boltzmann's constant $k_B$ is obtained by dividing the molar constant
$R$ of a perfect gas by Avogadro's number $\mathcal N$, such that
one obtains a quantity which refers to a single molecule:
\begin{eqnarray}
\label{kb}
 k_B=\frac{R}{\mathcal N}=1.381 \times 10^{-23}\, {\rm J}\, {\rm
 K}^{-1}.
\end{eqnarray}
The energy $k_B T$ gives the average thermal energy at the standard
temperature: $k_BT =4 \times 10^{-21}$ J. The constant $k_B$ was not
introduced by Boltzmann but by Planck in his famous presentation
on December 14, 1900, on black-body radiation, at the same time he
presented Planck's constant~$h$!

\subsubsection{An analysis of the solution of Langevin's equation.}

\label{OrnsteinUhl}
The method presented in section (\ref{langevin}) is the one that
Langevin gave in his original paper. A more modern formulation proceeds from
the time-correlation functions of the stochastic force $X$ in
canonical form,
\begin{eqnarray}
\label{lang1}
\langle X\rangle={0},\,\,\langle X(t)X(t')\rangle=A \delta(t-t'),
\end{eqnarray}
where $A$ is a coefficient to be determined and $\delta(t-t')$ is
the
Dirac distribution. The generalization to $d$ dimensions is
\begin{eqnarray}
\nonumber
\langle\vec{X}\rangle&=&\vec{0},\\
\langle X_i(t)X_j(t')\rangle&=&A \delta_{ij}\delta(t-t'),
\label{lang2}
\end{eqnarray}
where $\delta_{ij}$ is the Kronecker symbol and $i,j=1,\cdots d.$
	
We can easily integrate the linear equation for the velocity
\begin{eqnarray}
\label{newtonvitesse}
m\frac{{\rm d} {\vec v}}{{\rm d}t}=-\mu {\vec v} +\vec{X}.
\end{eqnarray}
The solution is
\begin{eqnarray}
\label{vitesseint}
{\vec v}(t)={\vec v}(0)\, e^{-\frac{\mu}{m}t}
+\frac{1}{m}\int_0^{t}{\rm d}t'\, \vec X(t')\,
e^{-\frac{\mu}{m}(t-t')}.
\end{eqnarray}
Therefore by taking the square of the velocity and by using formula
(\ref{lang2}), we find the average value of kinetic energy at time $t$
\begin{eqnarray}
\label{vitessec}
\frac{1}{2}m\langle{\vec v}^{\,2}(t)\rangle=\frac{Ad}{4\mu}
\left(1-e^{-2\frac{\mu}{m}t}\right)+\frac{1}{2}m{\vec v}^{\,2}(0)e^{-2\frac{\mu}{m}t}.
\end{eqnarray}
We then see that this energy relaxes towards a constant value at large
time, i.e., at equilibrium.  From the theorem of equipartition of
kinetic energy,
\begin{eqnarray}
\label{theq}
\frac{1}{2}m\langle{\vec v}^{\,2}(t)\rangle_{t\to\infty}=\frac{d}{2}k_BT,
\end{eqnarray}
we deduce the important identity
\begin{eqnarray}
\label{A}
A={2}\mu k_BT.
\end{eqnarray}

We then have
\begin{eqnarray}
\label{vitessecarree}
\langle{\vec v}^{\,2}(t)\rangle=\frac{d k_BT}{m}+
\left({\vec v}^{\,2}(0)-\frac{d k_BT}{m}\right)e^{-2\frac{\mu}{m}t}.
\end{eqnarray}

A second stage consists in integrating equation (\ref{vitesseint}) to
obtain the displacement $\vec{r}(t)-\vec{r}(0)$.  Then taking the
square, and the stochastic average by means of formulae (\ref{lang2}),
we obtain after some calculation,
\begin{eqnarray}
\label{depcarre}
\nonumber
\langle\left[\vec{r}(t)-\vec{r}(0)\right]^2\rangle&=&
2d D \left[t-\frac{m}{\mu}\left(1-e^{-\frac{\mu}{m}t}\right)\right]\\
&+&\left({\vec v}^{\,2}(0)-\frac{d
k_BT}{m}\right)\left(\frac{m}{\mu}\right)^2\left(1-e^{-\frac{\mu}{m}t}\right)^2,
\end{eqnarray}
where $D=k_BT/\mu$, as before. The derivative $u$ considered by
Langevin is then given by
\begin{eqnarray}
\label{u}
\nonumber
u&=&\frac{1}{2}\frac{{\rm d}}{{\rm
d}t}\langle\left[\vec{r}(t)-\vec{r}(0)\right]^2\rangle\\
&=&d\frac{k_BT}{\mu}-{\vec v}^{\,2}(0)
\frac{m}{\mu}e^{-\frac{\mu}{m}t}+\left({\vec v}^{\,2}(0)
-\frac{d k_BT}{m}\right)\frac{m}{\mu}e^{-2\frac{\mu}{m}t}.
\end{eqnarray}

Notice first that these results at large $t$, or $t\gg
\tau=m/\mu$, go asymptotically to those of thermal equilibrium and
to the associated motion of diffusion, as expected.
One remarks then the role played by the initial velocity in memory
effects and in the approach to equilibrium.  A very special value of
${\vec v}^{\,2}(0)$ is that of equipartition $\frac{d k_BT}{m}$.  For
this value only, the average quadratic velocity in
(\ref{vitessecarree}) becomes {\it invariant} in time, $\langle{\vec
v}^{\,2}(t)\rangle=\frac{d k_BT}{m}, \forall t$.  The average
quadratic displacement (\ref{depcarre}) then takes Ornstein's simple
form (\ref{delta2comp}), and the quantity $u$ (\ref{u}) takes the form
predicted by Langevin in (\ref{solution}), with a determined value for
$C$.  One consistently obtains the same result by using for
${\vec v}^{\,2}(0)$ its most probable value, meaning its {\it thermal
average at equipartition}. One can then understand  (but only {\it a
posteriori}) the consistency of  Langevin's approach when he inserted the
identity (\ref{equi}) in the middle of the derivation. That amounted to
chosing the peculiar
boundary condition ${\vec v}^{\,2}(0)=\frac{d k_BT}{m}$, which
enforces {\it stationary} equipartition!

On the other hand, if one gives to the initial quadratic velocity
${\vec v}^{\,2}(0)$ a value which is different from that of
equilibrium, the relaxation will occur in a bit more complex way, as
we showed in the above results.
	
The regime at short times, $\frac{m}{\mu}\gg t$, also naturally
depends on the initial conditions.  In fact, by developing in series
(\ref{depcarre}) one finds the expected ballistic regime
\begin{eqnarray}
\label{depcarrecourt}
\nonumber
\langle\left[\vec{r}(t)-\vec{r}(0)\right]^2\rangle={\vec v}^{\,2}(0)\, t^2+{\mathcal O} (t^3),
\end{eqnarray}  
that naturally cross-checks with (\ref{ball}) if one takes once again
the value at equipartition.

\subsubsection{Microscopic model}
\label{modele}
The force proposed by Langevin, $-\mu v+X$, can only be an
approximation to the underlying molecular reality, made up of
innumerable collisions where multiple correlations, due to
interactions between molecules, exist at very short time scales.  The
stochastic term $X$ in (\ref{lang1},\ref{lang2}) is a white noise
without memory, i.e., it neglects temporal correlations.

Moreover, the hydrodynamic form of the friction term, $-\mu v$, is a
description that pertains to the continuous limit, which requires extremely
frequent collisions of the particle in suspension.  The mass $m$ of the
particle must then be large enough so that the characteristic time
$\tau=m/\mu$ is large compared to the inverse frequency of collisions.

To give an idea of the origin of Langevin's equation (\ref{newton})
and of its parameters $\mu$ and $A$ (\ref{A}), it is natural to
consider the simplest model, where the collisions of a particle in
suspension occur with a surrounding perfect gas, and thus itself without
interaction.

One can therefore consider a perfect gas of identical particles with
mass $m'$, a particle density $n'$, at temperature $T$, and colliding the  particle of large mass $m$
in suspension. To simplify, we
consider the gas in one dimension, where the equations for the particle-gas elastic
collisions are particularly simple.\footnote{J. W. S. Rayleigh was apparently the first to address this problem in one dimension, in
{\it Dynamical Problems in Illustration of the Theory of Gases}, {\it Phil. Mag.} {\bf 32}, pp. 424-445 (1891);
in {\it Scientific Papers by Lord Rayleigh}, Vol. III, {\bf 183}, pp. 473-490, Dover Publications, New-York (1964).
He establishes in particular the evolution equation for the velocity distribution of the large particle and its solution,
as well as the convergence of the latter in the steady state to a Maxwellian.

In {\it The Motion of a Heavy Particle in
an Infinite One Dimensional Gas of Hard Spheres}, {\it Z. Warscheinlichkeitstheorie verw. Geb.} {\bf 17}, pp. 181-219 (1971),
R. Holley establishes rigorously the (weak) convergence of the velocity or position distributions of the heavy particle to the
respective {\it Ornstein-Uhlenbeck} processes, in the limit where the mass ratio $m/m'\to \infty$. This applies to gas particles
with a Poisson distribution in position space and an arbitrary distribution in velocity space, provided that the latter
distribution is
symmetric and has four moments.}  One then finds that
the equation for the momentum variation of the test particle is
similar to Langevin's equation, with the explicit
coefficients\footnote{One calculates, in the process of discrete
collision, the average momentum variation $\langle
\frac{{\rm d}p}{{\rm d}t}\rangle=-\mu\langle v \rangle$ as well as 
the fluctuations $\langle \frac{{\rm d}p(t)}{{\rm d}t}\frac{{\rm
d}p(t')}{{\rm d}t'}\rangle -\langle\frac{{\rm d}p(t)}{{\rm d}t}\rangle
\langle\frac{{\rm d}p(t')}{{\rm d}t'}\rangle= A\delta(t-t')+\cdots$,
and finds by comparison the values (\ref{modlange}) of parameters $\mu$
and $A$ for Langevin's equation.  See the article from B. Derrida and
\'E. Brunet in {\it Einstein aujourd'hui}, \'eds. M. Leduc and M. Le Bellac,
Savoirs actuels, EDP Sciences/CNRS \'Editions (2005).}
\begin{equation}
\label{modlange}
\mu =4 n' \sqrt{\frac{2m'k_BT}{\pi}},\,\,\,\,A=8n'k_B T\sqrt{\frac{2m'k_B T}{\pi}};
\end{equation}
$\mu$ and $A$ thus verify (\ref{A}).

It is then particularly interesting to rewrite these terms as a
function of {\it molar sizes} that characterize the perfect gas.  One
introduces as well the gas pressure\footnote{In one dimension, the
pressure $p'$ is equivalent to a force, because the boundaries of the
``box'' containing the gas are simple points.} $p'$, which responds to the
equation of perfect gases $p'=n'k_BT$, which gives
\begin{equation}
\label{modlangevin}
\mu=4 p' \sqrt{\frac{2\mathcal M}{\pi R T}}
,\,\,\,\,A=\frac{2RT}{\mathcal N} 4p'\sqrt{\frac{2\mathcal M}{\pi R
T}},
\end{equation}
where $\mathcal M=\mathcal N m'$ is the molar mass of the gas.

A similar problem can be solved in three dimensions.\footnote{M. S. Green, {\it Brownian Motion in a Gas of Noninteracting Molecules},
{\it J. Chem. Phys.} {\bf 19}, pp. 1036-1046 (1951). In a bibliographic note, Green cites Smoluchowski for having
discussed the same
three-dimensional case {\it ``in one of the earliest papers in which the true nature
of Brownian motion was understood.''} Green adds: {\it ``His method was admittedly approximate and the formula which he obtained for the friction
constant was the same as ours in its dependence on the temperature of the gas, the mass of the particle and the mass, and
concentration of the molecules, but with a different, and incorrect numerical constant.''}
He further cites H. A. Lorentz, in {\it Les Th\'eories Statistiques en Thermodynamique}, B. G. Teubner, Leipzig (1912),
for having obtained the correct formula by a method which was equivalent to his own modification of Rayleigh's method. See also: J. L. Lebowitz,
{\it Stationary Nonequilibrium Gibbsian Ensembles}, {\it Phys. Rev. } {\bf 114}, pp. 1192-1202 (1959); D. D\"urr, S. Goldstein, J. L. Lebowitz, {\it A
Mechanical Model of Brownian Motion}, {\it Commun. Math. Phys.} {\bf
78}, pp. 507-530 (1981).} The friction coefficient $\mu$ and Brownian coefficient $A$ are in this case
\begin{equation}
\label{modlange1}
\mu =\frac{8}{3} \pi a^2 n' \sqrt{\frac{2 m'k_BT}{\pi}},\,\,\,\,A=2 k_B T \mu,
\end{equation}
where the Brownian particle's radius $a$ now enters through the area term $\pi a^2$. In terms of molar parameters one therefore finds:
\begin{equation}
\label{modlangevin1}
\mu=\frac{8}{3} \pi a^2 p' \sqrt{\frac{2 \mathcal M}{\pi R T}}
,\,\,\,\,A=\frac{2RT}{\mathcal N} \frac{8}{3}\pi a^2 p'\sqrt{\frac{2 \mathcal M}{ \pi R
T}}.
\end{equation}

\subsubsection{Discontinuity in Nature and the existence of Brownian motion}

The explicit results above, in their last forms (\ref{modlangevin}) or (\ref{modlangevin1}),
rigorously state that the Sutherland-Einstein equation (\ref{ES}),
$D=\frac{RT}{\mu\mathcal N}$, reflects the existence of molecules.

In fact, the friction coefficients $\mu$ can be expressed independently
from Avogadro's number $\mathcal N$, and depend only on the ideal gas constant
$R$ and the {\it macroscopic} parameters of the
surrounding gas, like the pressure $p'$, temperature $T$, and molar
mass $\mathcal M$ (and also on the Brownian particle's radius in space dimension $d\geq 2$.).  On the other hand, the variance $A$ of
the Langevin
stochastic force, which controls diffusion, continues
to depend on $\mathcal N$ and vanishes when Avogadro's number goes to
infinity.

In the same way, the limit of the diffusion coefficient
$D=\frac{RT}{\mu\mathcal N}$, when Avogadro's
number goes to infinity, $\mathcal N \to \infty$, is of course {\it
zero}, i.e., the {\it Brownian motion would cease immediately if Nature
were continuous!}  An entire branch of mathematics might perhaps never have seen
the light of day.

\subsection{Jean Perrin's experiments}
\subsubsection{The triumph of the ``Molecular Hypothesis''}
Jean Perrin is often cited as the one who established
the Einstein-Smoluchowski-Sutherland theory by his beautiful experiments.
He was also an outstanding promotor of atomistic ideas.  His book,
{\it Atoms},\footnote{J. Perrin, {\it Les Atomes}, F\'elix Alcan,
Paris (1913); r\'e\'edition Champs Flammarion (1991); English translation: {\it Atoms},
transl. by D. Ll. Hammick, Ox Bow Press, Woodbridge (1990).}  which
contains a detailed description of his experiments on Brownian motion,
is highly recommended.  It begins:\smallskip

{\small ``Molecules: Some twenty-five centuries ago, before the close of the lyric
period in Greek history, certain philosophers on the shores of
the Mediterranean  were already teaching that changeful matter is made up of
indestructible particles in constant motion; atoms  which chance or
destiny has grouped in the course of ages into the forms or substances
 with which we are familiar.  But we know next to nothing of these early
  theories, of the works of Moschus of Sidon, of Democritus of Abdera, or of his
  friend Leucippus. No fragments remain that might enable us to judge of what in
  their work was of scientific value. And in the beautiful poem, of a much later date, wherein
   Lucretius expounds the teachings of Epicurus, we find nothing that enables us to grasp what facts or
   what theories guided Greek thought.''}\medskip

He further expounded on the idea that non-differentiable continuous
functions, such as the trajectory of Brownian motion, were as
completely natural as differentiable functions, objects of all
prior studies.  In the preface of {\it Atoms}, by considering the very
irregular surface of a colloid and by making the analogy with the
shape of Brittany's coast, he announced with a dazzling geometric
intuition the ideas of Lewis Fry Richardson on Hausdorff anomalous
dimensions, which would later be developed by Beno\^{\i}t
Mandelbrot.\footnote{B. Mandelbrot, {\it Fractal Objects}, (3\`eme
\'ed.), followed by {\it A Survey of Fractal Language}, Flammarion,
Nouvelle Biblioth\`eque scientifique (1989).}

Regarding Brownian motion, we find as well:\smallskip

{\small ``We are still in the realm of experimental reality when, under
 the microscope, we observe the Brownian movement agitating each small
particle suspended in a fluid.  In order to be able to fix a tangent to the
trajectory of such a particle, we should expect to be able to establish, within at least
approximate limits, the
direction of the straight line joining the positions occupied by a particle
at two very close successive instants.  Now, no matter how many
experiments are made, that direction is found to vary absolutely irregularly
 as the time between the two instants is decreased.
 An unprejudiced
observer would therefore come to the conclusion that he was dealing with a
 function without derivative, instead of a curve to which a tangent could be drawn.''}\medskip

Further along we read:\smallskip

{\small ``It is impossible to fix a tangent, even approximately, to any
point on a trajectory, and we are thus reminded of those
continuous functions\footnote{``{\it Continuous} because it is not possible to regard the grains as
passing from one position to another without
cutting any given plane having one of those positions on each side of it.''[original note]} without
derivative that mathematicians had imagined.  It would be incorrect to regard
such functions as mere mathematical curiosities, whereas Nature suggests them as much as
 differentiable functions.''}\medskip

These remarks would stimulate the research of the young mathematician
Norbert Wiener.\footnote{N. Wiener, {\it I am a Mathematician},
Doubleday, Garden City, NY (1956). He writes: ``The Brownian motion was nothing new
as an object of study by physicists. There were fundamental papers by Einstein
and Smoluchowski that covered it, but whereas these papers concerned what was
happening to any given particle at a specific time, or the long-time statistics
of many particles, they did not concern themselves with the mathematical properties of the
curve followed by a single particle.

Here the literature was very scant, but it did include a telling
comment by the French physicist Perrin in his book {\it Les Atomes}, where
he said in effect that the very irregular curves followed by particles in the Brownian motion
led one to think of the supposed continuous non-differentiable curves of the mathematicians.
He called the motion continuous because the particles never jump over a gap and non-differentiable
because at no time do they seem to have a well-defined direction of movement.''}

The beautiful experiments  of 1908-1909 by Perrin and his students, on
emulsions of gum-resin (``gamboge'') or of mastic, are described in detail in his
review article {\it Brownian Motion and Molecular Reality}, which
 appeared in 
Annales de Chimie et Physique in 1909,\footnote{J. Perrin, {\it Ann. Chim.  Phys.}
{\bf 18}, pp. 1-114 (1909); available online at http://gallica.bnf.fr/.} and
  the results are published in several {\it Notes aux Comptes
Rendus.} The same material is also summarized in his book {\it Atoms.}

Perrin began by verifying the exponential distribution of the density
of $n$ particles in a suspension, as a function of the height $h$ in a
gravitational field $g$, a formula that generalizes the barometric
formula for the atmosphere.  Perrin writes it in the form
\begin{equation}
\label{perrin}
\frac{2}{3}W \ln \frac{n_0}{n}=\phi (\rho-\rho_0)g h,
\end{equation}
where $\phi$ is the volume of each grain, $\rho$ and $\rho_0$ are the
 mass per unit volume of the grains and of the inter-granular liquid,
respectively, and last but not least, $W$ is the average kinetic
energy per particle (with $W=\frac{3}{2}\frac{RT}{\mathcal
N}=\frac{3}{2}{k_BT}$).\smallskip

He writes: {\small ``I indicated this equation at the time of my first
experiments ({\it Comptes Rendus}, May 1908).  I have since learned that
Einstein and Smoluchowski, independently, at the time of their beautiful theoretical
researches of which I shall speak later, had already seen that
the exponential repartition is a necessary consequence of the
equipartition of energy.  Beyond this it does not seem to have occurred that
 in this sense, an {\it experimentum crucis} could be obtained, deciding for or against
the molecular theory of the Brownian movement.''}

He continues: {\small ``If it is possible to measure the magnitudes other than $W$
which enter into
 this equation, one can see whether it is verified and whether the value it
indicates for $W$ is  the same as that which has been approximately assigned to the
molecular energy.  In the event of an affirmative answer, the origin of the Brownian movement
 will be established, and the laws of gases, already extended by
van 't Hoff to solutions, can be regarded as still valid even for emulsions
with visible grains.''}\medskip

He built as well an apparatus for fractioned centrifugation to produce
emulsions of uniform size, a key element of his success.  Using three
independent processes to measure the radius of particles, one of which
went via Stokes' law, he could verify the validity of the latter for particles in
suspension.  It was in fact one of the weak points of the theoretical
proofs, because the continuity conditions required by hydrodynamics
were far from being clearly fulfilled in the case of small spheres in
 very active Brownian motion.

Finally, by ingenious and patient observations, he could verify
the law of rarefaction of density (\ref{perrin}).\footnote{J. Perrin,
{\it C. R. Acad. Sci. Paris} {\bf 146}, 967 (1908); {\bf 147}, 475
(1908).} Thanks to the value of $W$ (independent of all
experimental conditions except the temperature), he verified the famous
law of equipartition of energy, and obtained a first estimate of
Avogadro's number, $\mathcal N = 7.05\times 10^{23}$, compared with
the present accepted value $\mathcal N = 6.02\times 10^{23}$.

\subsubsection{Einstein's formulae}
Perrin turned next to Einstein's formulae for Brownian diffusion:
{\small ``...another approach was possible and was proposed by
Einstein, in conclusion to his beautiful theoretical works, of
which I must now speak.''}  Further on, he adds: {\small ``It's fair
to recall that, almost at the same time as Einstein and by another
route, Smoluchowski arrived at a formula a bit different in his
remarkable work on {\it A kinetic theory of Brownian motion} [Bulletin
de l'Acad. des Sc. de Cracovie, July 1906, p. 577] where one finds,
besides very interesting observations, an excellent history of work
before 1905.''}\medskip

In {\it Atoms} he stresses that:\footnote{{\it Atoms, op. cit.}, chapter IV.}\smallskip

{\small Einstein and Smoluchovski have defined the activity of the
Brownian movement in the same way. Previously, we had been obliged to
determine the ``mean velocity of agitation'' by following as
nearly as possible the path of a grain. Values so obtained were always a few
 microns per second for grains of the order of a micron.\footnote{``Incidentally this
 gives the grains a kinetic energy $10^5$ times too small.'' [original note]}}

{\small But such evaluations of the activity are {\it absolutely wrong}. The trajectories are confused and
 complicated so often and so rapidly that it is impossible to follow them; the trajectory actually
 measured is very much simpler and shorter than the real one. Similarly, the apparent mean speed
 of a grain during a given time varies {\it in the wildest way} in magnitude and direction, and does not
 tend to a limit as the time taken for an observation decreases [...].}

{\small Neglecting, therefore, the true velocity, which cannot be measured, and
 disregarding the extremely
 intricate path followed by a grain during a given time, Einstein and Smoluchowski
 chose, as the magnitude
 characteristic of the agitation, the rectilinear segment joining the strarting
 and end points; in the mean, this line will clearly be longer the more active the agitation.
 The segment will be the {\it displacement} of the grain in the time considered.}\medskip

He begins his review by recalling the early work of Exner, prior to the
publication of Sutherland-Einstein-Smoluchowski formula for the average quadratic displacement
(\ref{x2}), and in which one can see ``at least one presumption of
partial verification for the formula in question.''

Soon after the publication of this formula, verification was
quickly tried by The Svedberg, who thought he achieved
it.\footnote{Th. Svedberg, {\it Studien zur Lehre von den kolloidalen
L\"osungen}, {\it Nova Acta Reg. Soc. Sc. Upsaliensis}, {\bf 2},
1907.} Perrin made a sharp criticism of these results, and declared him
 ``a victim of an illusion,'' regarding his description of
Brownian trajectories ``as regularly modulated in amplitude and
with well defined wavelength!''\footnote{One must add  that
The Svedberg won the Nobel Prize in Chemistry in 1926 for his invention
of the ultracentrifuge, the same year as Perrin won the Noble Prize in Physics for his work on Brownian motion!}

Victor Henri's results, published in {\it Comptes Rendus} in 1908, were
obtained from a better founded cinematographic study of Brownian
motion of natural latex grains.  The average displacement varied as 
the square root of time, but the coefficient was three times too
large.\footnote{Perrin then noted almost mischievously: ``As far as I could
judge from the conversation, a current of opinion was produced among
the French physicists community that closely followed these questions,
and which really shocked me, proving to me how much the credit that we
give to theories is limited, and at what point we see them as
instruments of discovery rather than as true demonstrations. Without
hesitating, they admitted that Einstein's theory was incomplete or
inexact. On the other hand, there was no reason to renounce placing
the origin of Brownian motion in molecular agitation, because I just
showed by an experiment that a diluted emulsion behaves as a very dense
perfect gas in which the molecules had a weight equal to the grains of
the emulsion.  They limited themselves to assuming that a few unjustified complementary
hypotheses slipped into Einstein's reasoning.''}

\begin{figure}[htbp]
\begin{center}
\includegraphics[angle=0,width=.5\linewidth]{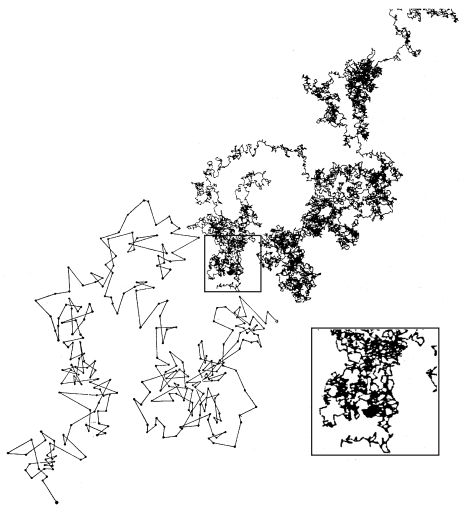}
\end{center}
\caption{{\it Brownian motion.
Bottom left: Strong magnification, showing the discretized aspect
of sequential recordings of the position of a particle in suspension,
observed by Jean Perrin and his collaborators.  Bottom right:
Magnification showing the self-similarity of the continuous Brownian
curve.}}
\label{fig.perrin}
\end{figure}

Having prepared grains with known diameter, Perrin asked his student
Chaudesaigues to verify the law of Brownian displacement by direct
observation, sequenced every thirty seconds, with gamboge grains 
of radius $0.212$ $\mu$m.\footnote{M. Chaudesaigues, {\it
C. R. Acad. Sci. Paris}, {\bf 147}, 1044 (1908); {\it Dipl\^ome
d'\'Etudes}, Paris (1909).} This was completed by similar measurements 
by M. Dabrowski\footnote{J. Perrin and Dabrowski, {\it
C. R. Acad. Sci. Paris}, {\bf 149}, 477 (1909).} on mastic grains, and
gave the famous diagrams of random positions that one can find in
Jean Perrin's book.  (See figure \ref{fig.perrin}.)

The conclusion was ``the rigorous exactness of the formula proposed by
Einstein,'' and ``that some unknown complication or unknown cause of
systematic error oddly affected the results of Victor Henri.''  They
then deduced a new average value of Avogadro's number, $\mathcal
N=7.15\times 10^{23}$.  A wonderful verification was at last made of
``Maxwell's irregularity law,'' that is, of the Gaussian distribution (\ref{gauss})
of the Brownian particle's position in a plane orthogonal to gravity.

Jean Perrin did not stop there, but turned to {\it rotational
Brownian motion.}  Einstein himself did not really think that
his predictions (\ref{delta2''}) were experimentally verifiable, because the speed of rotation seemed to be
too large to be observable.  In
fact, for grains of 1 $\mu$m in diameter, the rotation is about 1 degree per hundredth of second.
Perrin could then prepare spheres with larger diameter,
from 10-15 $\mu$m up to 50 $\mu$m, and he succeeded
in preparing them in suspension in a 27\% solution of urea.  In this
case the angular speed falls to a few degrees per minute.  The spheres carried
inclusions of different refractive indices, which made their rotation
observable under a microscope!  The result was a spectacular
verification of Einstein's second formula (\ref{delta2''}), this time
for grains 100~000 times heavier than the small grains of gamboge
first studied.\footnote{J. Perrin, {\it C. R. Acad. Sci. Paris}, {\bf
149}, 549 (1909).}  On 11 November 1909, Einstein wrote to Perrin:
``I  would not have considered a measurement of rotations as feasible. In my eyes it was only a
pretty triffle''.\footnote{Quoted in J. Stachel, {\it Einstein's Miraculous
Year} (Princeton University Press, Princeton, New Jersey, 1998).}\medskip

Perrin received the Nobel Prize in 1926 for his work on Brownian
motion.  His book, {\it Atoms}, one of the most finely written
physics books of the 20th century, contains a postmortem, in the great
classic style, about the fight for establishing the reality of
molecules:
\smallskip

{\small ``La th\'eorie atomique a triomph\'e. Encore nombreux nagu\`ere, ses adversaires enfin conquis
renoncent l'un apr\`es l'autre aux d\'efiances qui, longtemps, furent l\'egitimes et sans doute utiles.
C'est au sujet d'autres id\'ees que se poursuivra d\'esormais le conflit des instincts de prudence et d'audace dont
l'\'equilibre est n\'ecessaire au lent progr\`es de la science humaine.''}

{\small ``The atomic theory has triumphed. Its opponents, who until recently were numerous, have been
convinced and have abandoned one after the other the sceptical position that was for a long time
 legitimate and likely useful. Equilibrium between the instincts towards caution and towards boldness is
 necessary to the slow progress of human science; the conflict between them will henceforth be waged in
 other realms of thought.''}\medskip

To conclude this section, let us return for a last time to Einstein.  One reads in his autobiographical notes:\footnote{{\it Albert Einstein: Philosopher-Scientist},
{\sc The Library of Living Philosophers}, Vol. VII, Paul Arthur Schilpp  Ed., Open Court, La Salle, Illinois,
 3rd Edition (2000).}\smallskip

{\small ``The agreement of these considerations with experience together with Planck's determination
of the true molecular size from the law of radiation (for high temperatures) convinced the sceptics, who were quite
numerous at that time (Ostwald, Mach) of the reality of atoms. The antipathy of these scholars towards atomic theory
can undubitably be traced back to their positivistic philosophical attitude. This is an interesting example of the
fact that even scholars of audacious spirit and fine instinct can be obstructed in the interpretation of facts by
philosophical prejudices. The prejudice --which has by no means died out in the meantime-- consists in the faith
that facts by themselves can and should yield scientific knowledge without free conceptual construction.
 Such a misconception is possible only because one does not easily become aware of the free choice of such concepts,
 which, through verification and long usage, appear to be immediateley connected with the empirical material.''}\medskip

Let us finally mention Ostwald's magnanimous concession: In 1908 he refers to the Brownian motion results and
says they {\it ``entitle even the cautious scientist to speak of the experimental proof for the atomistic
constitution of space-filled matter''.} In 1910, he is the first person to nominate Einstein for the Nobel Prize
(for special relativity).


\section{Measurements by Brownian fluctuations}

Jumping ahead a century, we observe how the theory of Brownian
fluctuations, whose construction we just described, today finds spectacular applications in physics applied
to biology.  We will give an example from the physics of DNA.

\subsection{Micromanipulation of DNA molecules}
\subsubsection{The interest of DNA for physicists}

Physicists are interested today in DNA for several reasons.  First,
it is a remarkable polymer for its length, reaching several
centimeters, and for its monodispersity (the DNA of the virus
bacteriophage-$\lambda$, for example, always has 48502 base pairs in the
identical sequence).  DNA is an important subject in polymer physics
because it can be easily shaped by bio-molecular tools and it can be
directly observed and manipulated.  A fluorescent intercalation placed
between base pairs (such as ethidium bromide) permits the observation,
under a microscope and by fluorescence, of {\it single} DNA
molecules in solution.

\subsubsection{Experimental realization of a micro-manipulation}

One can also micro-manipulate molecules individually.  The
techniques of micro-manipulation of isolated bio-molecules have
developed considerably during the past few years, thanks to an ever-growing
number of tools: optical or magnetic ``tweezers,'' atomic force
microscopes, glass micro-fibers, and also hydrodynamic flow
observations.

A recent example consists in pulling a single DNA molecule to measure
its extension as a function of the force, which allows one to measure
various important mechanical parameters of the DNA chain.

\begin{figure}[tb]
\begin{center}
\includegraphics[angle=0,width=.6\linewidth]{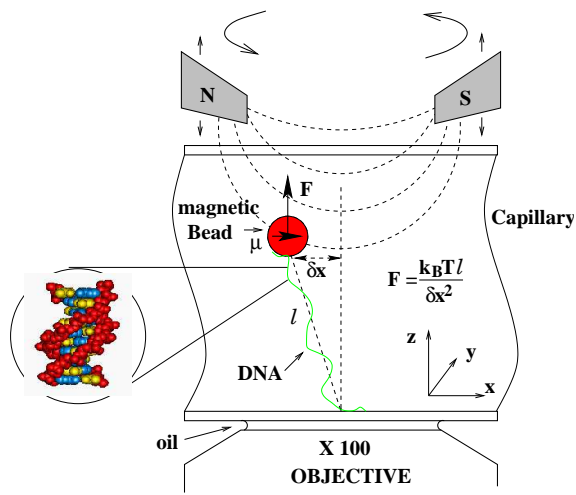}
\end{center}
\caption{{\it Micro-manipulation of a DNA molecule by ``magnetic tweezers''.}
}
\label{fig.forceex}
\end{figure}

In ``magnetic tweezers'' (figure \ref{fig.forceex}), a magnetic
bead is placed in the field of a magnet; the bead is attracted towards
regions with a high gradient field, and one can move the magnets or
rotate them.  This allows one to pull the DNA or to twist it, creating as
well torsions, or super-coilings, that are a part of topological
configurations for biological functions.

We give a brief overview of forces playing a role in biology, and of
the specific problems related to their smallness.

\subsubsection{Biological interaction forces and thermal agitation forces.} 

The interaction forces in biological systems are typically generated
by hydrogen or ionic bonds, as well as by van der Waals interactions
that shape nucleic acids and proteins.  Their order of magnitude is
typically obtained by dividing $k_BT$, the order of magnitude of
the ``quantum of energy'' provided by the hydrolysis of the ATP
in ADP\footnote{ATP: adenosine triphosphate, universal biological
``fuel,'' made of one sugar, ribose, and of one base, adenine,
and of three phosphate groups; ADP: adenosine diphosphate, is the
degraded version after losing a phosphate group under enzymatic
action and release of energy.} (10 $k_BT$), by the
characteristic size of biological objects, of the order of a
nanometer (nm).  We then find the {\it picoNewton}:

$$\ds\frac{k_BT}{1{\rm nm}}=4\underset{\substack{\Vert\\
10^{-12}\text{ N}}}{\text pN}.$$
\begin{figure}[htbp]
\begin{center}
\includegraphics[angle=0,width=.69\linewidth]{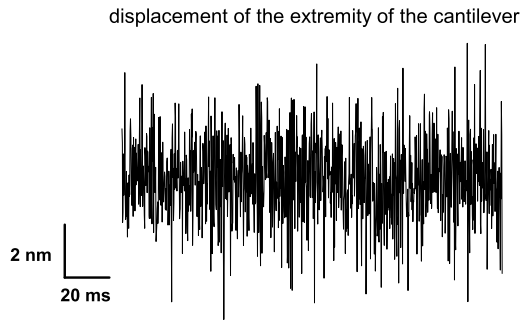}
\end{center}
\caption{{\it The track of the random displacement, in a liquid, of the tip
of the cantilever in an atomic force  microscope. It executes a one-dimensional Brownian motion.
(Kindly provided by Pascal Silberzan and Olivia du Roure, Curie Institute.)}}
\label{fig.cantilever}
\end{figure}

Such a force is the one typically needed to stretch a DNA molecule.
As it is extremely small, it is not easy to detect with standard measuring
devices.

The smallest measurable forces are in principle limited by the {\it
thermal agitation} of the measuring device (see figure
\ref{fig.cantilever}).  This thermal agitation generates Langevin's
stochastic force seen above, whose value depends on the coefficient of viscous
friction of the object, and also on the temporal window of
observation.  We have: $$\la X^2_{\text{Langevin}}\ra
=2k_BT\;6\pi\eta\; a \, \delta f,$$ where $\eta$ is the medium's
viscosity, $a$ the radius of a spherical bead taken as an example, and $\delta f$
the observed frequency range.  For example, for $a=1.5\;\mu {\rm m}$, in
water (viscosity $\eta\simeq 10^{-2}$ Poise = $10^{-3}$ Pa/s), the average force over a period of a second is
$X_{\text{Langevin}}\sim 15\underset{\substack{\Vert\\ (10^{-15}\text{
N})}}{f\text{N}}$, i.e., 15 femtoNewtons.

Astonishingly, Brownian fluctuations can be used directly to measure
forces of biological origin!

\subsection{Measurement of force by Brownian fluctuations}
\begin{figure}[tb]
\begin{center}
\includegraphics[angle=0,width=.6\linewidth]{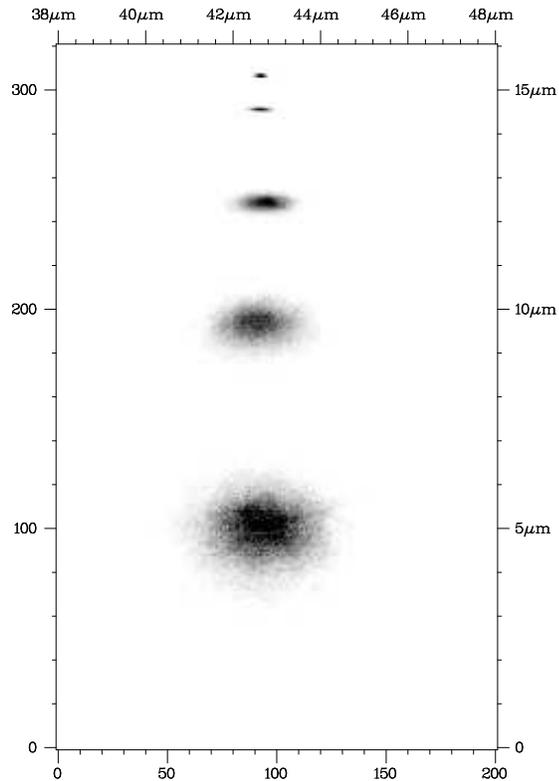}
\end{center}
\caption{{\it Brownian cloud of a bead's fluctuating position in the vertical plane $(Ox,Oz)$,
for different applied forces.  The larger the force, the more a molecule is stretched, and the more the
Brownian fluctuations are constrained.  (Figure kindly provided by Vincent Croquette, Statistical Physics Laboratory,
ENS, Paris.)}}
\label{fig.nuagebrownien}
\end{figure}

This technique of measuring a force is largely inspired by the method
proposed by Einstein\footnote{A. Einstein, {\it Investigations on the
Theory of the Brownian Movement}, R. F\"urth {\it Ed.},  A. D. Cowper {\it Transl.},
Dover Publications, p. 24 (1956).}  for measuring the elastic constant
of a spring by means of {\it Brownian fluctuations}.  When we apply
 a force upon a small magnetic bead in a gradient field, the
stretched molecule and the bead form a minuscule pendulum of length
$\ell$ (figure \ref{fig.forceex}).  The bead is animated by Brownian
motion, connected to the thermal agitation of surrounding
water molecules.  The small magnetic pendulum is thus perturbed
from its equilibrium position by Langevin's random force.  It is then
brought back towards equilibrium by the pulling force exerted by the
DNA (figure \ref{fig.nuagebrownien}).

As we will show in detail further along, the pendulum possesses a
transverse elastic constant $k_{\perp}$ that is directly related to
the pulling force $F$ by $k_{\perp}=F/\ell$.  If we call $x$ the
position of the bead with respect to its equilibrium position in the direction
perpendicular to the force $\vec F$, the theory states $$F=k_BT \ell
/\la x^2\ra,$$ where $\la x^2\ra$ is the average quadratic fluctuation
of $x$.  To measure the pulling force on a DNA molecule, one simply
measures the length $\ell$ and the average quadratic fluctuation $\la
x^2\ra$!  This is reminiscent of Einstein's formula (\ref{x2}), as
well as of the surprise of being able to deduce Avogadro's number from
it.

To measure such fluctuations, one must follow the movements of the
bead during a given amount of time, just as in Jean Perrin's
experiments of 1908 on Brownian motion.  Today, a computer program
analyzes in real time the images on a video of the bead observed via a
microscope, and determines its positions in a three-dimensional space
with a precision of 10 nm (figure \ref{fig.nuagebrownien}).  Such
precision is obtained through a technique of image correlation.

This sort of Brownian measurement has several advantages:\\
- One gauges the force by absolute measurement of position fluctuations;\\
- There is no contact with the bead, therefore it is non-invasive;\\
- The range of values of $x$ is between $\mu$m to nm, the force goes from a dozen
femtoNewtons to hundreds of picoNewtons.\\ 
The drawback is its slowness: to accumulate sufficient fluctuations and to have 
reliable statistics, a minute of recording is needed for a force of
1pN, and more than an hour for 10fN.

We shall now describe the theory of measurement by Brownian
fluctuations.

\subsection{Theory}
\subsubsection{Equilibrium and fluctuations}
\vspace{1ex}
\begin{figure}
\begin{center}
\includegraphics[angle=0,width=.32\linewidth]{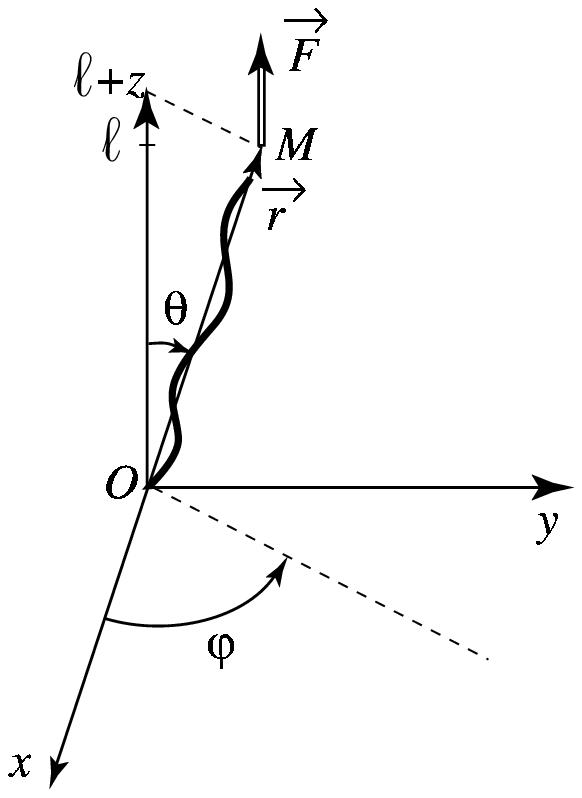}
\end{center}
\caption{{\it Axis of a DNA chain fluctuating around
the vertical position; the extremity $M$  moves from the equilibrium position
 $(0,0,\ell)$ in presence of $F$ towards the random position $(x,y, \ell + z)$.}}
\label{fig.Force}
\end{figure}

One considers  a DNA chain of length $\ell_0$ with one
extremity fixed at the origin $0$, while the other extremity $M$ is
determined by $\overrightarrow{{OM}}=\vec r$ (see figure \ref{fig.Force}).  A
force $\vec F$ acts on the extremity $M$ along the direction of the 
$Oz$ axis.  At equilibrium, the chain is parallel to the $Oz$ axis and
is elastically stretched up to a length $\ell$ dependent on $F$.
The Brownian fluctuations, originating from the shocks between the bead that is
attached to the DNA chain and the molecules of the solution, induce small displacements $(x, y, z)$
that one can consider as
perturbations of the macroscopic equilibrium
position (0,0,$\ell$).  The extremity $M$ is thus shifted from its
equilibrium position $(0, 0,\ell)$ (in the presence of $F$) to a random
position $(x, y, \ell + z)$.  Let $r=\vert\overrightarrow{{OM}}
\vert$ be the
radial distance between the extremities of the chain.  Because of elasticity,
the chain develops a {\it restoring radial force} $F_r(r)$.  At
equilibrium, one has $F_r(\ell)=F$, where $F$ is the external
force given experimentally.

In the presence of fluctuations, the radial distance is written
\begin{equation}
r=[(\ell+z)^2+x^2+y^2))]^{1/2},
\label{r}
\end{equation}
and the restoring force
\begin{eqnarray}
\label{Fxyz}
\vec   F_r=-\frac{\vec   r}{r}F_r(r)=
\begin{cases}
F_{rx}=-\ds\frac{x}{r}F_r(r)\\[.4cm]
F_{ry}=-\ds\frac{y}{r}F_r(r)\\[.4cm]
F_{rz}=-\ds\frac{\ell+z}{r}F_r(r).
\end{cases}
\end{eqnarray}

\subsubsection{Series expansions}

One writes the series expansion  of the distance $r$ for
$x, y, z$ small compared to $\ell$:
\begin{eqnarray}
\label{rdl}
r=[(\ell+z)^2+x^2+y^2]^{1/2}
=\ell+z+\cdots
\end{eqnarray}
An expansion to the  first linear order in $x,y,z$ will be sufficient, and from now on we 
will denote by $+\cdots$  all second order terms (of
$O(x^2,y^2,z^2)$) in the expansions.

The radial force $F_r(r)$ of  the DNA on the bead,
 depends only on the radial distance $r$; therefore, from (\ref{rdl}), it has
the series expansion:
\begin{eqnarray}
 F_r(r)&=&F_r\left[\ell+z+\cdots\right]
       =F_r(\ell)+z\frac{d\, F_r}{d\, r}(\ell)+\cdots.
       \label{Frdlo}
\end{eqnarray}
One can now easily determine the components (\ref{Fxyz}) of the radial force
by using (\ref{rdl}) and (\ref{Frdlo}):
\begin{eqnarray}
\nonumber
F_{rx}&=&-\frac{x}{r}F_r(r)=-\frac{x}{\ell}F_r(\ell)+\cdots,\\
\nonumber
F_{ry}&=&-\frac{y}{r}F_r(r)=-\frac{y}{\ell}F_r(\ell)+\cdots,\\
\nonumber
F_{rz}&=&-\frac{\ell+z}{r}F_r(r)
=-F_r(\ell)-z\frac{d\, F_r}{d\, r}(\ell)+\cdots.
\\
\label{Fxyzdl}
\nonumber
\end{eqnarray}

We finally note that at the equilibrium point the external force, $\vec
F= F \vec u_z$, exactly cancels the term $-F_r(\ell) \vec u_z$ of the
vertical component $F_{rz}\vec u_z$. 
Leaving aside terms of second order, our analysis leads us
to a fluctuating resultant force on the DNA~:
\begin{eqnarray}
\label{forcefluctuante}
{\vec   f}=F \vec   u_z+ {\vec F_r} =\begin{cases}
-\ds\frac{x}{\ell}F_r(\ell)\\[.4cm]
-\ds\frac{y}{\ell}F_r(\ell)\\[.4cm]
-z\ds\frac{dF_r}{d\ell}(\ell)
\end{cases}\quad {=-\nabla_{\vec   r}\,U.}
\end{eqnarray}

\subsubsection{Elastic energy}

The beauty of this approach is that one can determine the elastic
energy of the Brownian fluctuations of the DNA chain
without even knowing the analytic form of the elastic force.
In these expressions, it must be understood that the equilibrium length 
$\ell$ is determined by the external force, while the
fluctuating force (\ref{forcefluctuante}) is {\it linear}
in $x,y,z$, as expected from an expansion to first order.
A quadratic potential energy $U$ is associated to the force by 
${\vec f}=-{\nabla}_{\vec r}\, U$, given by the simple expression:
\begin{eqnarray}
U=\frac{1}{2}\left({x^2}+{y^2}\right)\frac{1}{\ell}F_r(\ell)
+\frac{1}{2}z^2\frac{d\, F_r}{d\, \ell}(\ell).
\label{U}
\end{eqnarray}

\subsubsection{Elastic constants}

One can write the energy $U$ (\ref{U}) as that of a three-dimensional anisotropic harmonic oscillator
with two elastic constants, $k_{\perp}$ and $k_{\parallel}$, corresponding to the perpendicular and parallel
directions, respectively,  with respect to the force:
\begin{eqnarray}
U=\frac{1}{2}k_{\perp}\left({x^2}+{y^2}\right)
+\frac{1}{2}k_{\parallel}\,z^2,
\label{Uk}
\end{eqnarray}
with
\begin{eqnarray}
\begin{cases}
k_\perp&=\ds\frac{F_r(\ell)}{\ell},\\[.4cm]
k_{\parallel}&=\ds \frac{d\, F_r}{d\, \ell}(\ell).
\label{k}
\end{cases}
\end{eqnarray}
As one can intuitively imagine, the transverse elastic constant, which
opposes lateral movements of the DNA molecule, is weaker than the
longitudinal elastic constant, which opposes mechanical stretching
of the DNA.

\subsubsection{Energy equipartition}

In {\it classical} statistical mechanics, we have seen the historically important
result about the {\it equipartition of energy}.  The theory
simply states that each {\it quadratic} degree of freedom has average energy 
$\frac{1}{2}k_B T$ exactly, where $k_B$ is Boltzmann's
constant and $T$ is the absolute temperature.  In the case of the harmonic energy
(\ref{Uk}), the theorem immediately gives us:
\begin{eqnarray}
\frac{1}{2}k_{\perp}\langle {x^2}\rangle=\frac{1}{2}k_{\perp}\langle {y^2}\rangle
=\frac{1}{2}k_{\parallel}\,\langle  z^2\rangle=\frac{1}{2}k_B T\, .
\label{equip}
\end{eqnarray}
Therefore we find, with the help of (\ref{k})
\begin{eqnarray}
\begin{cases}
k_\perp&=\ds\frac{F_r(\ell)}{\ell}= \frac{k_B T}{\langle {x^2}\rangle}\, ,\\[.4cm]
k_{\parallel}&=\ds \frac{d\, F_r}{d\, \ell}(\ell)=\frac{k_B T}{\langle {z^2}\rangle}\, .
\label{kfluc}
\end{cases}
\end{eqnarray}

\begin{figure}[htbp]
\begin{center}
\includegraphics[angle=0,width=.7\linewidth]{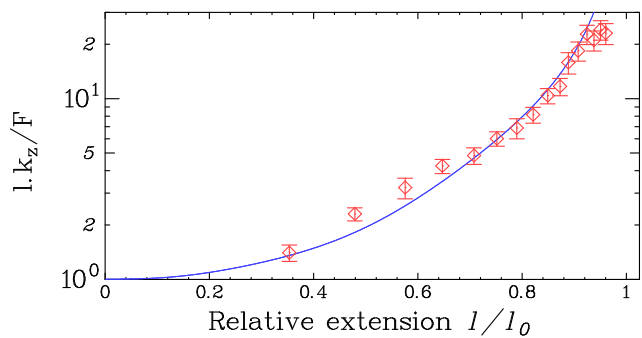}
\end{center}
\caption{{\it Dimensionless ratio $\frac{\ell
k_{z}}{F}=\frac{k_{\parallel}}{k_{\perp}}= \frac{\la {x^2}\ra }{\la {z^2}\ra}
=\frac{\ell}{F_r(\ell)} \frac{d\, F_r}{d\, \ell}(\ell),$ plotted as a
function of the length $\ell$ of the DNA chain (in units of maximum
length $\ell_0$).  The points correspond to the ratio of
experimental measurements of transverse ($\la {x^2}\ra$) and vertical ($\la {z^2}\ra $)  quadratic
Brownian fluctuations.  The curve
is theoretically predicted from the knowledge of $F_r(\ell)$, in a
model of a semi-flexible DNA chain, also known as the {\it Worm-like
Chain Model}.  We stress the remarkable agreement between experiment
and theory.  (Figure kindly provided by Vincent Croquette.)}}
\label{fig.force2tb}
\end{figure}

Because of the difference between the elastic constants, $k_{\perp} <
k_{\parallel}$, transverse fluctuations dominate over  longitudinal
ones: $\langle {x^2}\rangle=\langle {y^2}\rangle >
\langle {z^2}\rangle$, as one can see in figure \ref{fig.nuagebrownien}.  One
sees, for instance, that the fluctuations $\sqrt{\langle {x^2}\rangle}$
and $\sqrt{\langle {z^2}\rangle}$ are of the order of $2\,
\mu {\rm m}$ and of less than $1\, \mu {\rm m}$, respectively, for the second
Brownian cloud from the bottom.  Such Brownian fluctuations can be
directly measured optically, as can the length $\ell$, and
equation (\ref{kfluc}) allows a truly ingenious direct measurement of
the elastic force $F_r(\ell)$ and its derivative ${F_r^{'}(\ell)}$!
One can then compare the experimental results to the predictions of
theoretical models for the statistical description of the DNA
configurations (see figure \ref{fig.force2tb}).


\section{Potential theory and Brownian motion}

\centerline{{\it Et ignem regunt numeri}\footnote{Joseph Fourier's major work, {\it La th\'eorie
analytique de la chaleur}, was published in 1822, with {\it Et ignem
regunt numeri} as its motto ({\it Numbers rule fire}). See also Gaston Bachelard, {\it \'Etude sur l'\'evolution
d'un probl\`eme de physique, la propagation thermique dans les solides}, Librairie philosophique J. Vrin, Paris (1927).}}

\subsection{Introduction}
\subsubsection{Laplace's equation}
Potential theory concerns the equilibrium properties of
continuous bodies, like the distribution of electrostatic charges on
conductors, the distribution of the Newtonian potential in the classic
theory of gravitation, the distribution of temperature in Fourier's
theory of heat conduction, or in addition the distribution of positions
of a stretched elastic membrane.\footnote{One can cite
O. D. Kellogg's classic work {\it Foundations of Potential Theory},
Springer-Verlag (1929); Dover Books on Advanced Mathematics (1969).}

A deep relation exists between potential theory and the theory of
diffusion, and therefore also with Brownian motion.\footnote{See the
article  {\it Brownian Motion and Potential Theory}, by
R. Hersch and R. J. Griego, {\sc
Scientific American}, {\bf 220}, March 1969; translated into French in
{\it Le mouvement brownien et la th\'eorie du potentiel},  appearing in 1977 within the first
out-of-series of {\sc Pour la Science}.} We will first give an intuitive
illustration within the framework of Fourier's theory of heat
conduction.

The temperature of a body, $u(x,y,z; t)$ at the point ${x,y,z}$  and
at the instant $t$, follows the equation of heat
\begin{equation}
\label{chaleur}
\frac{\partial u}{\partial t}=D \Delta u,
\end{equation}
where, as in the case of Brownian motion, $D$ is the diffusion
coefficient, and $\Delta$ is
the Laplacian in three-dimensions
$\Delta=\frac{\partial^2}{\partial x^2}
+\frac{\partial^2}{\partial y^2}+\frac{\partial^2}{\partial z^2}$.
In general, the Laplacian in $d$ dimensions is:
\begin{equation}
\label{laplacien}
\Delta=\sum_{i=1}^{d}\frac{\partial^2}{\partial x_i^2},
\end{equation}
where $x_i$ are $d$-dimensional Cartesian coordinates. When
the temperature reaches equilibrium, the time dependence cancels,
and the temperature field is  described by Laplace's equation:
\begin{equation}
\label{laplace}
\Delta u=0.
\end{equation}
Any function with zero Laplacian is called {\it harmonic}.

Such a function, the {\it potential}, therefore can be seen as the
equilibrium solution of a diffusion process (at infinite time), which
is the first elementary relation we meet between potential theory
and Brownian diffusion.  To specify in our example the value of the
temperature everywhere, we must fix the initial conditions in case
one starts from an out-of-equilibrium situation.

In the case we will consider here, we want to directly study
equilibrium and the associated harmonic functions, or more generally the potential.
For that purpose one  must know either the position of the sources of 
the potential, or the boundary conditions on it, in a way
that will be made more precise in the following.

Giving the position of the sources is natural in the well-known theory
of the Newtonian or Coulomb potential, where the sources of the potential
are  masses or  electrostatic charges.  Imposing boundary
conditions on the potential is also possible, as is natural in
the case of heat conduction and temperature distribution, where one
gives the temperature distribution on the surface of a body to
determine the internal temperature distribution.
	
Such representations are mathematically equivalent.  Let us first
recall the elementary properties of the Newtonian or Coulomb potential that
will be useful for obtaining the finer properties of harmonic
functions.
To fix the ideas, we will adopt the familiar language of a Newton or
 Coulomb potential created by masses or electrostatic charges, but the
mathematical results of course will not depend on this choice.

\subsection{Newtonian potential}
\subsubsection{The potential created by a point source}

In order to consider the potential in a universal way, as for 
gravitation or electrostatics, the physical constants like the
universal gravitational constant $G$, or the electric permeability of
the vacuum, $\varepsilon_0$, are not indicated.  In general, we will
adopt the electrostatic language.

\begin{figure}[tb]
\begin{center}
\includegraphics[angle=0,width=.26\linewidth]{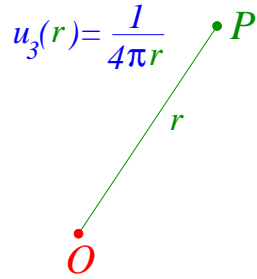}
\end{center}
\caption{{\it Newtonian potential in three dimensions.}}
\label{fig.newton}
\end{figure}

The potential at a point $P$ in three dimensions created by a unit charge
or mass
placed at the origin $O$ is
\begin{equation}
\label{potentiel}
u_3(r)=\frac{1}{4\pi r},\;\; r=|\overrightarrow {OP}|,
\end{equation}
where $r$ is the distance between $O$ and $P$ (figure \ref{fig.newton}).

The associated electric (or gravitational) field is
\begin{equation}
\label{champ}
\vec E_3(\vec r)=-\nabla_{\vec   r}\,u_3(r)=\frac{1}{4\pi} \frac{\vec r}{r^3},
\end{equation}
where $\vec r$ is the relative position vector $\vec r=\overrightarrow {OP}$.

In $d$ dimensions, the potential and field generalize to
\begin{equation}
\label{potentield}
u_d(r)=\frac{1}{(d-2)S_d}\frac{1}{r^{d-2}},
\end{equation}
and
\begin{equation}
\label{champd}
\vec E_d(\vec r)=-\nabla_{\vec   r}\,u_d(r)=\frac{1}{S_d} \frac{\vec r}{r^d},
\end{equation}
where $S_d=2\pi^{d/2}/\Gamma(d/2)$ is the surface of the unit sphere
in $\mathbb R^d$.

The two-dimensional case is more complicated, and leads to a
logarithmic potential,
\begin{equation}
\label{potentiel2}
u_2(r)=\frac{1}{2\pi}\log\frac{1}{r},
\end{equation}
\begin{equation}
\label{champ2}
\vec E_2(\vec r)=-\nabla_{\vec   r}\,u_2(r)=\frac{1}{2\pi} \frac{\vec r}{r^2}.
\end{equation}

\subsubsection{Laplace's equation and the Dirac distribution}

The Laplacian of the above potential $u_d(r)$ vanishes identically
everywhere in space, except at the origin: $\Delta
u_d(r)=0,\; r\neq 0$.  At $\vec r=\vec 0$ it is divergent,
and its value is given by a distribution, namely
\begin{equation}
\label{dirac}
\Delta u_d(r)=\frac{1}{(d-2)S_d}\Delta\frac{1}{r^{d-2}}=-\delta^d(\vec r),
\end{equation}
where $\delta^d(\vec r)$ is the Dirac distribution in $d$ dimensions, zero
everywhere except at the origin $\vec r=\vec 0$, where it is
singular (infinite).  This divergence is such that the integral
\begin{equation}
\label{diraci}
\int_{\mathbb R^d} f(\vec r)\delta^d(\vec r)\, {\rm d}^dr=f(\vec 0)
\end{equation}
yields the value at the origin of any test function $f(\vec r)$.

Equation (\ref{dirac}) is Poisson's equation, where the second
term represents the mass or charge density, i.e., the source of
the
potential.  In the case of a potential (\ref{potentiel}),
(\ref{potentield}) or (\ref{potentiel2}), such a source is a point,
at which a singular density appears.

In the elementary approach that follows, we shall not use this formalism.  Rather, we will follow the
elementary path that uses Gauss' theorem.\footnote{O.D. Kellogg, {\it
op.cit.}}

\subsubsection{Gauss' theorem}
Gauss' theorem says that the flux of an electric (or gravitational) field across
any closed surface $\Sigma$ is equal to the total
charge $Q(\Sigma)$ (or mass) enclosed by the surface:
\begin{eqnarray}
\label{thgauss}
\int_{\Sigma} \vec E \cdot\, {d \vec S}=Q(\Sigma).
\end{eqnarray}
This theorem can be proved in two stages. By linearity, since the case
of a distribution of charges can be treated by adding the fields, one
can reduce it to the case of a point charge.  Actually, if each one of
these fields satisfies Gauss' theorem, their sum will as well.

Next, for a point charge enclosed by the surface, we notice that the
flux of $\vec E$ is invariant when we deform the surface $\Sigma$
without crossing the charge.\footnote{We have, from the Green-Ostrogradski
theorem, that $\int_{\Sigma} \vec E \cdot\, {d \vec
S}-\int_{\Sigma'} \vec E \cdot\, {d \vec S}=\int_{\mathcal D} {\rm div}\vec
E \, d^3v=-\int_{\mathcal D} \Delta u \, d^3v=0$, where $\mathcal D$
is the domain  between the two surfaces $\Sigma$ and $\Sigma'$, and $u$ is the
potential. Indeed, we have the identities $\vec E=-\vec \nabla u$
and ${\rm div}(\vec \nabla u)=\Delta u=0$, because $u$ is harmonic in
the domain $\mathcal D$ without charges.}  We can thus restrict attention to a
sphere around the charge, for which Gauss' theorem is trivial.
Actually, because of the form (\ref{champ}) of the $1/r^2$ field with
spherical symmetry, the integral (\ref{thgauss}) on a sphere of a
radius $r$ is equal to the charge.

Gauss' theorem immediately generalizes to any number of dimensions.

\subsubsection{Potential generated by a sphere}

Let us consider the sphere $\mathcal S(a)$ of radius $a$ centered at the origin $O$.
Imagine that it carries a charge $Q$ uniformly
distributed over its surface.

The associated field $\vec E(r)$ is radial and with spherical
symmetry.  It satisfies Gauss' theorem (\ref{thgauss}).  If we
choose the surface $\Sigma$ as a sphere $\mathcal S(r)$ centered at
$O$, of radius $r>a$, i.e., exterior to $\mathcal S(a)$, we
have $Q(\Sigma)=Q$, and the flux of $\vec E(r)$ across $\Sigma$ is
simply, by spherical symmetry, $E(r) 4\pi r^2=Q$.  We then deduce that
$E(r)=\frac{Q}{4\pi r^2}$ is the same field that would be created by a
charge as if it was concentrated at the center of sphere.  If the
surface $\Sigma$ is chosen like a sphere $\mathcal S(r)$ of radius
$r<a$, i.e., inside  $\mathcal S(a)$, then $Q(\Sigma)=0$ and the
flux (\ref{thgauss}) is then zero.  By symmetry, we then deduce that
the field $\vec E$ is zero everywhere inside the sphere.

Let $u_{\mathcal S}(P)$ now be the potential created at a point $P$ by
the same sphere $\mathcal S(a)$ of radius $a$ with total charge $Q$,
uniformly distributed on the surface.  This potential has a spherical
symmetry, as does its associated field.  Outside the sphere, the field is
the same as that of a point charge $Q$ placed at the center, while
inside the sphere the field is zero.  The potential outside the sphere
is therefore the one, (\ref{potentiel}), created by a point charge placed at the center of
the sphere, while inside the sphere it is
constant, and by continuity equal to its value on the boundary. One thus has
\begin{equation}
\label{potentielsphere}
u_{\mathcal S}(P)=\frac{1}{4\pi}\frac{1}{r}
\vartheta(r-a)+\frac{1}{4\pi}\frac{1}{a}\vartheta(a-r),\;\; r=|\overrightarrow {OP}|,
\end{equation}
where $\vartheta$ is the Heaviside distribution
$ \vartheta(x<0)=0,\; \vartheta(0)=1/2,\; \vartheta(x>0)=1$.

\subsection{Harmonic functions and the Theorem of the Mean}

\subsubsection{Gauss' theorem of the Arithmetic Mean}

The property that two bodies or two charges attract one another with
equal and opposite forces, reflects itself in the potential. Actually
  the potential is symmetric with respect to the coordinates of the two points,
in such a way that the potential at $P$ of a charge $Q$ at $S$ is the
same as the potential at $S$ of a charge $Q$ at $P$.  From such a
simple fact follow theorems with important applications.  We derive
two of them, called Gauss' theorems of the Arithmetic
Mean.\footnote{O. D. Kellogg, {\it op. cit.}}\\
\begin{figure}[htbp]
\begin{center}
\includegraphics[angle=0,width=.26\linewidth]{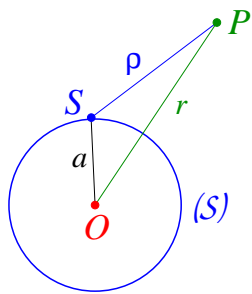}
\end{center}
\caption{{\it Newtonian potential (\ref{spherepoint}) created by a uniformly charged sphere of radius
$a$.}}
\label{fig.moyenne}
\end{figure}

The potential
\begin{eqnarray}
\label{spherepoint}
u_{\mathcal S}(P)&=&\frac{Q}{4\pi a^2}\int_{\mathcal S}\frac{d^2S}{ 4\pi \rho}, \; \rho=|\overrightarrow{SP}|
\end{eqnarray}
is the one at point $P$, created by all points $S$ on the surface of a
sphere $\mathcal S$ of radius $a$, and with uniform charge density
$\frac{Q}{4\pi a^2}$ (see figure
\ref{fig.moyenne}).  In (\ref{potentielsphere}) we just saw that outside 
the sphere the potential is equal to $\frac{Q}{4\pi r}$,
where $r$ is the distance $r=|\overrightarrow {OP}|$, while inside the sphere it is
constant and equal to $\frac{Q}{4\pi a}$.

But because of the exchange symmetry which we just mentioned, the
potential can also be interpreted as the arithmetic mean on the
surface of a sphere of the potential created by the same charge $Q$,
this time placed in $P$.

The equations (\ref{potentielsphere}) (\ref{spherepoint}) therefore
have the following interpretation:\\
\noindent a) The average on the surface of a sphere of the potential created by a
charge situated {\it outside} the sphere, and at a distance $r$ from its
center, is equal to the value (varying as ${1}/{r}$) of the potential at the
center of the sphere;\\
\noindent b) The average on the surface of
a sphere of the potential created by a charge in any position 
{\it inside} the sphere, is equal to the value (varying as ${1}/{a}$) of
the potential on the sphere, after concentrating the whole charge at
the center of the sphere.

Now let us suppose that we have a group of charges placed either
entirely on the outside of the sphere, or entirely on the inside. By adding up the
above results for each elementary charge, we find the following two generalizations:\\
\noindent{\bf a) Gauss' Theorem of the Arithmetic Mean:} the average on a surface of a
sphere of the potential created by charges situated entirely outside the
sphere is equal to the value of the potential at the center;\\
\noindent{\bf b) The Second Theorem of the Mean:} the average of the potential on a surface of a
sphere, created by charges situated entirely inside the
sphere, is independent of their distribution inside the
sphere, and it is equal to the total charge divided by the radius of the
sphere.\footnote{One can find the first theorem in Gauss' complete works,
{\it Allgemeine Lehrs\"atze}, vol. V, p. 222. The second theorem, less
known, can be found there too.}

\subsubsection{Harmonic functions}

Finally let us come back to harmonic functions, and consider a
function $u$ such that $\Delta u=0$ in some domain $\mathcal D$. Such
a harmonic function can be represented as a potential created inside
the domain $\mathcal D$ by a distribution of charges outside $\mathcal D$.  We
can then apply the first of Gauss' theorems, and obtain the {\it
mean-value theorem for harmonic functions}: {\it The average of a
harmonic function $u$ on a sphere $\mathcal S$ centered at a point P is equal
to the value of $u$ at $P$.}  For instance, in three dimensions:
\begin{equation}
\label{moyenne}
u(P) =\int_{\mathcal S} u(S)\frac{d^2S}{4\pi a^2},
\end{equation}
where $a$ is the radius of the sphere; the theorem can be generalized to any
number of dimensions.

The reciprocal is also true: any function that fulfills the
Theorem of the Mean on every sphere inside a given domain, is
harmonic inside that domain. This theorem is going to be the key
relation between  potential theory and Brownian motion.\footnote{A
proof of the Theorem of the Mean  can be obtained by vectorial
analysis.  We write the average $\la u\ra_{\mathcal S}$ of $u$ on the
surface of the $(d-1)$-sphere $\mathcal S$ of radius $a$ in $\R^d$, as
the flux of the vector $\ u(\vec r)\vec r/r^d$ on the surface of the
sphere:
\begin{eqnarray}
\label{fluxd}
\la u\ra_{\mathcal S}=\frac{1}{S_d \, a^{d-1}} \int_{\mathcal S} u(S) \, d^{d-1}S=\frac{1}{S_d}
\int_{\mathcal S} u(\vec r)\frac{\vec r}{r^d}.\, \vec n\, d^{d-1}S=-\int_{\mathcal S} u(\vec r)\vec \nabla u_d(r).\, \vec n\,d^{d-1}S,
\end{eqnarray}
where $S_d$ is the area of the unit sphere, $\vec n$ is the unit vector
normal to the surface of the sphere (and directed towards the exterior), 
and where we used (\ref{champd}).  We therefore use Green's
theorem in the volume $\mathcal D$ inside the sphere:
\begin{eqnarray}
\label{greenth}
\int_{\mathcal D} \left[u_d(r)\, \Delta u(\vec r) -u(\vec r)\, \Delta u_d(r)\right]d^dr
=\int_{\mathcal S} \left[u_d(r)\, \vec \nabla u(\vec r)-u(\vec r)\vec \nabla u_d(r)\right].\, \vec n\, d^{d-1}S.
\end{eqnarray}
We have $\Delta u(\vec r)=0$, because $u$ is harmonic, and from
(\ref{dirac}) we have $\Delta u_d(r)=-\delta^d(\vec r)$.  From the 
definition (\ref{diraci}) of Dirac distribution and by substituting
(\ref{fluxd}) in (\ref{greenth}), we have:
\begin{eqnarray}
\label{u0}
u(0)=\la u\ra_{\mathcal S}
+\int_{\mathcal S} \left[u_d(r)\, \vec \nabla u(\vec r)\right].\, \vec n\, d^{d-1}S.
\end{eqnarray}

As the Newtonian potential is constant on the sphere,
$u_d(r)=u_d(a)=\frac{1}{(d-2)S_d a^{d-2}}$,
the last flux integral is transformed into a volume
integral and it yields
\begin{eqnarray}
\label{greenc}
u_d(a)\int_{\mathcal S} \vec \nabla u(\vec r).\, \vec n\, d^{d-1}S
=u_d(a)\int_{\mathcal D} \Delta u(\vec r)\, d^dr=0,
\end{eqnarray}
because $u$ is a harmonic function by hypothesis.  We have then obtained the
Theorem of the Mean as expected:
\begin{eqnarray}
\label{u0f}
\la u\ra_{\mathcal S}=u(0).
\end{eqnarray}
}

\subsection{The Dirichlet problem}
A classic problem of potential theory is that of
Dirichlet.  One considers a domain $\mathcal D$ of the Euclidean space
$\mathbb R^d$ and its boundary  $\partial \mathcal D$.  The potential $u$
is given on the boundary by means of a given continuous function $f$:
\begin{eqnarray}
\label{dirichlet}
\Delta u &=& 0 \;\;  {\rm inside} \;\;  \mathcal D,\\
u&=&f \;\; {\rm on} \;\; \partial \mathcal D.
\label{bord}
\end{eqnarray}

For instance, the Dirichlet problem in the case of heat conduction
is to determine the equilibrium temperature inside a
conducting body $\mathcal D$, once the distribution $f$ of the temperature
 along the boundary $\partial \mathcal D$ is given.

It is here that Brownian motion comes into play, to provide an
entirely probabilistic representation of the solution.

\subsection{Relation between potential theory and Brownian motion}
\subsubsection{Newtonian potential and probability density}
The first relation, which contains the kernel of all the others,
is obtained simply by considering  the Gaussian probability density 
(\ref{gauss}),\footnote{For this subject one can consult the book of
K. L. Chung, {\it Green, Brown, and Probability \& Brownian Motion on
the Line}, World Scientific, Singapore (2002).} which represents
 the probability density of finding a Brownian particle at
a point $\vec r$ at time $t$, knowing that the particle was at the
origin at time $t=0$.  In $d$ dimensions, formula (\ref{gauss})
generalizes to
\begin{equation}
\label{gaussd}
P(\vec r;t)=\frac{1}{(4\pi D t)^{d/2}}
\exp{\left(-\frac{r^2}{4Dt}\right)},
\end{equation}
where $r$ is the distance from the origin.

By integrating $P(\vec r ; t)$ over the time variable $t$ one obtains 
\begin{equation}
\label{gaussdint}
D\int_{0}^{+\infty} P(\vec r ;t)\, {\rm d}t=\frac{1}{(d-2)S_d}\frac{1}{r^{d-2}}=u_d(r).
\end{equation}
For a unit diffusion coefficient $D=1$, the total
Brownian probability density of arriving at $\vec r$ at any time is
then exactly equal to the Newtonian potential created at $\vec r$ by a
unit charge or mass.

Let us look now at the Dirichlet problem from a more general point of view.

\subsubsection{Discrete random walks and the Dirichlet problem}

This problem was considered in the 1920's with the work of Phillips and
Wiener,\footnote{H. B. Phillips and N. Wiener, {\it J. Math. Phys.},
{\bf 2}, pp. 105-124 (1923).} and of Courant, Friedrichs and
Lewy.\footnote{R. Courant, K. Friedrichs and K. Lewy, {\it Math. Ann.},
{\bf 100}, pp. 37-74 (1928).}  They obtained a probabilistic
representation of the solution of the Dirichlet problem
(\ref{dirichlet}, \ref{bord}), in the form of an approximate sequence
of random walks on a $d$-dimensional cubic lattice $\varepsilon
\mathbb Z^d$, of lattice spacing $\varepsilon$.

More precisely, one considers random walkers $w=\{w_n, n\in
\mathbb N\}$ on the lattice $\varepsilon \mathbb Z^d$, at discrete
times $n=0,1,2,\cdots$, all starting from the initial point $w_0=P$
 in  domain  $\mathcal D$
and diffusing away.  When the walkers ultimately reach the
boundary, one measures the value of the function $f$ at that point on
the boundary. One repeats the process and then takes the {\it average}
of the values of the function $f$ over all first contact points on
the boundary reached  by random walkers that started from $P$.

We can formally write the averaging operation as
\begin{eqnarray}
\label{dirichletsol}
u_{\varepsilon}(P) =\sum_{{\{w:\; w_0=P\}}} \frac{1}{(2d)^{\tau_{\mathcal D}}}f(w_{\tau_{\mathcal D}}),
\end{eqnarray}
where the sum is over all random walks $w=\{w_n, n\in \mathbb N\}$
on the lattice $\varepsilon \mathbb Z^d$, at discrete times
$n=0,1,2,\cdots$, leaving the initial point $w_0=P$ and diffusing
towards the boundary. In (\ref{dirichletsol}), $\tau_{\mathcal D}$ is
the first instant at which the boundary $\partial \mathcal D$ is reached
by the random walker, and $w_{\tau_{\mathcal D}}$ its position on
the boundary at this instant.  The sum must be normalized by the inverse of the total number
 $(2d)^{\tau_{\mathcal D}}$ of walks of length $\tau_{\mathcal D}$, in a way to
be a probability measure on the set of discrete random walkers.

To extend the result in the continuum, one next takes the limit of the
lattice spacing $\varepsilon$ to $0$.  The result $\lim_{\varepsilon
\to 0} u_{\varepsilon}(P)=u(P)$ is then the value of $u$ at point $P$,
which is the solution of the Dirichlet problem in $\mathbb R^d$.

In the language of heat theory for instance, the temperature at point $P$ is the
average of the temperature at the boundary, evaluated after {\it random
walking} towards it! In mathematics, a standard notation of the average (\ref{dirichletsol})
is
\begin{eqnarray}
\label{dirichletdisc}
u_{\varepsilon}(P) =\int f(w_{\tau_{\mathcal D}}) \Pi_P^{\varepsilon}{(dw)},
\end{eqnarray}
where $\Pi_P^{\varepsilon}$ is the probability measure on discrete
random walks in $\varepsilon \mathbb Z^d$ started at $P$.

\subsubsection{Norbert Wiener}
A first attempt to define  integral calculus over a function space
 was made by Daniell (circa 1920).\footnote{ P.J. Daniell, {\it Ann. Math.} {\bf  21}, 203 (1920).}
 A few years later,
 Norbert Wiener introduced a measure in function space which is rigorous
 from a mathematical point of view  (it is a
 {\it bona fide} Borel measure), and  which  made it possible
 to define and calculate an integral over a space of functions.

Wiener had indeed known
Einstein's theory since his visit to Cambridge in 1913.  At 19, he
came to study logic with Bertrand Russell, who suggested that he go
listen to Hardy, the mathematician, and read Einstein!

So, motivated also by his reading of Perrin, Wiener constructed, in his fundamental article of 1923, ``Differential
Space,''\footnote{N. Wiener, {\it J. Math. Phys.}, {\bf 2}, pp. 131-174
(1923).} a probability
measure for Brownian paths in $\mathbb R$ (then in $\mathbb R^d$). The basic
idea was to directly construct on the space of continuous functions $w(t)$ of
a single real variable (representing the position as a function of
time),  a probability measure such that the changes of the positions $w(t_i)=x_i$, $i=0,\cdots,n$, over
disjoint time intervals, $[t_{i-1},t_{i}]$, $i=1,\cdots,n$,  have a joint Gaussian probability distribution,
\begin{equation}
\label{gaussgene}
P(\{x_i\};\{t_i\})=\prod_{i=1}^{n}\frac{1}{[4\pi D (t_{i}-t_{i-1})]^{1/2}}
\exp{\left[-\frac{(x_{i}-x_{i-1})^2}{4D(t_{i}-t_{i-1})}\right]}.
\end{equation}
This is a direct generalization of the Brownian displacement distribution (\ref{gauss1}).

 Wiener obtained  his measure  by using an
  explicit mapping of the space  ${\mathcal C}$  of
 continuous functions into the interval (0,1) (minus a set of
 Lebesgue measure zero).  This mapping allows to pull-back the ordinary
  Lebesgue measure on the space  ${\mathcal C}$.  In   this
 language, the  Brownian  motion  has the following
 probabilistic interpretation: a Brownian path
  corresponds to  the random choice
 of an element  of  the measured set  ${\mathcal C}$
  ({i.e.}, a continuous function), endowed with the ``Wiener measure''.

Nowadays, this measure is indeed  universally called {\it Wiener measure} in mathematical circles, while physicists
 prefer to speak of functional integrals, even though, like
{\it Monsieur Jourdain}, they really calculate with the Wiener measure when they perform their formal
calculations!\footnote{This is true in perturbation theory. See, e.g.,
in the case of polymer theory, B. Duplantier,
{\it Renormalization and Conformal Invariance for Polymers}, in
{\it Proceedings of the Seventh International Summer School on Fundamental Problems in Statistical
Mechanics}, Altenberg, Germany, June 18-30, 1989, H. van Beijeren Editor, North-Holland, Amsterdam (1990).}

The integral over such a measure is called the {\it Wiener average}.
It is denoted ${\mathcal W}({dw})$ here and more
precisely ${\mathcal W}_P({dw})$ for a Brownian motion $w$ started at
$P$.  It corresponds to the continuous limit for $\varepsilon\to 0$ of the
measure $\Pi^{\varepsilon}(dw)$ on random walks on the
discrete lattice $\varepsilon\mathbb Z^d$, introduced in the preceding section.

Once that construction was made, Wiener verified that the measure of the subset
of differentiable functions vanishes, in agreement with Perrin's intuition, and
that the support of the measure is given by H\"older functions (of order at
least $1/2-\epsilon, \epsilon >0$). In subsequent years, he further developed
the very broad ramifications of his theory.\footnote{N. Wiener, {\it Acta. Math.} {\bf 55},  117 (1930); 
R. E. A. C. Paley and N. Wiener, {\it Fourier
Transforms in the Complex Domain}, Amer. Math. Soc. Colloq. Publ.,
{\bf 19}, New-York (1934);
N. Wiener,  {\it Generalized Harmonic Analysis and Tauberian Theorems},
 MIT Press, Cambridge, Mass. (1964).}

In a study written in 1964 on Wiener and functional integration, Mark
Kac highlighted the profound originality of Wiener during his time,
and in counterpoint, the difficulty for mathematicians to
understand his approach:\footnote{M. Kac, {\it Bull. Amer. Math. Soc.}, {\bf
72}, pp. 52-68 (1964).}

{\it ``Only Paul L\'evy, in France, who had
himself been thinking along similar lines, fully appreciated their
significance.''}

The next steps were indeed made by Paul L\'evy, in his great work on Brownian motion,
{\it Processus stochastiques et
mouvement brownien} (1948).\footnote{Paul L\'evy, {\it Processus
stochastiques et mouvement brownien}, Gauthier-Villars, Paris (1965).}
Since then, the blooming of the subject in mathematics was such that one can only
make an extremely limited citation list. We refer the interested reader to
the introductory article of J.-F. Le Gall' for a first
journey into
the Brownian world of mathematics,\footnote{J.-F. Le Gall, {\it
Introduction au mouvement brownien}, Journ\'ees annuelles de la
Soci\'et\'e Math\'ematique de France, 28 janvier 1989, three expos\'es
on Brownian motion (J.-F. Le Gall: {\it supra}, G. Ben Arous: {\it
Grandes d\'eviations et noyau de la chaleur}, B. Duplantier: {\it Le
mouvement brownien en physique, les polym\`eres et leur relation avec
les ph\'enom\`enes critiques}).} and to D. Revuz and M. Yor's book for
a more thorough visit.\footnote{D. Revuz and M. Yor, {\it Continuous
Martingales and Brownian Motion}, Berlin-Heidelberg: Springer
(1991); second edition, 1994.}\medskip

The connection
 between  the Wiener path measure  for Brownian motion and path integrals is perhaps best intuitively understood by
 considering the multiple distribution (\ref{gaussgene}) for a set of successive equal time intervals,
 $t_{i}=\frac{i}{n}t, i\in \{1,n\}$. One conditions the path, normalized to start at the origin $x=0$ at time $t=0$,
  to be at times $t_i=\frac{i}{n}t, i\in \{1,n\}$ within intervals ${\rm d}x_i$ of the set of points
 $x_{i}$ in $\mathbb R$, and one then takes  the {\it formal} limit
 $n\to\infty$:
\begin{eqnarray}
\label{wienerexpl}
\nonumber
{\mathcal W}({dw})&=&\lim_{n\to\infty}\prod_{i=1}^{n}{\rm d}x_i P(\{x_i\};\{t_i\}))\\
\nonumber
&=&\lim_{n\to\infty}\prod_{i=1}^{n}{\rm d}x_i\prod_{i=1}^{n}\frac{1}{(4\pi D \, {t}/{n})^{1/2}}
\exp{\left[-\frac{(x_{i}-x_{i-1})^2}{4D\,{t}/{n}}\right]}\\
&=&{\mathcal D}w \exp\left(-\frac{1}{4D}\int_0^t \left(\frac{{\rm d}w(t')}{{\rm d}t'}\right)^2 {\rm d}t' \right),
\end{eqnarray}
now with a continuum ``Lebesgue'' measure on paths,
\begin{eqnarray}
\label{intexpl}
\nonumber
{\mathcal D}w=\lim_{n\to\infty}\prod_{i=1}^{n}\frac{{\rm d}x_i}{(4\pi D \,{t}/{n})^{1/2}}.
\end{eqnarray}
This notation is marvellously appealing to physicists, since one recognizes in the exponential in (\ref{wienerexpl}) the
Boltzmann-Gibbs weight associated with the classical kinetic energy of the particle.
As Marc Kac noted,\footnote{M. Kac, {\it Probability and related Topics in the Physical Sciences},
  Interscience, New York (1959).}

  {\it ``The disadvantages of such an approach from the purely
  mathematical point of view are obvious, although it is appealing on formal grounds''.}

In $d$ dimensions, the formal equivalence between Wiener's measure and functional integrals
is simply obtained by using the $d$-dimensional Gauss distribution, so that
\begin{eqnarray}
\label{wienerexpld}
{\mathcal W}({dw})
&=&{\mathcal D}{w} \exp\left(-\frac{1}{4D}\int_0^t \left(\frac{{\rm d}{\vec w}(t')}{{\rm d}t'}\right)^2
{\rm d}t' \right),\\
\nonumber
{\mathcal D}{w}&=&\lim_{n\to\infty}\prod_{i=1}^{n}\frac{{\rm d}^d x_i}{(4\pi D \,{t}/{n})^{d/2}}.
\end{eqnarray}

The rigorous connection
 between  the Wiener path integral  and Brownian motion
 is further illuminated  by the Feyman-Kac formula that allows one to  write explicit
  path integral  representations
 for the  solutions of parabolic differential equations, corresponding to Brownian motion
 in presence of a general potential,\footnote{M. Kac, {\it Probability and related Topics in the Physical Sciences},
  {\it op. cit.}; L. S. Schulman,  {\it Techniques and Applications of Path Integration},
  John Wiley and Sons, New York (1981); F. W. Wiegel, {\it Introduction to Path Integral Methods in Physics
 and Polymer Science},  World Scientific, Singapore (1986); J. Zinn-Justin, {\it Quantum Field
Theory and Critical Phenomena}, 4th Edition, {International Series of Monographs on Physics} {\bf 92}, Oxford University Press
(2002).} the case pioneered by Smoluchowski.

 When
 formally continued to imaginary time, the  Feyman-Kac formula
 provides an expression for  the Green function of the
 Schr\"odinger equation, thus leading to the celebrated path integral
 representation of Quantum Mechanics invented by Feynman in 1948.\footnote{R. P.  Feynman,
 {\it Rev. Mod. Phys.} {\bf 20}, 367  (1948); R. P.  Feynman and A. R. Hibbs,
 {\it Quantum Mechanics and Path Integrals}, McGraw-Hill, New York (1965).}

\subsubsection{S. Kakutani}
\begin{figure}[htbp]
\begin{center}
\includegraphics[angle=0,width=.45\linewidth]{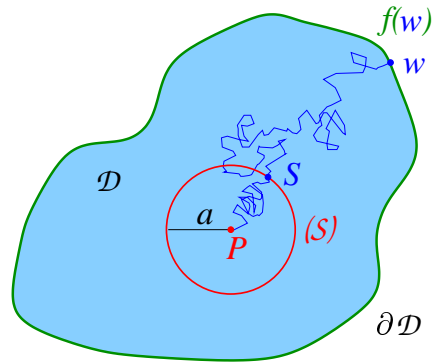}
\end{center}
\caption{{\it 
The Dirichlet problem in a domain $\mathcal D$, and its Brownian
representation.  The point $w=w_{\tau_{\mathcal D}}$ is the point of
first contact of a Brownian motion that started at $P$ with the boundary
$\partial \mathcal D$, at the instant ${\tau_{\mathcal D}}$
of first exit from the domain $\mathcal D$.  The point $S$ is the
point of first passage across the surface of the sphere $\mathcal S$.}}
\label{fig.domain}
\end{figure}

The existence of the Wiener measure and Wiener integral allowed for some
very important progress by
S. Kakutani in 1944-1945.\footnote{S. Kakutani, {\it Proc. Imp. Acad. Japan}, {\bf
20}, pp. 706-714 (1944).}  He showed that by substituting an integral
with the Wiener measure ${\mathcal W}$ in the formula
(\ref{dirichletdisc}) with the discrete measure
$\Pi^{\varepsilon}$ indeed solved the Dirichlet problem in continuous space $\mathbb R^d$.
Thus we have Kakutani's formula
\begin{eqnarray}
\label{dirichletprob}
u(P) =\int f(w_{\tau_{\mathcal D}}) {\mathcal W}_P^{}{(dw)}.
\end{eqnarray}
That means that {\it the potential at any point $P$ is given by the average of
the potential chosen at random on the boundary by a Brownian motion
started at $P$} (figure \ref{fig.domain}).

In the following section we give an elementary demonstration of this
result.

\subsubsection{Demonstration}

In probability theory, the quantity $u(P)$ defined by equation
(\ref{dirichletprob}) is called the {\it expectation value} associated to the
point $P$, because it represents the expectation for a random sampling of the value $f$ 
on the boundary, by a process of Brownian diffusion from $P$.

We want to verify that this expectation value fulfills the two conditions
(\ref{dirichlet}) and (\ref{bord}).

The second condition is easy to verify: if the point $P$ is on the
boundary $\partial \mathcal D$, any Brownian motion $w$ coming from
$P$ is immediately stopped on the boundary at  $w_{\tau_{\mathcal
D}}=P$, therefore $u(P)=f(P)$ for $P$ on $\partial \mathcal D$, as
expected.

Moreover, if the Brownian motion leaves from an internal point $P$,
close to a point $P_0$ of the boundary, it is (almost) certain (in 
a probabilistic sense) that the motion will meet the boundary in
a neighborhood of $P_0$, and that the expectation value $u(P)$ will
be close to the value $f(P_0)$ of $f$ in $P_0$. Kakutani's
solution has the right properties of regularity near the boundary, 
under the condition that the latter has a sufficiently
regular geometry and that the ``temperature'' $f$ on the boundary is a
continuous function.

The continuity of the expectation value $u$, with respect to point $P$,
is equally clear: a small displacement of $P$ will only slightly modify
the Brownian trajectories diffusing from $P$, as well as their
subsequent exploration of the boundary.

We will now establish the first property, (\ref{dirichlet}), i.e., that the
expectation value $u(P)$ (\ref{dirichletprob}) is a harmonic function,
by showing that it satisfies the equivalent  property
(\ref{moyenne}) on all spheres centered in $P$.

We draw a sphere $\mathcal S$ of radius $a$ centered at $P$  and
contained inside the domain $\mathcal D$ (figure
\ref{fig.domain}).  The aim is to show that the Brownian expectation value
$u(P)$ obtained by leaving from any point $P$ is equal to the average of
Brownian expectation values $u(S)$ obtained from any point $S$ on the surface
of the sphere $\mathcal S$.

In order to move beyond the boundary $\partial \mathcal D$ of
the domain, a Brownian motion must cross the sphere $\mathcal S$ at least once.
Calling $S$ the first crossing point of the sphere (figure
\ref{fig.domain}), and $u(P/S)$ the expectation value obtained for all Brownian
motions coming from $P$ and first crossing
$\mathcal S$ at the point $S$.

As there is no preferential direction for Brownian motion, each
point of $\mathcal S$ can be met first with equal probability.
  One distinguishes the average for
Brownian motions starting at $P$ in two steps: the choice of the point of
first passage $S$, and diffusion across $S$, with the expectation value
$u(P/S)$.  By averaging the averages, one has the result that $u(P)$ must
be equal to the average of $u(P/S)$ on the sphere, i.e., in
mathematical terms:
\begin{eqnarray}
\label{moyenne1}
u(P) =\int_{\mathcal S} u(P/S)\frac{d^2S}{4\pi a^2}.
\end{eqnarray}

The last thing to show is that the expectation value $u(P/S)$, obtained by
leaving from $P$ and passing through $S$, is the same as the expectation value 
$u(S)$, obtained by simply starting from $S$ on the sphere.  It is
here that a very important property of Brownian motion comes into
play: the motion at an instant $t$ only depends on the position at
that instant and not on previous motions.  Somehow, there is an absolute
loss of memory, where only the present instant and position are
important: Brownian motion is {\it Markovian}.  In
probability theory, one speaks generally as well of a {\it Markov
process} when the future dynamic of a process is not influenced by its
previous states.  The future behavior of a Brownian particle leaving from
$S$, or passing through $S$ knowing that it began at $P$, does not differ.
It follows that $u(P/S)=u(S)$, which ends the proof of the Theorem of
the Mean (\ref{moyenne}).

\subsection{Recurrence properties of Brownian motion}

We give an illustration of a non-trivial probabilistic property of
Brownian motion, which is deduced from potential theory, that is
its {\it recurrence properties}.

\subsubsection{Brownian motion in one dimension}

Let us consider now the one-dimensional real line $\mathbb R$ and
 points $x$ of a domain $\mathcal D$, here the line segment
$\mathcal D=[0,R]$, where $R$ is a positive number.  Let us search for
the harmonic function $u(x)$ that satisfies the simple Dirichlet 
problem: $u(0)=0,\;u(R)=1$. In one dimension, the Laplacian
(\ref{laplacien}) is simply the second derivative, so the
harmonic equation (\ref{dirichlet}) becomes ${\rm d}^2 u(x)/{\rm
d}x^2=0$.  The solution is simply linear in $x$: $u(x)=x/R$; it evidently
satisfies the required conditions at the boundaries.

Let us consider Kakutani's solution for the Dirichlet problem by
Brownian mathematical expectation.  The boundary
$\partial \mathcal D$ of the segment $\mathcal D=[0,R]$ is
made up of two points: $\partial
\mathcal D=\{0,R\}$.  The function $f$ with Dirichlet conditions
(\ref{bord}), takes the values on the boundary: $f(0)=0,\;f(R)=1$.
According to Kakutani's result, the value $u(x)=x/R$ of the harmonic
function $u$ is the average of the function $f$ obtained from random sampling
by means of a Brownian motion starting at $x$.  The case of the first
Brownian exit from the segment $\mathcal D=[0,R]$ at point $x=0$
gives a value $f=0$, and at point $x=R$ the value $f=1$.  The
Brownian expectation of $f$ is thus exactly the probability
for the Brownian motion to first exit from the segment $[0,R]$ at the endpoint
$R$ rather than at $0$, or else the probability, starting from $x$, to attain $R$ before
$0$. The complementary probability to attain $0$
before $R$ is thus $p_R(x)=1-u(x)=1-x/R.$

Let us now keep the point $x$ fixed while taking the limit $R\to \infty$,
so that the segment $\mathcal D$ extends to the positive real axis
$\mathbb R^{+}$. We see that $p_{R\to \infty}(x)\to 1$, and this is
for all $x$. The probability $p_{\infty}(x)$,  for a
Brownian motion started at $x$,  to reach the origin $0$ before
leaving to infinity is therefore identically equal to {\it one}.

The Brownian motion, wherever it leaves from, passes by the origin
(quasi-) certainly\footnote{In {\it continuous} probability theory, an event
with probability $1$ is only said to be ``quasi-certain'' or ``almost
surely true,'' contrary to the common language.  The reason is that in the case of
events forming a continuum, it
can always exist a non-empty set of irreducible events where the
prediction is not realized,
which is still of zero measure in the sense of measure theory,
and therefore of zero probability.  One cannot forgo the consideration of zero-measure sets,
hence go beyond the {\it ``almost surely''} {(\it a.s.)} probabilistic description.}.  Since spatial and temporal origins
were arbitrary in our demonstration, the following property was
established: {\it a Brownian motion in one dimension passes through
all points on the real axis, infinitely many times}.  One says it is {\it
recurrent} in one dimension.

This property did not appear as evident a priori from  the
probability theory side.  Thanks to the relation to potential theory,
 it has been obtained by simply solving a
second order differential equation!  Einstein would surely not have thought
of this in 1905, although who knows?\smallskip

Now we will generalize the above study to two and  then to $d$ dimensions.

\subsubsection{The two-dimensional case}

This time, we consider the planar annular domain
$\mathcal D$, which is that between two concentric
circles $\mathcal C_1$ and $\mathcal C_2$ centered at the origin
$O$, of respective radii $\rho_1$ and $\rho_2$, with $\rho_1 <
\rho_2$.  The boundary of the domain $\mathcal D$ is then made of 
two circles, $\partial \mathcal D=\mathcal C_1 \cup \mathcal C_2$.  We
pose the Dirichlet problem in the annular domain $\mathcal D$:
\begin{eqnarray}
\label{dirichlet2}
\Delta u &=& 0 \;\;  {\rm inside} \;\;  \mathcal D,\\
\label{bord2}
u&=&0 \;\;{\rm on}\;\; \mathcal C_1,\;\;  u=1 \;\;{\rm on}\;\; \mathcal C_2.
\end{eqnarray}
By using the two-dimensional Newtonian potential $u_2(r)$
(\ref{potentiel2}), it is easy to see that the solution to the Dirichlet 
problem is spherically symmetric and at a distance $r$ from the center
evaluates to:
\begin{eqnarray}
u_{2}(r; \rho_1, \rho_2)&=&\frac{u_2(r)-u_2(\rho_1)}{u_2(\rho_2)-u_2(\rho_1)}
\label{sol2}
=\frac{\log{r}-\log{\rho_1}}{\log{\rho_2}-\log{\rho_1}},\;\;  \rho_1 \le r \le \rho_2.
\end{eqnarray}
Actually, this function obviously satisfies (\ref{bord2}) and is
harmonic in the annular domain $\mathcal D$, because the potential
$u_2(r)$ (\ref{potentiel2}) is harmonic too (except at the origin, which indeed does
not belong to $\mathcal D$).\smallskip

Let us come now to Kakutani's representation of the solution to
the Dirichlet problem.  In a manner similar to the one-dimensional case
in the preceding paragraph, $u_{2}(r; \rho_1, \rho_2)$ (\ref{sol2})
represents the probability that a Brownian motion, started at a
distance $r$ from the center, hits the outer circle $\mathcal C_2$
before hitting the inner circle $\mathcal C_1$.

As in the preceding paragraph, let us fix the distance $r$ and the
internal circle $\mathcal C_1$, and push the boundary of the outer
circle $\mathcal C_2$ to infinity.  By taking $\rho_2 \to \infty$ in
 formula (\ref{sol2}), we see that by continuity the probability
that the Brownian motion goes to infinity is $u_{2}(r; \rho_1,
\infty)= 0$, for all $r$ and $\rho_1$ finite. It means that the 
Brownian motion reaches the disk of radius $\rho_1$ with probability
$1$, whatever its point of departure outside of the disk. Since the
initial departure time is arbitrary too, likewise the origin in the
plane, one then concludes that {\it a two-dimensional Brownian motion
passes through neighboring points of any point infinitely often}.
It is then {\it recurrent in two dimensions}, just as it is in one
dimension.

It is equally interesting to fix $r$ and $\rho_2$ in (\ref{sol2}), and
to take the limit of an infinitesimal circle around the origin, i.e.,
$\rho_1 \to 0$.  We then find that by continuity $u_{2}(r;
\rho_1= 0, \rho_2) = 1$.  The probability that a Brownian motion
starting at a distance $r \neq 0$ from the origin, moves away from
the origin up to a distance $\rho_2 > r$ without having
visited the origin at $\rho_1=0$, is then equal to $1$.
In other words, a Brownian motion that does not  leave from the origin
$O$ avoids the origin with probability $1$, without ever
being able to pass through it.

We deduce an apparently paradoxical result: {\it in two dimensions,
any Brownian motion passes through a given point with zero
probability, but it passes through immediate neighboring points
infinitely often with probability $1$!}

Such a double result is due to the presence in the expectation
(\ref{sol2}) of one function, the logarithm, that diverges both at
short distance for $\rho_1 \to 0$, and at long distance for $\rho_2
\to \infty$.  This is peculiar to two dimensions and announces
exceptional properties known as {\it conformal invariance} in two
dimensions, which will be described in the following section.

In $d>2$ dimensions, a simple power law controls the Newtonian
potential $u_d(r)$ (\ref{potentield}), and only a divergence at short
distance appears.  We will see the consequences of such a divergence
on the recurrence properties of Brownian motion.

Let us mention however that these properties only constitute the ``tip
of the iceberg'': the singular character of the potential at short
distances is the source of divergences in quantum field
theories, which led to the creation of  {\it renormalization
theory}, whose consequences have been quite fruitful in the
physics of elementary particles and in statistical
mechanics.\footnote{For this subject, one can consult the text {\it
Renormalization} from S\'eminaire Poincar\'e 2002, {\it in}
B.~Duplantier \& V.~Rivasseau, {eds.}, {\it Poincar\'e Seminar 2002},
Progress in Mathematical Physics, Vol. 30, Birkh\"auser, Basel
(2003); see also the monograph by J. Zinn-Justin, {\it Quantum Field
Theory and Critical Phenomena}, 4th Edition, {International Series of Monographs on Physics} {\bf 92}, Oxford University Press
(2002).} Actually, the intersection of Brownian motions\footnote{G. F.
Lawler, {\it Intersection of Random Walks} (Birkh\"{a}user, Boston,
1991).}  provides the random geometric mechanism that underlies any
interacting field theory.\footnote{ K. Symanzyk, in{\it \ Local
Quantum Theory}, edited by R. Jost (Academic Press, London, New-York
(1969)).}  This equivalence is fundamental in the theory of
polymers\footnote{P.-G. de Gennes, {\it Phys. Lett.} {\bf A38},
339-340 (1972); J. des Cloizeaux, {\it J. de Physique} {\bf 36},
281-291 (1975).} and also in the rigorous theory of second order phase
transitions.\footnote{M. Aizenman, {\it Phys. Rev. Lett.} {\bf 47},
1-4, 886 (1981); {\it Commun.  Math. Phys.} {\bf 86}, 1-48 (1982);
D. C. Brydges, J. Fr\"{o}hlich, and T. Spencer, {\it Commun.
Math. Phys.} {\bf 83}, 123-150 (1982); G. F.  Lawler, {\it Commun.
Math. Phys.} {\bf 86}, 539-554 (1982).}  But {\it ``Revenons \`a nos moutons.''}\footnote{From
{\it La farce de Maistre Pierre Pathelin} (c. 1460), meaning  ``Let's get back to our main subject''.}

\subsubsection{The $d$-dimensional case}

We are now well enough equipped to pass to the $d$-dimensional case,
for $d > 2$.  Let us consider two concentric hyperspheres, $\mathcal
S_1$ and $\mathcal S_2$, centered at origin $O$, and of respective
radii $\rho_1$ and $\rho_2$, with $\rho_1 < \rho_2$.  The boundary
of the domain $\mathcal D$ is then made of the two spheres
$\partial \mathcal D=\mathcal S_1 \cup \mathcal S_2$.  Let us state 
the Dirichlet problem (\ref{bord})
\begin{eqnarray}
\label{dirichletd}
\Delta u &=& 0 \;\;  {\rm inside} \;\;  \mathcal D,\\
\label{bordd}
u&=&0 \;\;{\rm on}\;\; \mathcal S_1,\;\;  u=1 \;\;{\rm on}\;\; \mathcal S_2.
\end{eqnarray}
Here again, in using this time the $d$-dimensional Newtonian potential
$u_d(r)$ (\ref{potentield}), it is easy to see that the spherically
symmetric solution of the Dirichlet problem at a distance $r$ from the
center, is:
\begin{eqnarray}
\label{sold}
u_{d}(r; \rho_1, \rho_2)&=&\frac{u_d(r)-u_d(\rho_1)}{u_d(\rho_2)-u_d(\rho_1)}
=\frac{{r^{2-d}}-{\rho_1^{2-d}}}{{\rho_2^{2-d}}-{\rho_1^{2-d}}},\;\;  \rho_1 \le r \le \rho_2.
\end{eqnarray}
This function satisfies (\ref{bordd}); it is harmonic in the annular
$d$-dimensional domain $\mathcal D$, because the potential $u_d(r)$
(\ref{potentield}) is harmonic too (except at the origin, which does not belong
to $\mathcal D$).\smallskip

Finally let us apply the probabilistic result: $u_{d}(r; \rho_1,
\rho_2)$ (\ref{sold}) is the probability that a Brownian motion,
starting from a given point at a distance $r$ from the center, meets
the outer sphere before the internal sphere.

First, let us take in (\ref{sold}) the limit $\rho_2 \to
\infty$, at $r$ and $\rho_1$ fixed.  As the dimension $d$ is here greater than
 $2$, one has $\rho_2^{2-d}\to 0$.  The probability for
the Brownian motion to escape to infinity, $u_{d}(r; \rho_1, \rho_2 \to
\infty)$, is by continuity the limit $u_{d}(r; \rho_1,
\infty)=1-(\rho_1/r)^{d-2}$, which is {\it finite}.

This result shows that {\it in all spaces of at least three
dimensions Brownian motion is not recurrent}, because the space is
larger than that in one or two dimensions.  We say that it is {\it
transient}.  Such a result, very important in probability theory,
was obtained in an elegant and simple manner via potential
theory.\smallskip

The complementary probability at a distance $\rho_1 \le r$, $p_{d}(r; \rho_1, \infty)=1-u_{d}(r;
\rho_1, \infty)$, that is, of visiting a neighborhood of the origin, is then equal to
$(\rho_1/r)^{d-2}$. In the usual physical case, $d=3$, one finds
$p_{3}(r; \rho_1,
\infty)={\rho_1}/{r}$, for $\rho_1 \le r$.

One can generalize the definition of  $p_{d}(r; \rho_1,
\infty)$ to the whole  space, by giving it the value $1$ inside 
the sphere of radius $\rho_1$, that is for $r \le \rho_1$. Such a
generalized function is called
{\it potential capacity} of the sphere of  radius $\rho_1$. The potential capacity of an ensemble $\mathcal
B$ is an important concept in classic potential theory; it is a
harmonic function outside $\mathcal B$, equal to $1$ inside $\mathcal
B$, and zero at infinity.  It is then the probability that a single
particle, animated by Brownian motion and leaving from a given point,
will reach $\mathcal B$.\smallskip

Research in this domain allowed the discovery of important
generalizations, both for the theory of Brownian motion and potential theory.
We have seen that the equivalence between them rests on the
Markovian property of Brownian motion.  Similarly, a generalized potential
theory can be associated to any
``standard'' Markov process.

We see therefore the profound relation that exists between the
mathematical theory of potential, invented in the 17th century by
Newton, then developed by Laplace, Poisson and Green, and Brownian
motion, observed during the same era, but understood only in the 20th
century, thanks to Sutherland, Einstein, Smoluchowski, Perrin and Langevin in
physics, Bachelier, Wiener, L\'evy, Kakutani and many others in mathematics.

\section{The fine geometry of the planar Brownian curve}
\subsection{The Brownian boundary}
\begin{figure}[tb]
\begin{center}
\includegraphics[angle=0,width=.35\linewidth]{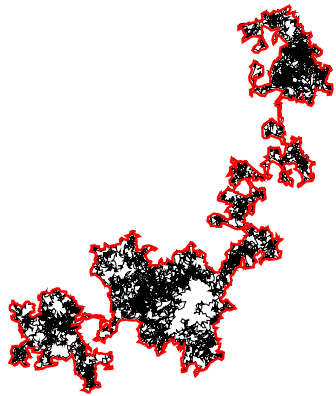}
\end{center}
\caption{{\it Boundary or outer envelope curve of a planar Brownian path.}}
\label{fig.brownianfrontier}
\end{figure}

In this last part, we are interested in the geometry of the Brownian
curve in the plane. By  Brownian curve, or Brownian path, we mean
the random curve traced by a Brownian motion on the plane. We can see
a typical representative in figure \ref{fig.brownian}. In particular, we
will consider the {\it boundary} of such a curve.  It is the outer envelope 
of the Brownian curve.  We observe that it is an extremely
irregular curve, {\it fractal} in Mandelbrot's sense (figure
\ref{fig.brownianfrontier}).\footnote{See the classic works of
Beno\^{\i}t Mandelbrot, {\it Les objets fractals~: forme, hasard et
dimension, survol du langage fractal}, Champs, Flammarion (1999), and
{\it The Fractal Geometry of Nature}, Freeman, New-York (1982).}

\begin{figure}[htb]
\begin{center}
\includegraphics[angle=-90,width=.75\linewidth]{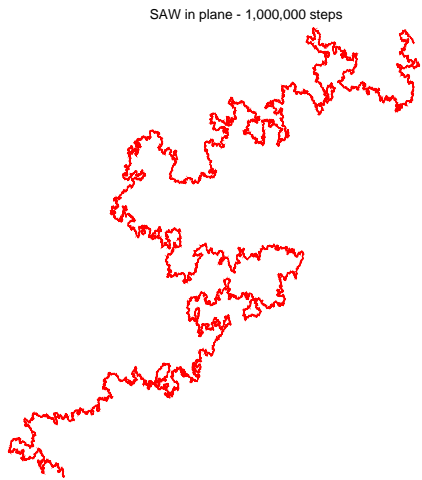}
\end{center}
\caption{{\it A self-avoiding walk in the plane, made of 1 million steps! (Kindly provided by
T. G. Kennedy, University of Arizona.)}}
\label{SAW}
\end{figure}

From a series of accurate numerical simulations, Mandelbrot made the
conjecture in 1982 that such a boundary is the continuous limit of a
particular random walk, the {\it self-avoiding walk} (SAW) (figure
\ref{SAW}). That is a process where the random walker cannot visit any
point of his own path twice.  To define it, one considers {\it a
priori} the ensemble of all possible random paths of a given length (with and without
self-intersections) on, say, a square lattice, and select among them
the small subset of all the paths that do not
self-intersect. Those are then weighted with a uniform measure.\footnote{See the monographs: P.-G. de Gennes,
{\it Scaling Concepts in Polymer Physics}, Cornell University Press
(1979); J. des Cloizeaux and G. Jannink, {\it Polymers in Solution,
their Modeling and Structure} (Clarendon, Oxford University Press,
1989).}

The resulting conjecture is that the fractal dimension or Hausdorff
dimension of the Brownian boundary is equal to $D_H=4/3$, like that
which was calculated by the Dutch theoretical physicist Bernard
Nienhuis in 1982 for a two-dimensional self-avoiding
random walk.\footnote{B. Nienhuis, {\it Phys. Rev. Lett.} {\bf 49},
pp. 1062-1065 (1982); {\it J. Stat. Phys.} {\bf 34}, pp. 731-761 (1984); {\it
Phase Transitions and Critical Phenomena}, edited by C.  Domb et
J. L. Lebowitz, (Academic Press, London, 1987), Vol. 11; see also  M. den Nijs, J. Phys. A {\bf 12}, pp. 1857-1868 (1979);
Phys. Rev. B {\bf 27}, pp. 1674-1679 (1983).}  The fractal
dimension $D_H$ is here defined in an non-rigorous way, as
follows.  We cover the fractal object of size $R$ by small disjoint
disks of radius $\varepsilon$, and we count the number $n$ of these
disks.  In general, this number grows with a power law in $R$ and
$\varepsilon$, $n \propto
\left(R/\varepsilon\right)^{D_H}$. We then see that $D_H$ generalizes
the notion of Euclidean dimension of regular sets to the case of very
irregular sets.

Nienhuis used a representation of statistical mechanics, known as
the Coulomb gas, a precursor to the methods of conformal invariance or
of conformal field theories that in 1984 would enter the theory of
two-dimensional critical phenomena, thanks to the
work of Belavin, Polyakov, and Zamolodchikov.\footnote{A. A.  Belavin,
A. M.  Polyakov and A. B.  Zamolodchikov, {\it Nucl. Phys.} {\bf
B241}, pp. 333-380 (1984). One can find an introduction in the book by
C. Itzykson and J.-M. Drouffe, {\it Th\'eorie statistique des champs},
tome 2, EDP Sciences/CNRS \'Editions (2000); English version: {\it Statistical Field Theory},
Vol. 2, Cambridge University Press, Cambridge (1989).  For further reading, see
J. L. Cardy, in {\it Phase Transitions and Critical Phenomena},
edited by C.  Domb and J. L. Lebowitz, (Academic Press, London, 1987),
Vol. 11; J. L. Cardy, {\it Conformal Invariance and Statistical
Mechanics}, in ``Fields, Strings, and Critical Phenomena,'' Les
Houches Summer School 1988, edited by \'E. Br\'ezin and
J. Zinn-Justin, North-Holland, Amsterdam (1990); Ph. Di Francesco,
P. Mathieu and D. S\'en\'echal, {\it Conformal Field Theory},
Springer-Verlag, New-York (1997).}
\begin{figure}[htb]
\begin{center}
\includegraphics[angle=0,width=.45\linewidth]{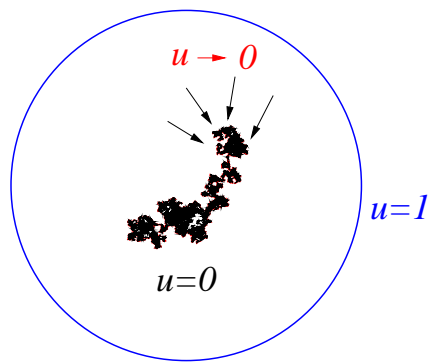}
\end{center}
\caption{{\it Dirichlet  
problem associated to a planar Brownian path. The latter serves as an
electrode where the potential vanishes.}}
\label{fig.potentialu}
\end{figure}

It was already envisioned in the
mid-eigh\-ties that the critical properties of planar Brownian paths, whose conformal invariance was
 well-established, could be the opening gate to rigorous studies
of two-dimensional critical phenomena, as advocated in particular by Michael Aizenman.\footnote{It is perhaps
interesting to note that P.-G. de Gennes originally studied polymer
theory with the same hope of understanding from that perspective the broader class of critical phenomena. It turned out to be historically the converse:
the Wilson-Fisher renormalization group approach to spin models with $O(N)$ symmetry yielded in 1972 the polymer critical exponents
as the special case of the $N \to 0$ limit.

Michael Aizenman, in a
seminar in the Probability Laboratory of the University of Paris VI
in 1984, insisted on the importance of the $\zeta_2$ exponent governing in two dimensions
the non-intersection
probability up to time $t$, $P_2(t) \approx t^{-\zeta_2}$, of two Brownian paths,
and promised a good bottle of Bordeaux wine for its resolution.
A famous Ch\^ateau-Margaux 1982  was finally savored, in company
of M. Aizenman, by G. Lawler, O. Schramm, W. Werner and the author in 2001.} A family of exponents $\zeta_n$,
governing the non-intersection
properties of $n$ Brownian paths, played a crucial role. They are
defined as follows: Let $B_1(t), B_2(t),\cdots$ be independent planar Brownian motions started
from distinct points at $t=0$. Then the probability $P_n(t)$ that their traces, $B_1([0,t]), \cdots, B_n([0,t])$, are disjoint is
scaling as $t^{-\zeta_n}$ as $t\to +\infty$ for constant exponents $\zeta_n$.

The precise values of these exponents
were originally conjectured in 1988 from conformal invariance and numerical studies.\footnote{
B. Duplantier and K.-H. Kwon, {\it Phys. Rev. Lett.} {\bf 61}, pp. 2514-2517 (1988).} They correspond to a conformal
field theory with ``central charge'' $c=0$.
The planar intersection exponents $\zeta_n$, for $n\geq 2$,
are given by\footnote{For $n=1$ the non-trivial value $\zeta_1=1/8$ gives the exponent associated with the
probability $P_1(t)$ that a Brownian path $B_1([0,t])$ does not disconnect its own tip at $t$, $B_1(t)$, from infinity.}
\begin{equation}
\label{interexp}
\zeta_n=\frac{4n^2-1}{24}\, .
\end{equation}
Interestingly
enough, however, their analytic derivation  resisted  attempts by standard ``Coulomb-gas'' techniques, which
proved successful for self-avoiding walks, but not for simple walks. A
 derivation of these Brownian intersection exponents $\zeta_n$ (and a heuristic demonstration of Mandelbrot's related
conjecture),
inspired by some probabilistic structure of conformal invariance obtained in the meantime by
Lawler and Werner,\footnote{G. F.  Lawler and W. Werner, {\it Ann.
Probab.} {\bf 27}, pp. 1601-1642 (1999).} was given by the author in 1998 in
the framework of theoretical physics, by means of the formalism of {\it ``2D quantum gravity''}
in conformal field theory.\footnote{B. Duplantier, {\it Phys. Rev. Lett.}
{\bf 81}, pp. 5489-5492 (1998); {\it ibid.} {\bf 82}, pp. 880-883 (1999),
arXiv:cond-mat/9812439.}

The same results for the Brownian intersection exponents and Mandelbrot's conjecture were at last rigorously proved in
the framework of
probability theory in 2000 by Greg Lawler, Oded Schramm and Wendelin
Werner,\footnote{G. F.  Lawler, O. Schramm, and W. Werner, {\it Acta
Math.}  {\bf 187}, (I) pp. 237-273, (II) pp. 275-308 (2001),
arXiv:math.PR/9911084, arXiv:math.PR/0003156; {\it Ann. Inst. Henri
Poincar\'e} PR {\bf 38}, pp. 109-123 (2002), arXiv:math.PR/0005294; {\it
Acta Math.}  {\bf 189}, pp. 179-201 (2002), arXiv:math.PR/0005295; {\it
Math. Res. Lett.} {\bf 8}, pp. 401-411 (2001), math.PR/0010165.}  by means
of a conformally invariant stochastic process invented by Schramm, the
{\it ``$SLE$''} ({\it ``Stochastic Loewner Evolution''}), which is itself based on
Brownian motion.\footnote{O. Schramm, {\it Israel Jour. Math.} {\bf
118}, pp. 221-288 (2000).  The ${\rm SLE}_{\kappa}$ process, and its path,
are generated by the Loewner equation, describing the
evolution of the Riemann's conformal map which maps the unit disc, slit by the random
path, onto itself. This map erases the path and maps its two sides
onto the boundary of the unit
disc, with the tip under the form of a Brownian motion characterized by a diffusion
coefficient $\kappa$. This is the case of the so-called {\it radial} SLE.
Another case is that of the {\it chordal} SLE, where the conformal map
acts on the slit complex half-plane. See the recent book by G. F. Lawler,
{\it Conformally Invariant Processes in the Plane}, Mathematical Surveys and Monographs, AMS, Vol. {\bf 114} (2005).}
Despite the simplicity of formula (\ref{interexp}), prior to SLE, its proof by conventional probabilistic techniques
had been out of reach.

 W. Werner has been awarded the Fields Medal on August 22nd, 2006, at the International
 Congress of Mathematicians in Madrid,{\it ``for his contributions to the development of stochastic Loewner evolution,
 the geometry of two-dimensional Brownian motion, and conformal field theory.''}\footnote{One can find the official press
 release, and the laudatio given by Chuck Newman at the {\it International
 Congress of Mathematicians} in Madrid,
 on August 22, 2006,  at the web addresses: www.mathunion.org/General/Prizes/2006/WernerENG.pdf, and
 http://icm2006.org/AbsDef/ts/Newman-WW.pdf/.}

We are not going to describe this work in detail here,\footnote{For
further details, see the article for the general public by Wendelin
Werner, {\it Les chemins al\'eatoires}, published in {\sc Pour La Science}
in August 2001.\\
For the SLE process, consult: the notes from W. Werner's courses,
{\it Random Planar Curves and Schramm-Loewner Evolutions}, Lectures
Notes from the 2002 Saint-Flour Summer School, Springer
L. N. Math. {\bf 1840}, pp. 107-195, (2004), arXiv:math.PR/0303354; the
 book by G. F. Lawler, {\it Conformally Invariant Processes
in the Plane}, {\it op. cit.}, as well as the article of W. Kager and
B. Nienhuis, {\it A Guide to Stochastic Loewner Evolution and its
Applications}, {\it J. Stat. Phys.} {\bf 115}, pp. 1149-1229 (2004),
arXiv:math-phys/0312056.
 
For the link of SLE with quantum gravity, see: B. Duplantier, {\it Conformal Fractal Geometry and Boundary Quantum Gravity}, in {\it
Fractal Geometry and Applications, A Jubilee of Beno\^{\i}t
Mandelbrot}, Proceedings of Symposia in Pure Mathematics, AMS,
Vol. {\bf 72}, Part {\bf 2},
edited
by M. L. Lapidus and F. van Frankenhuijsen, pp. 365-482 (2004);
arXiv:math-phys/0303034.

See also: B. Duplantier, {\it Conformal Random Geometry}, in Les Houches, Session LXXXIII, 2005,
{\it Mathematical Statistical Physics}, A. Bovier, F. Dunlop, F. den Hollander, A. van Enter \& J. Dalibard eds.,
Elsevier B. V. (2006), arXiv:math-phys/0608053.

For the link of SLE with conformal field theories, see in mathematics: R. Friedrich and W. Werner,
{\it C. R. Acad. Sci. Paris S\'er. I Math.}
{\bf 335}, pp. 947-952 (2002), arXiv:math.PR/0209382; {\it Commun.
Math. Phys.}, {\bf 243}, pp. 105-122 (2003), arXiv:math-ph/0301018;
W. Werner, {\it Conformal restriction and related questions},
Proceedings of the conference {\it Conformal Invariance and Random
Spatial Processes}, Edinburgh, July 2003, arXiv:math.PR/0307353;
W. Werner and G. F. Lawler, {\it Probab. Th. Rel. Fields} {\bf 128},
pp. 565-588 (2004), arXiv:math.PR/0304419; W. Werner, {\it
C. R. Ac. Sci. Paris S\'er. I Math.} {\bf 337}, pp. 481-486 (2003),
arXiv:math.PR/0308164; see also J. Dub\'edat,
arXiv:math.PR/0411299; {\it J. Stat. Phys.} {\bf 123}, pp. 1183-1218 (2006),
arXiv:math.PR/0507276; in physics: M. Bauer and D. Bernard, {\it Phys. Lett.} {\bf B543}, pp. 135-138 (2002),
arXiv:math-ph/0206028; {\it Commun. Math. Phys.} {\bf 239}, pp. 493-521
(2003), arXiv:hep-th/0210015; {\it Phys. Lett.} {\bf B557}, pp. 309-316
(2003), arXiv:hep-th/0301064; {\it Ann. Henri Poincar\'e} {\bf 5},
pp. 289-326 (2004), arXiv:math-ph/0305061; Proceedings of the conference
{\it Conformal Invariance and Random Spatial Processes}, Edinburgh,
July 2003, arXiv:math-ph/0401019; M. Bauer, D. Bernard and J. Houdayer, {\it J.  Stat. Mech. Theor. Exp.}   P03001 (2005),
arXiv:math-ph/0411038;
M. Bauer and D. Bernard, arXiv:cond-mat/0412372; M. Bauer, D. Bernard and K. Kyt\"ol\"a, {\it J. Stat. Phys.}
{\bf 120}, pp. 1125-1163
(2005), arXiv:math-ph/0503024; K. Kyt\"ol\"a, {\it Rev. Math. Phys.} {\bf 19}, pp. 1-55 (2007), arXiv:math-ph/0504057; M. Bauer and D. Bernard, {\it Phys. Rep.} {\bf 432},
 pp. 115-221
(2006), arXiv:math-ph/0602049;
Ilya A. Gruzberg, {\it J. Phys. A: Math. Gen.} {\bf 39}, pp. 12601-12655 (2006), arXiv:math-ph/0607046.} but we will look instead at the generalization of the
results on the geometry of Brownian motion, and at the {\it
multifractal} nature of its boundary. The latter actually reveals a
structure made of a continuum of fractal subsets that we will
describe.

In continuity with the previous part, we will focus on  the
{\it potential theory associated with the neighborhood of a planar Brownian
path}.  We will show how the fine geometry of the Brownian boundary
 appears as an essential component of the solution to the Dirichlet
associated electrostatic problem.
\subsection{Potential theory in the neighborhood of a Brownian curve}
\subsubsection{Brownian Dirichlet problem}
Let us then consider a planar Brownian path $\mathcal B$ enclosed by a large circle,
and the associated Dirichlet problem where the
potential $u$ has the value $u=0$ on the boundary $\partial
\mathcal B$ of the Brownian curve, and $u=1$ on the circle (figure
\ref{fig.potentialu}).  The Brownian path serves as an electrode
creating the potential, and by electrostatic induction, its boundary
will charge itself. This {\it a priori} appears as a rather complex problem,
since the Brownian curve is completely random!

Far from the Brownian curve, the potential will depend on the global
geometry of the system, and in particular on the presence of the 
outer
circle that acts as an external electrode.  Let us imagine for a
moment that this circle is pushed towards infinity.  Seen from intermediate regions located very far from
the Brownian curve (and from the outer
circle), the Brownian electrode would
then appear to be confined to a point.  Its potential will then coincide with that of a point charge equal
to the total charge carried by the boundary of the Brownian curve, i.e.,  the
logarithmic Newtonian potential $u_2(r)$ (\ref{potentiel2}).

On the other hand, close to the Brownian curve, the geometry of the
boundary is crucial.  The potential vanishes exactly on the boundary
$\partial \mathcal B$, and the natural question here is its analytic
behavior in the neighborhood of $\partial \mathcal B$, i.e., the way
in which it goes to $0$.  As the geometry of the boundary is
particularly wild, the way the potential vanishes is as well.

However, the random Brownian curve hides at its heart a fundamental structural regularity 
connected to its {\it conformal invariance},
and one can in fact describe the potential close to the Brownian
path in a way which is  probabilistic, but {\it universal}.

\subsubsection{Conformal invariance}

A {\it conformal} map $\Phi$ of the plane is a bijection of
the plane into itself that preserves {\it angles} between curve
intersections. To any analytic function
$\Phi(z)$ in the complex plane can be associated one such conformal
map.  Locally, i.e., infinitesimally close to the image $\Phi(z)$
of any point $z$ in complex coordinates, a conformal map is
the composition of a local dilation (by a factor of
$|\Phi^{\prime}(z)|$), and of a rotation around $\Phi(z)$ (by an angle
$\arg \Phi^{\prime}(z)$).  This is why angles are locally
conserved.

Let us come back for a moment to the Brownian representation of the
general Dirichlet problem in a domain $\mathcal D$ (figure
\ref{fig.domain}). An auxiliary Brownian motion issued from an
arbitrary point $P$, stops upon touching the boundary $\partial
\mathcal D$, and its  Wiener integral represents the potential
$u(P)$.  Let us imagine the domain $\mathcal D$ to be transformed by a
conformal map $\Phi$ into a domain ${\mathcal D}^{\prime}
=\Phi(\mathcal D)$, while the Brownian trajectory $\mathcal B$ is
transformed into a curve $\Phi(\mathcal B)$, which is thus stopped upon touching
 the boundary $\partial {\mathcal D}^{\prime}=\Phi(\partial
\mathcal D)$.  Paul L\'evy showed that $\Phi(\mathcal B)$ is still the
trajectory of a Brownian motion, after a time reparameterization: this is
the property of {\it conformal invariance of planar Brownian
motion.}\footnote{Paul L\'evy, {\it Processus stochastiques et mouvement
brownien}, Gauthier-Villars, Paris (1965).}

Let us then consider the new potential $u^{\prime}(P^{\prime})$ at a
transformed point $P^{\prime}=\Phi(P)$, i.e.,
the solution to the
Dirichlet problem in the transformed domain $\mathcal D^{\prime}$.
Since all geometric objects that represent the potential were
transformed by $\Phi$, and since the transformed auxiliary Brownian
path is still Brownian, the result is that its Wiener integral,
$u^{\prime}(P^{\prime})$, does not change.  The potential
$u^{\prime}(P^{\prime})$ is then equal to the potential $u(P)$, that
is the solution to the Dirichlet problem in the original domain
$\mathcal D$, and thus there is an {\it invariance of potential under a
conformal map}.\smallskip

In the case we are focusing on here, that is of the
Dirichlet problem of  a potential $u(P)$ in the neighborhood  of a planar Brownian curve
(figure \ref{fig.potentialu}), the Brownian representation of the 
potential introduces a second auxiliary Brownian motion that diffuses
from the point $P$, while avoiding the original Brownian curve (figure
\ref{fig.potential2u}).  As we just saw, the two Brownian paths are
statistically conformally invariant and this probabilistic geometric
problem is invariant under any conformal map in the plane.

\subsubsection{The role of angles}

Conformal maps preserve angles in the plane, and
this is why the latter will play an essential role in the description of
the potential close to the Brownian boundary.

Let us first consider the simple problem of a potential existing in an
angular sector of the plane.  More precisely, let us consider an open
angle $\theta$ centered at a point $w$ (figure \ref{fig.angle}).  One
easily shows, by using the singular conformal map  of the complex
plane that opens the angle $\theta$ into a flat angle,
$\Phi(z)=z^{\pi/\theta}$, that the potential $u(z)$ varies at
any point $z$ close to $w$ like
\begin{eqnarray}
\label{pa}
u(z) \approx r^{\pi/\theta},
\end{eqnarray}
where $r$ is the distance from $w$, $r=|z-w|$.
For a flat angle, $\theta=\pi$, and we again find a linear behavior as a 
function of the distance, corresponding to a constant electric field
close to a straight line.

\subsection{Multifractality}
\subsubsection{Distribution of potential}

\begin{figure}[htb]
\begin{center}
\includegraphics[angle=0,width=.45\linewidth]{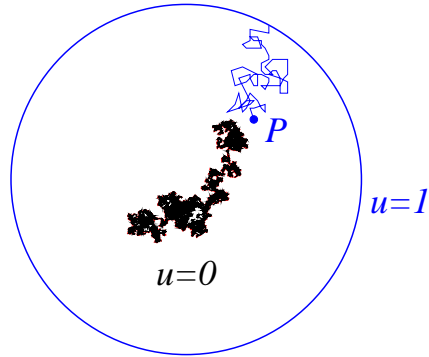}
\end{center}
\caption{{\it The Dirichlet potential $u$ created at point $P$ by a Brownian curve
(center), and vanishing on the boundary of the latter, is
represented by a second auxiliary Brownian motion, that
diffuses from $P$ towards the exterior, while completely avoiding the
first motion.}}
\label{fig.potential2u}
\end{figure}

\begin{figure}[htb]
\begin{center}
\includegraphics[angle=0,width=.4\linewidth]{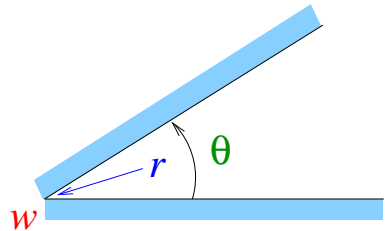}
\end{center}
\caption{{\it Angular sector with apex $w$ and angle $\theta$.}}
\label{fig.angle}
\end{figure}

Let us come back finally to the initial question of the distribution of
the potential in the region close to a Brownian curve $\mathcal B$ (figures
\ref{fig.potentialu} and \ref{fig.potential2u}).  Its boundary
$\partial \mathcal B$ is a fractal curve without a microscopic scale,
and the irregularities of this curve go down to the infinitesimally
small.  Among all these irregularities, it is natural, from the point
of view of potential theory and of conformal invariance, to
look for those that are locally like ``angles''.  Actually, such a
distribution of angles and the distribution of the associated potential are
invariant under a conformal map.  They are then stable in the
class of all Brownian curves which are obtained by conformal maps of a single realization
of a Brownian curve.

\begin{figure}[t]
\begin{center}
\includegraphics[angle=0,width=.4\linewidth]{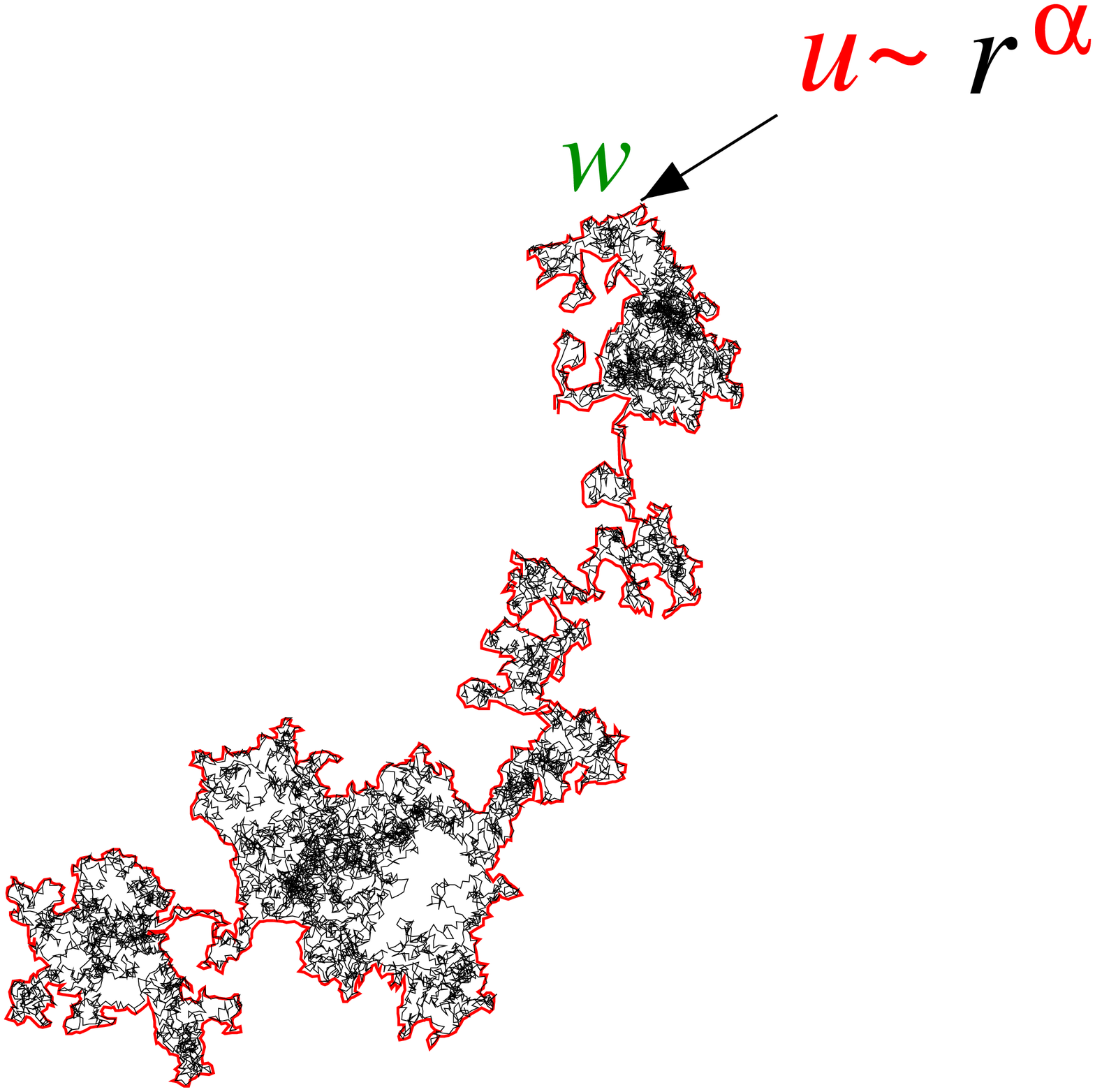}
\end{center}
\caption{{\it Singular behavior in $r^{\alpha}$ of the potential close to a point $w$ of type $\alpha$.}}
\label{fig.brownianpotentialu}
\end{figure}

We can then classify the points $w$ of the boundary $\partial \mathcal
B$ according to the properties of variation of the potential $u(z)$ when a
point $P$ with complex coordinate $z$ approaches $w$ on the boundary.  We say
that a point $w$ is of type $\alpha$ if
\begin{eqnarray}
{u}\left( z \to w\right) \approx r ^{\alpha },
\label{h}
\end{eqnarray}
in the limit where the distance $r=|z-w|\to 0.$ (Figure
\ref{fig.brownianpotentialu}.)

By comparing the property (\ref{h}) to the form (\ref{pa}) of the
potential of an angle, we see that an exponent $\alpha$ corresponds,
from the point of view of the potential, to an {\it equivalent
electrostatic angle} $\theta$ such that
\begin{eqnarray}
\alpha=\frac{\pi}{\theta}.
\label{pah}
\end{eqnarray}

The behavior is as if an angle $\theta=\pi/\alpha$ existed locally on
the boundary.\footnote{The presence of a local singularity exponent $\alpha$
does not necessarily mean that $\theta=\pi/\alpha$ is a geometric angle,
because the surroundings of a point $w$
on a random fractal object will in general screen the  potential, and reduce
the equivalent electrostatic angle with respect to a possible geometric angle.}  The
angular domain being such that $0 \leq \theta
\leq 2\pi$, the domain of the exponents $\alpha$ is $1/2 \leq \alpha <
\infty$, which  is rigorously supported by  a theorem of A. Beurling.  The
domain where $\alpha$ is close to $1/2$ corresponds to $\theta$ close
to $2\pi$, which is a completely open angular sector, and thus
to the presence of an extremely thin {\it needle} on the boundary.
The domain where $\alpha$ is very large corresponds to $\theta$ close
to $0$, thus to a very narrow angular sector, and one then
speaks of a {\it fjord}.

Now, let ${\partial\mathcal B}_{\alpha}$ be the set of points of type
$\alpha$ on the boundary.  To measure the probability of  finding such 
points of type $\alpha$, we introduce the Hausdorff dimension of the set
${\partial\mathcal B}_{\alpha}$,
\begin{eqnarray}
f\left(\alpha\right)={\rm  dim}\left({\partial\mathcal B}_{\alpha}\right).
\label{fa}
\end{eqnarray}

This defines the {\it multifractal spectrum} $f\left(\alpha\right)$ of
the potential distribution.  Such a spectrum is conformally invariant
in two dimensions, because in any conformal map the local
exponents $\alpha=\pi/\theta$ of the potential are themselves
invariant.\footnote{The local definitions of the exponent $\alpha$ and
of $f(\alpha)$ as given in (\ref{h}) and (\ref{fa}) are only
heuristic, since the way of taking limits was not explained.
For any given point $w$ on the boundary of a random fractal object, in
general no stable local exponents $\alpha$ exist, such that they are
obtained by a ``simple limit'' to the point. One then proceeds in
another way. Define the {\it harmonic measure} $\omega(w, r)$ as
the probability that the Brownian motion leaving from any point on the
outer circle (therefore from infinity), touches the frontier
$\partial\mathcal B$ for the first time inside a ball centered at
$w$ and of radius $r$.  (This harmonic measure is similar to the
Brownian representation of the potential $u(P)$, which is just the harmonic
measure of the outer boundary of $\mathcal D$ seen from
a point $P$). Next, we define the set ${\partial\mathcal B}_{\alpha,
\eta}$ of points on the boundary $\partial\mathcal B$, $w=w_{\alpha,
\eta}$, for which there exists a decreasing series of radii $r_j,
j\in \mathbb N$ tending towards $0$, such that $r_j^{\alpha+\eta}\leq
\omega(w, r_j)\leq r_j^{\alpha-\eta}$.  The multifractal spectrum
$f(\alpha)$ is then globally defined as the limit  $\eta \to 0$
of the  Hausdorff dimension of the set ${\partial\mathcal
B}_{\alpha, \eta}$, i.e.,
$$f(\alpha)=\lim_{\eta \to 0}{\rm dim}
\left\{w\, : \exists\
\{r_{j}\to 0,\, j\in \mathbb N\} : r_j^{\alpha+\eta}\leq \omega(w,
r_j)\leq r_j^{\alpha-\eta}\right\}.$$}
\medskip

From a historical point of view, the concept of {\it
multifractality} was introduced by B. Mandelbrot in
1974,\footnote{B. B. Mandelbrot, {\it J. Fluid.\ Mech.} {\bf 62},
pp. 331-358 (1974).} in the context of the phenomenon of turbulence in hydrodynamics,
then by H. Hentschel, I. Procaccia, U. Frisch and
G. Parisi.\footnote{H. G. E.  Hentschel and I. Procaccia, {\it Physica
D} {\bf 8}, pp. 435-444 (1983); U. Frisch and G. Parisi, {\it Proceedings
of the International School of Physics ``Enrico Fermi,''} course
LXXXVIII, edited by M. Ghil (North-Holland, New York, 1985) p. 84.}
It was then further developed at the University of Chicago by T
.C. Halsey {\it et al}.\footnote{T. C. Halsey, M. H. Jensen,
L. P. Kadanoff, I. Procaccia and B. I. Shraiman, {\it Phys. Rev. \ A}
{\bf 33}, pp. 1141-1151 (1986); {\it ibid.} {\bf 34}, 1601 (1986).  }  It
corresponds to the existence of a continuous set of fractal dimensions
$f(\alpha)$, that are functions of a continuum of exponents $\alpha$.

\subsubsection{The Brownian multifractal spectrum}

One of the first properties is that the global Hausdorff dimension of a
multifractal object is always the maximum of its multifractal spectrum.
Thus, for the boundary of a Brownian curve,
\begin{eqnarray}
D_H=\sup_\alpha f(\alpha)=\frac{4}{3},
\label{mandelbrot}
\end{eqnarray}
because of Mandelbrot's conjecture, which we mentioned above.

The complete spectrum $f(\alpha)$ for the Brownian curve was
calculated in 1998 by a method called ``quantum
gravity''.\footnote{B. Duplantier, {\it Phys. Rev. Lett.} {\bf 82}, 
pp. 880-883 (1999), arXiv:cond-mat/9812439.}  One uses a representation of
the same problem on a random surface where the metric fluctuates,
instead of the normal Euclidean plane.  The geometric and
probabilistic laws are largely simplified by the ``quantum''
fluctuations of the metric, and the singular behavior of the Brownian
Dirichlet problem is directly accessible!

Next, one can obtain the multifractal spectrum in the plane
$\mathbb R^2$, thanks to a fundamental relationship between critical
exponents in the plane and on a random surface, a formula known  by the initials
``KPZ,'' discovered originally in 1988 by three Russian physicists,
V. G. Knizhnik, A. M. Polyakov, and A.~B. Zamolodchikov.\footnote{V. G. Knizhnik, A. M.
Polyakov and A.~B.  Zamolodchikov, {\it Mod. Phys. Lett.} {\bf A 3},
pp. 819-826 (1988). See also F. David, {\it Mod. Phys. Lett.} {\bf A 3}, pp. 1651-1656 (1988);
J. Distler and H. Kawai, {\it Nucl. Phys.} {\bf B321}, pp. 509-527 (1988).}  We do not have space here to further
develop this method.\footnote{B. Duplantier, {\it Conformal Fractal
Geometry and Boundary Quantum Gravity}, in {\it Fractal Geometry and
Applications, A Jubilee of Beno\^{\i}t Mandelbrot}, Proceedings of
Symposia in Pure Mathematics, AMS, Vol. {\bf 72}, Part {\bf 2}, edited
by M. L. Lapidus and F. van Frankenhuijsen, pp. 365-482 (2004);
arXiv:math-phys/0303034; see also V. Fateev, A. Zamolodchikov,
Al. Zamolodchikov, {\it Boundary Liouville Field Theory I. Boundary
State and Boundary Two-point Function}, arXiv:hep-th/0001012;
I. K. Kostov, B. Ponsot and D. Serban, {\it Nucl. Phys.} {\bf B683},
pp. 309-362 (2004), arXiv:hep-th/0307189; I. K. Kostov, {\it Nucl. Phys.}
{\bf B689} pp. 3-36 (2004), arXiv:hep-th/0312301; {\it Proceedings of the
Conference ``Lie theory and its applications in physics - 5,''} Varna,
Bulgaria (2003), arXiv:hep-th/0402098, and references therein.}

We find the formula
\begin{eqnarray}
\label{falpha}
 f(\alpha)=\alpha+b-\frac{b\alpha^2}{2\alpha-1},\,\,\,\,\,\,
  b=\frac{25}{12}.
\end{eqnarray}
\begin{figure}[tb]
\begin{center}
\includegraphics[angle=0,width=.7\linewidth]{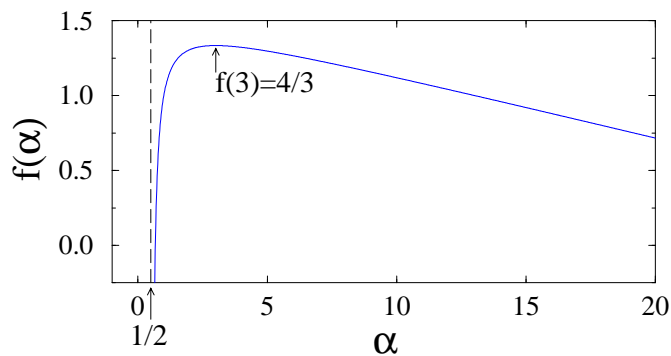}
\end{center}
\caption{{\it Multifractal function $f(\alpha)$ of the Brownian frontier.}}
\label{fig.flambda0}
\end{figure}

This curve is drawn in figure \ref{fig.flambda0}.  The definition domain 
is the half-line $(1/2, +\infty)$.  One verifies that the
maximum of $f$ is  at  $4/3$, in agreement with
Mandelbrot's conjecture (\ref{mandelbrot}) for the fractal dimension
of the boundary.  It corresponds to a value of $\alpha=3$, or to a 
typical electrostatic angle of $\pi/3$.

Moreover, one can calculate by the same method the multifractal
spectrum of the potential close to a self-avoiding random walk,\footnote{B. Duplantier, {\it Phys. Rev. Lett.} {\bf 82},
pp. 880-883 (1999), arXiv:cond-mat/9812439.} and one finds a
spectrum which is identical to that of a Brownian curve, fully confirming
the identity of the Brownian frontier to a self-avoiding walk in the scaling limit.

One also predicts  by this heuristic method that  the
spectra of a Brownian curve and of a critical percolating
cluster  are
identical\footnote{B. Duplantier, {\it Phys. Rev. Lett.} {\bf 82},
pp. 3940-3943 (1999), arXiv:cond-mat/9901008; M. Aizenman, B. Duplantier 
and A. Aharony, {\it Phys. Rev. Lett.} {\bf 83}, pp. 1359-1362 (1999),
arXiv:cond-mat/9901018.}.  It then follows  that both the Brownian frontier and the
external perimeter of a critical percolation cluster coincide with the scaling limit of a
self-avoiding walk, which further extends Mandelbrot's conjecture.

Let us mention that the works of Lawler, Schramm, and Werner contain
also in principle the necessary information to calculate the spectrum
of a Brownian potential.  In a rigorous approach using SLE, these
authors identify the boundary with that of the ${\rm
SLE}_{6}$ process, conjectured also to be an ${\rm SLE}_{8/3}$ and the scaling
limit of a self-avoiding polymer.

This curve $f(\alpha)$, also called the harmonic measure spectrum,
then solves the problem of the potential distribution  close to
a Brownian path in a probabilistic sense, since it gives
the fractal dimension of the set of points where the potential
varies in a specific way, namely as $r^{\alpha}$.

Other values of $b$ in
(\ref{falpha}) ($b=\frac{25-c}{12} \geq 2$, where $c \leq 1$ is the
``central charge'' of the associated conformal theory)  generate the multifractal spectra of the potential or
 harmonic measure of  conformally invariant random
curves in the plane.\footnote{B. Duplantier, {\it Phys. Rev. Lett.} {\bf 84},
pp. 1363-1367 (2000), arXiv:cond-mat/9908314; {\it J. Stat. Phys.} {\bf
110}, pp. 691-738 (2003), arXiv:cond-mat/0207743.}  These are the SLEs describing the boundaries of critical clusters
in two-dimensional statistical models, such as Ising  or
Potts models. For an ${\rm SLE}_{\kappa}$, with
 $0\leq \kappa <\infty$, one simply sets in (\ref{falpha})
\begin{equation}
c=\frac{1}{4}(6-\kappa)\left(6-\frac{16}{\kappa}\right),\;\;\;\;
{b}=1+\frac{1}{8}\left(\kappa+\frac{16}{\kappa}\right).
\label{b'}
\end{equation}

\subsection{Generalized multifractality}
\subsubsection{Logarithmic spirals}
Until now we have considered variations of the potential only.  We can
also study the form of the equipotential lines.  As the potential
follows the properties of conformal invariance of the Brownian curve,
it is now necessary first to determine the geometric forms that are
conserved by such invariance.

These are the {\it logarithmic spirals} that play a particular role in
 potential theory in two dimensions.  One such spiral centered at  
the origin is defined by the logarithmic variation of the polar angle
$\varphi$ as a function of the distance $r$ from the origin:
\begin{eqnarray}
\nonumber
\varphi = \lambda \ln\, r\ , 
\end{eqnarray}
where $\lambda$ is a real positive or negative parameter.

When we apply a conformal map  $\Phi$, around the center it is equivalent to a 
dilation  $r \to |\Phi^{\prime}(0)|\,
r$, composed with a rotation.  The dilation transforms the angle
$\varphi=\lambda \ln\, r$ into $ \lambda \ln\, r + \lambda
|\Phi^{\prime}(0)|$, which thus amounts to a local rotation of the spiral,
whose geometrical shape is thereby locally conserved.

\begin{figure}[tb]
\begin{center}
\includegraphics[angle=0,width=.5\linewidth]{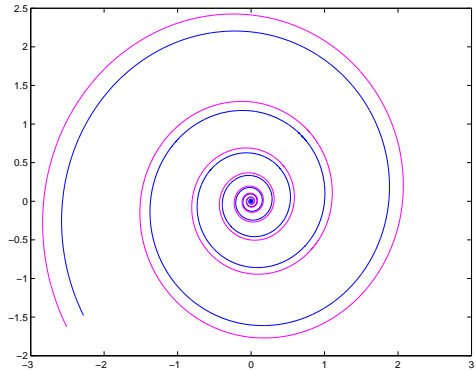}
\end{center}
\caption{{\it Double logarithmic spiral.}}
\label{fig.spiral2}
\end{figure}

In  the potential theory considered here, the Brownian frontier is equipotential by construction.  There
exists a multitude of points where such equipotential boundary will
locally roll onto itself in a double logarithmic spiral, as shown in 
figure \ref{fig.spiral2}.

\subsubsection{Mixed multifractal spectrum}

We come then to Ilia Binder's idea from his 
thesis\footnote{I. A. Binder, {\it Harmonic Measure and Rotation of
Simply Connected Planar Domains}, PhD Thesis, Caltech (1997).} in 1997
defining a generalized multifractality.  One looks for a set
${\partial\mathcal B}_{\alpha,\lambda}$ of points $w$ of the boundary
${\partial\mathcal B}$, where the potential varies like  $r^{\alpha}$, and the boundary spirals 
at a given rate $\lambda$.  These conditions can be heuristically written for a 
point $z$ close to $w$:
\begin{eqnarray}
\nonumber
{u}\left( z \to w\in {\partial\mathcal B}_{\alpha,\lambda }\right)
&\approx& r ^{\alpha },\\
\varphi\left( z \to w\in {\partial\mathcal B}_{\alpha,\lambda
}\right) &\approx& \lambda \ln\, r\ , 
\label{ha}
\end{eqnarray}
in the limit $r=|z-w| \to 0$.  The Hausdorff dimension $f\left(\alpha,
\lambda\right)={\rm dim}\left({\partial\mathcal B}_{\alpha,\lambda
}\right)$ then defines the {\it mixed multifractal spectrum}, which is
conformal invariant because under a conformal map both
$\alpha$ and $\lambda$ are locally invariant.

\begin{figure}[ht]
\begin{center}
\includegraphics[angle=0,width=0.7\linewidth]{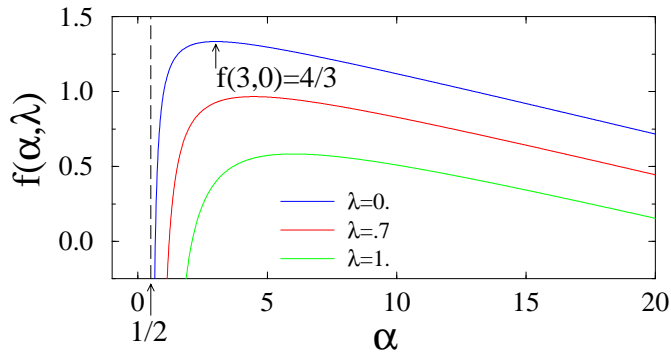}
\end{center}
\caption{{\it 
Universal multifractal spectrum $f(\alpha,\lambda)$ of a Brownian path
for different values of spiral rate $\lambda$. The maximum
$f(3,0)=4/3$ is the Hausdorff dimension of the Brownian frontier.}}
\label{fig.ff0}
\end{figure}

With Ilia Binder, we computed such a mixed spectrum for  a Brownian
motion, by the quantum gravity method.\footnote{B. Duplantier and I. A. Binder, {\it
Phys. Rev. Lett.} {\bf 89}, 264101 (2002); arXiv:cond-mat/0208045.} It satisfies an exact scaling law
\begin{eqnarray}
\label{scalinglaw}
 f(\alpha,\lambda)=(1+\lambda^2) f\left(\frac{\alpha}{1+\lambda^2}\right)-b \lambda^2\ ,
\end{eqnarray}
which gives from (\ref{falpha})
\begin{eqnarray}
f(\alpha,\lambda)=\alpha+b-\frac{b\alpha^2}{2\alpha-1-\lambda^2},\,\,\,\,\,\,
b=\frac{25}{12}.
\label{falphalambda}
\end{eqnarray}
Its domain of definition is ${\alpha}\geq \frac{1}{2}({1+\lambda^2})$,
according to a theorem of  Beurling.  Different spectra are represented
in figure \ref{fig.ff0}.

Since this function does not depend on the sign of $\lambda$, spiral
rotations in  positive and negative directions are equiprobable, as
expected.  One recovers the harmonic spectrum $f(\alpha)$
as the maximum
$$f(\alpha)=f(\alpha,\lambda=0)={\sup}_{\alpha}f(\alpha,\lambda).$$
By symmetry, the most probable situation for a point on the boundary is the
absence of spiral rotation, i.e.,  $\lambda =0$.
\begin{figure}[htb]
\begin{center}
\includegraphics[angle=0,width=0.7\linewidth]{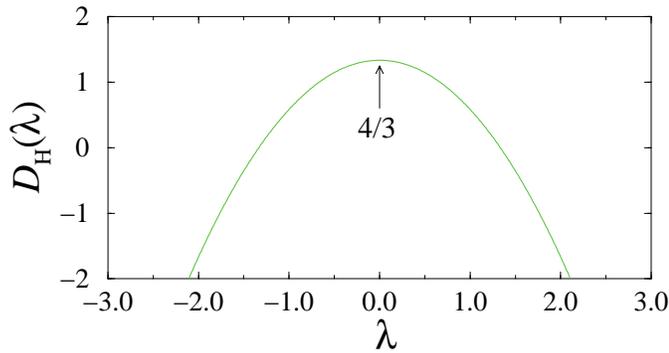}
\end{center}
\caption{\it Fractal dimension $D_H(\lambda)$ of spirals of type $\lambda$ along the Brownian boundary.}
\label{fig.DHlambda}
\end{figure}

One can then also consider only the fractal dimension
$D_H(\lambda)$ of the points on the boundary, which are the tips of 
logarithmic spirals of type $\lambda$. For this, we take the maximum of the
mixed spectrum with respect to the other variable, $\alpha$:

$$D_H(\lambda)={\sup}_{\alpha}f(\alpha,\lambda)=\frac{4}{3}-\frac{3}{4}\lambda^2.$$

This fractal dimension then has the form of a parabola as a function of
$\lambda$, whose maximum is still the global Hausdorff dimension of
the boundary, $D_H=4/3$ (figure \ref{fig.DHlambda}).\medskip

Let us add a few final remarks.\medskip

The quantum gravity calculations can be generalized to the whole class of
conformally invariant curves on the plane, and to Schramm's SLE process.
The spectra are given by the same formulae (\ref{falpha}) and
(\ref{falphalambda}) for different values of the parameter $b$. For the ${\rm SLE}_{\kappa}$ process, one substitutes:
$${b}=1+\frac{1}{8}\left(\kappa+\frac{16}{\kappa}\right)=\frac{1}{2\kappa}\left(2+\frac{\kappa}{2}\right)^2,\;\; \kappa \in \mathbb R^{+}.$$

Lastly, these multifractal results, originally found heuristically in
theoretical physics, can in principle be rigorously proved in the
general probabilistic framework of ${\rm
SLE}_{\kappa}$.\footnote{I. A. Binder and B. Duplantier, in
preparation (2007); see also: D. Beliaev, {\it Harmonic Measure on Random
Fractals}, PhD thesis, KTH, Stockholm, Sweden, 2005; I. Rushkin {\it et al.}, {\it J. Phys. A: Math. Gen.}
{\bf 40}, pp. 2165-2195 (2007).}  The application of this general result to the case of the Brownian
 and percolation cluster frontiers is then obtained by  identifying those boundaries to that of the ${\rm SLE}_6$
process  (thanks to  works by Lawler,
Schramm, and Werner and also by S. Smirnov\footnote{S. Smirnov, {\it
C. R. Acad. Sci. Paris S\'er. I Math.} {\bf 333}, pp. 239-244 (2001).},
and V. Beffara\footnote{V. Beffara, arXiv:math.PR/0204208, {\it Ann. Prob.} {\bf 32} (3) pp. 2606-2629 (2004); arXiv:math.PR/0211322.}), while,
from a rigorous point of view, the similar identification of the
 scaling limit of
a
self-avoiding walk to a ${\rm SLE}_{8/3}$ process, although certainly true,
remains to be proven!\footnote{G. F. Lawler, O. Schramm and W. Werner,
{\it On the Scaling Limit of Planar Self-Avoiding Walk}, in {\it
Fractal Geometry and Applications, A Jubilee of Beno\^{\i}t
Mandelbrot}, Proceedings of Symposia in Pure and Applied Mathematics,
AMS, Vol. {\bf 72}, Part {\bf 2}, edited  by M. L. Lapidus et
F. van Frankenhuijsen, pp.  339-384 (2004), arXiv:math.PR/0204277.}\medskip

Here we pause in 2005 at the end of the path started in 1827 by Robert Brown
with his observations at the microscope, and by Einstein in 1905 with his theory of
Brownian fluctuations.  The new paradigm of stochastic paths could be
today the SLE, or Stochastic Loewner Evolution, generated itself by
Brownian motion on the boundary of a planar domain, and its rather
extraordinary conformal  invariance properties in the Euclidean plane.
This process brought us to the shores of two-dimensional quantum
gravity, where the SLE stochasticity seems to call for fluctuations
of the metric, hence ``quantum gravity.''  In some sense, we are brought back to the work
of Einstein, whose 1916 general
relativity theory explained how gravitation is equivalent to a change of
metric.  Now it is Statistical Mechanics that stands in the breach, let us wish for fruitful
developments!\\

\noindent{\it Acknowledgements:}

I would like to thank John Stachel, Editor of the {\it Complete Papers of Albert Einstein}, for his kind permission to 
reproduce some material of the Editorial Notes.

I thank Bruce H. J. McKellar and Rod W. Home for kindly
providing first-hand information regarding the work of W. Sutherland, and Hector Giacomini for information about the
Besso-Einstein correspondence. I also wish to thank Jean-Pierre Kahane for
stimulating exchanges
about R. Brown and A. Brongniart, Giovanni Gallavotti about G. Cantoni, as well as
Andr\'e Roug\'e about L. Bachelier. I also thank Christian Van den Broeck and Chris Jarzynski
for providing up-to-date information about their latest works, and for interesting discussions about Brownian
motion, the Maxwell demon and the second principle.

I wish to thank Emmanuel Guitter for creating the figures, and Michel
Bauer, Denis Bernard, Vincent Pasquier, and Jean-Louis Sikorav for useful comments.

 Special
 thanks go to Thomas C. Halsey, David Kosower, Robert S. MacKay, Kirone and Shamlal Mallick, and Larry S. Schulman for their careful and critical
 proofreading of this article,  to Thomas C. Halsey for introducing me to the concept of multifractality
 almost 20 years ago, and to Kirone Mallick and Ugo Moschella for their precious help during the completion of this article.

Last, but not least, I wish to express my gratitude to Emily Parks (and to Graziano Vernizzi!)
who carried out the redoubtable task of translating the original text from French into English.\\

\noindent\author{Bertrand {\sc Duplantier}\\
Service de Physique Th\'eorique\\ Orme des Merisiers\\
CEA/Saclay\\ F-91191 Gif-sur-Yvette Cedex, France\\
e-mail: {\bf Bertrand.Duplantier@cea.fr}
}

\end{document}